\documentclass[journal]{new-aiaa}
\pdfoutput=1
\usepackage[utf8]{inputenc}
\usepackage[english]{babel} % generate random text - blindtexst language setting
\usepackage{blindtext}      % generate random text
\usepackage[dvipsnames,table]{xcolor}
\usepackage{graphicx}
\usepackage{amsmath}
\usepackage[version=4]{mhchem}
\usepackage{siunitx}
\usepackage{longtable,tabularx}
\usepackage{comment}
\usepackage{etoolbox}
\usepackage{subfig}
\usepackage{multirow}
\usepackage{tikz}
\usepackage{hyperref}
\usepackage{cleveref}
\usepackage[section]{placeins}
\usepackage{appendix}
\usepackage{newtxtext,newtxmath}

\usepackage[sort&compress]{natbib}

\newcommand{\evec}[1]{\mathbf{#1}}
\renewcommand{\Re}{\mathbb{R}}
\newcommand{\trans}{^{\mathsf{T}}}
\definecolor{Gray}{gray}{0.9}

\newcommand{\ak}[1]{{\color{black}{#1}}}
\title{Control-oriented model reduction for\\ minimizing transient energy growth in shear flows}
\author{Aniketh Kalur\footnote{Graduate Student. AIAA Student Member.} and Maziar~S.~Hemati\footnote{Assistant Professor. AIAA Senior Member.}}
\affil{Aerospace Engineering and Mechanics, University of Minnesota, Minneapolis, MN 55455, USA.}

\begin{document}
 \maketitle
 
 \begin{abstract}
   A linear non-modal mechanism for transient amplification of perturbation energy is known to trigger
   sub-critical transition to turbulence in many shear flows. 
   Feedback control strategies for minimizing this transient energy growth
   can be formulated as convex optimization problems based on linear matrix inequalities.
   Unfortunately, solving the requisite linear matrix inequality problem
   can be computationally prohibitive within the context of
   high-dimensional fluid flows.
   In this work, we investigate the utility of control-oriented reduced-order
   models to facilitate the design of feedback flow control strategies that
   minimize the maximum transient energy growth.
   An output projection onto proper orthogonal decomposition modes is
   used to faithfully capture the system energy.
   Subsequently, a balanced truncation is performed to reduce the state dimension,
   while preserving the system's input-output properties.
   The model reduction and control approaches are studied within the context of
   a linearized channel flow with blowing and suction actuation at the walls.
   Controller synthesis for this linearized channel flow system becomes
   tractable through the use of the proposed control-oriented reduced-order models.
   Further, the resulting controllers are found to reduce
   the maximum transient energy growth compared with more
   conventional linear quadratic optimal control strategies. 
 \end{abstract}
 
\noindent \textit{Keywords:} Reduced-order model; transient energy growth; channel flow; feedback flow control; modal decomposition.

\section*{Nomenclature}

%\noindent(Nomenclature entries should have the units identified)

{\renewcommand\arraystretch{1.0}
\noindent\begin{longtable*}{@{}l @{\quad=\quad} l@{}}
$\Theta_{max}$ & maximum transient energy growth \\
$\Phi_r$   & matrix of $r$ dominant proper orthogonal decomposition modes\\
$(\alpha,\beta)$ & streamwise and spanwise wave number pair \\
$A$  & linear time invariant state matrix of plant \\
$\bar{A}$ & reduced-order state matrix after balanced-truncation\\
$B$  &  linear time invariant input matrix of plant  \\
$\bar{B}$ & reduced-order input matrix after balanced-truncation\\
$E$  &   energy of the full-order plant \\
%%$\tilde{E}$ & reduced-order approximation of full-order energy ($E$)\\
$\bar{E}$ & reduced-order approximation of $E$\\
$K$ & controller feedback gain \\
$Re$ & Reynolds number \\
$T_s$ & matrix of $s$ dominant balanced modes\\
$U_{cl}$ & channel laminar base flow center-line velocity\\
$h$ & channel half-height\\
%%$LMI-ROM$ & linear matrix inequality controller developed on ROM and applied on the full-order plant\\
$n$ & dimension of the full-order plant\\
%$Q$  & state weight matrix used to quantify energy~(E)\\
%$\tilde{Q}$ & low-rank approximation of $Q$\\
$q_u,q_l$ & wall-normal velocities at upper and lower wall respectively\\
$r,s$ & number of proper orthogonal decomposition modes and balanced modes, respectively.\\
%%$U_{cl}$ & center-line velocity in channel flow\\
$\evec{u}$ & input vector\\
%%$W_c$ & controllability Gramian\\
%%$W_o$ & observability Gramian\\
$\evec{x}$ & state vector\\
%$\bar{\evec{x}}$, $\bar{\evec{y}}$ & plant state vector and output vector of the reduced-order models\\
$\evec{z}$ & proper orthogonal decomposition coefficients
\end{longtable*}}

% =======================
% Introduction
% =======================
 \section{Introduction}
 \label{Intro}
 The transition of flows from a laminar to turbulent regime has been extensively studied, and remains a topic of continuing interest. It is well known that turbulent flows exhibit specific detrimental effects on systems, e.g., an increase in skin friction drag in wall-bounded shear flows~\citep{Schmid2001}.
 % Therefore, imparting control to delay or suppress the transition to turbulence has numerous benefits: economic, environmental and engineering related.
% Many studies have been carried out to examine transition mechanisms in shear flows.
 %
It has been observed that transition to turbulence in many shear flows occurs at a Reynolds number~($Re$) much below the critical $Re$ predicted by linear (modal) stability analysis of a steady laminar base flow~\citep{PatelJFM1969,ReddyJFM1993}.
This sub-critical transition is associated with non-modal amplification mechanisms that cause small disturbances to grow before undergoing an eventual modal decay, based on the linear analysis~\citep{Trefethen1993,HenningsonPOF1994,Schmid2007}.
This transient energy growth~(TEG) of disturbances can trigger non-linear instabilities and
lead to bypass transition by driving the flow state beyond
the region of attraction of the laminar equilibrium profile~~\citep{Trefethen1993,Chapman2002}.

% \msh{this paragraph suffers from a ``source dump''.  What is the story?  Currently, this is just a list of citations and what they did (independent of one another).}
% Past studies have investigated strategies for reducing TEG and the transient amplification of
% flow disturbances can have a direct effect on delaying and suppressing transition.
% Consequently,
Numerous investigations have considered the possibility of reducing TEG and delaying transition by employing feedback control techniques.
Excellent reviews of past works can be found in~\citep{Bewley2001,Bagheri2011,KimAnnRev2006}.
Linear quadratic optimal control has been a common approach for such applications.
The full-information linear quadratic regulator~(LQR) has been investigated in various capacities,  
and has been shown to increase transition thresholds within the channel flow system~\citep{BewleyJFM1998,ilakAIAA2008,MartinelliPoF2011}.
% In~\citep{ilakAIAA2008}, the authors demonstrate an increase in the transition threshold of
% a localized perturbation in a channel flow through the use of a full-state feedback
% linear quadratic regulator~(LQR).\msh{there are many other studies that used LQR and
% full-state feedback strategies, dating back well before this paper.  Find better references for this.}
%
% Other investigations have sought to extend the benefits of linear quadratic optimal control strategies
% to practical settings in which full-state information is not available for feedback.
%
Observer-based output-feedback controllers have also been invoked with some success~\citep{BewleyJFM1998,Hogberg2003,Joshi1999}.
In~\citep{Joshi1999}, the authors successfully design a linear quadratic Gaussian~(LQG)
controller on a reduced-order Poiseuille flow system---combining the
optimal LQR law with optimal state estimates from a Kalman filter.
The study in~\citep{BewleyJFM1998} investigates LQG/$\mathcal{H}_2$ and $\mathcal{H}_\infty$
controllers, with particular emphasis on the use of
appropriate transfer function norms to minimize the energy of flow perturbations in a channel flow system.
The study in~\citep{Hogberg2003} further investigates LQG control of the channel flow system,
giving careful attention to the design of the Kalman filter for optimal state estimation for use in observer-based feedback control.
More recently, it has been shown that observer-based feedback strategies may exacerbate
TEG in the controlled system, due to an adverse coupling between the fluid dynamics and
the control system dynamics~\citep{Hemati2018}.
As a potential alternative, static output feedback formulations of the LQR problem have shown some
promise in overcoming the limitations of LQG strategies under certain flow conditions within
the channel flow system~\citep{YaoAIAA2018,YaoAIAA2019}.
Interestingly, although all of these investigations have shown the promise of feedback control for
reducing TEG and delaying transition, it stands that linear quadratic optimal control techniques
and related synthesis approaches do not necessarily minimize---nor even reduce---TEG.
Indeed, in the case of LQR, the objective function to be minimized is the balance of
integrated perturbation energy and input energy---not explicitly the TEG itself.
Thus, the most commonly employed controller synthesis approaches aim to achieve
an objective that does not necessarily address the TEG problem directly.

%~\citep{Bagheri2011,joshiJFM1997,BewleyJFM1998,MartinelliPoF2011,WhidborneIEEE2007}.

% For instance,~\citep{joshiJFM1997} apply linear system theory to two-dimensional channel flow and develop a constant gain with integral compensator to stabilize two-dimensional finite amplitude perturbations. The study in~\citep{BewleyJFM1998} presents a robust control approach using transfer function norms to minimize the energy of flow perturbation using wall-transpiration in a channel flow. In~\citep{KimAnnRev2006}, the authors study various feedback control methods associated with linear systems approach including optimal control methods like linear quadratic regulator~(LQR) and linear quadratic Gaussian~(LQG).
 %
 %Static-output feedback LQR controllers have also shown promise and have improved TEG performance compared to observer based controllers such as LQG~\citep{YaoAIAA2019}.
 %
  
 %

 Given the central role of TEG and non-modal instabilities in the transition process,
 it seems that a more appropriate objective function for transition control would be
 to minimize the maximum transient energy growth~(MTEG), as proposed in~\citep{WhidborneIEEE2007}.
 This objective has direct connections with notions of \textit{worst-case} or
 \textit{optimal perturbations}, which correspond to disturbances
 that result in the maximum TEG~\citep{ButlerPOF1992}.
 The objective also has connections with optimal forcing functions
 determined from input-output analysis~\cite{JovanovicJFM2005},
 though these types of ``persistent disturbances'' will not be considered in the present work.
 %A more targeted control strategy for MTEG minimization was proposed in.
 %
 The minimum-MTEG optimal control problem---which can be specified for either
 full-state or output feedback~\citep{WhidborneIEEE2007}---can be
 posed as a linear matrix inequality~(LMI).
 The LMI constitutes a
 convex optimization problem that can be solved using standard methods,
 such as interior-point methods~\citep{BoydLMI1994}.
 However, the specific LMI problem that arises for MTEG minimization
 is computationally intractable for high-order systems, such as fluid flows;
 the memory requirements associated with existing solution methods scale
 as system order to the sixth power~\citep{MartinelliPoF2011}.
 %
 % Hence, further work is needed to make this approach computationally tractable
 % for the high-dimensional systems of interest in flow control applications.

 Despite the computational challenge, control laws that minimize the MTEG can be
 desirable over linear quadratic optimal control techniques.
 MTEG-minimizing controllers have been found to outperform LQR controllers in reducing TEG within a
 channel flow configuration~\cite{MartinelliPoF2011}.
% however, we note here that the study in~\cite{MartinelliPoF2011}  may not have demonstrated the full promise of their control synthesis approach.
 %
To achieve this, a modal truncation was performed to obtain a reduced-order model~(ROM)
that would make controller synthesis tractable for the linearized channel flow system~\cite{MartinelliPoF2011}.
In spite of the noteworthy performance reported in their study,
 it is well-established that modal truncation methods tend to yield ROMs
 that are poorly suited for controller synthesis~\cite{barbagallo2012,Jones2015}.
 (A demonstration of this point is given in Appendix~\ref{appendix:modal_truncation}.)
 Thus, it may be possible to synthesize MTEG-minimizing controllers with even better performance
 than those reported in~\cite{MartinelliPoF2011} by exploiting an appropriately tailored
 \emph{control-oriented} model reduction strategy.
 Various ROM approaches have been studied in the literature (see~\cite{Taira2017,DawsonAnnRev2017} for
 excellent reviews of such techniques within the context of fluid dynamics).
 However, developing ROMs that facilitate the design of MTEG-minimizing
 controllers requires a tailored approach.
 Control-oriented ROMs within this context must address a dual need:
 (i)~approximate the perturbation energy to concisely and adequately
 describe the energy-based control objective; and
 (ii)~reduce dimensionality to faithfully represent the input-output dynamics, as needed for
 computationally tractable controller synthesis.
 Identifying states that contribute substantially to both the perturbation energy and
 a system's input-output properties (e.g.,~controllability and observability) is a non-trivial task~\cite{Rowley2005}.
 A projection onto proper orthogonal decomposition~(POD) modes is known to be optimal
 for capturing the energy of a given signal;
 however, projection-based model reduction based on POD modes
 often fails to capture a system's input-output dynamics,
 making such models poor candidates for controller synthesis~\cite{ilakPOF2008a,DawsonAnnRev2017}.
 In contrast, balanced truncation can be performed to obtain ROMs that
 retain a system's input-output properties~\cite{Moore1981,Antoulas2006}.
 The balanced truncation procedure can be shown to be equivalent to
 a Petrov-Galerkin ROM based on a projection onto a subspace spanned by
 a reduced set of balanced modes.
 For high-dimensional fluid flow systems, the balanced POD~(BPOD) and related methods
 provide efficient tools for computing these balanced modes~\citep{Rowley2005,Willcox2002,Antoulas2006}.
Alternative control-oriented model reduction techniques can also be devised to retain a system's input-output properties,
 e.g.,~using ideas from robust control~\citep{Jones2015,Antoulas2006}.
 %
 % BPOD
 % used to project the system onto most energetic modes of the system. POD and other model reduction methods like Galerkin projections, eigensystem realization algorithm~(ERA), balanced truncation, Balanced proper orthogonal decomposition~(BPOD are summarized in~\citep{DawsonAnnRev2017}. In \citep{Rowley2005}, BPOD was introduced to approximate controllability and observability Gramians for systems with large number of outputs, the author proposes to do this using output projection.
 % A good model for simulation need not be appropriate for feedback control~\citep{Jones2015}.
 %Similarly one ROM can be more favorable to control design than the other, our results validate~(although not theoretically) this statement. We also perform a case study using modal truncation as a limiting example in the appendix~\ref{appendix:modal_truncation}, to illustrate importance of precise energy approximations and appropriate mode selection while designing ROMs for fluid flow applications. %None of the above mention 
% Most ROM methods do not explicitly address the challenge of retaining the most controllable, observable modes along with capturing the physical energy of the system.

 The contribution of the present study is to investigate the utility of
 control-oriented model reduction for designing MTEG-minimizing controllers.
 The study focuses on full-information control of a linearized channel flow system,
 but the lessons learned can be generalized to other flows and control architectures as well.
 %
% In this work we develop ROMs that facilitate energy based controller synthesis, specifically, controllers for reducing the MTEG; we also investigate controller performance on the linearized channel flow system. %The straightforward application of the above mentioned ROM methods is not ideal, because the LMI formulation minimizes the maximum TEG of the full-order model~(FOM).
 %
 The ROMs in this work use POD and balanced truncation in conjunction. %--- with the aim to facilitate the development of controller that target reduction in maximum TEG, especially in high dimensional systems.
 First, we perform an output-projection of the full-state onto a set of dominant POD modes.
 Subsequently, a balanced truncation is performed to reduce the state-dimension,
 while retaining the most controllable and observable modes that contribute to the input-output dynamics.
 As we will see, this dual approach results in control-oriented ROMs that can yield effective MTEG-minimizing controllers.
 The resulting models are able to represent the perturbation energy in terms of a small number of POD modes,
 thus providing a convenient approximation of the objective function.
 Further, state-dimension is reduced to a computationally tractable level, while retaining the
 most controllable and observable states that are critical for effective controller design.

 The organization of the paper is as follows:
 Section~\ref{sec:Control_design} summarizes preliminaries on TEG and the relevant controller synthesis strategies,
 including the synthesis of LQR controllers and the LMI-based synthesis of MTEG-minimizing controllers.
 The control-oriented model reduction approach is then introduced in Section~\ref{sec:COROM}.
 Section~\ref{Results} presents all of the results, which are based on the linearized channel
 flow system described in Section~\ref{sec:model}.
 ROMs are analyzed in an open-loop context in Section~\ref{sec:Fresp_perf}.
 Controller performance of ROM-based controller designs is evaluated in Section~\ref{sec:control_perf},
 which also includes a comparative analysis of the flow response to LQR and MTEG-minimizing control laws.
 Finally, Section~\ref{sec:conclusion} summarizes the findings and contributions of this study.

\section{Reduced-order models and controller synthesis}
\label{control}

Consider the linearized dynamics of flow perturbations $\evec{x}(t)$ about a steady laminar base-flow,
\begin{equation}
  \dot{\evec{x}}(t) = A\evec{x}(t) + B\evec{u}(t) \label{eq:linsys1}
\end{equation}
where $\evec{x}$ is a vector of $n$ state variables,
$\evec{u}$ is a vector of $p$ control inputs, and $t$ is time.
%
%\msh{will need to update for consistency with $\Re$ and $\mathbb{C}$ in various sections.}
%
Specific details on arriving at such a representation for a channel flow
setup will be described in Section~\ref{sec:model}.
The free-response of the system to an initial perturbation $\evec{x}(t_o)=\evec{x}_o$ at time $t_o$ is given
in terms of the matrix exponential $\evec{x}(t)=\mathrm{e}^{A(t-t_o)}\evec{x}_o$.
The perturbation energy for this system is defined as $E(t):=\evec{x}\trans(t) Q\evec{x}(t)$,
where $Q=Q\trans>0$.
In this study, we are primarily interested in the maximum transient energy growth~(MTEG),
\begin{align}
\Theta_{\text{max}}=\max _{t \geq t_0}\max _{E(t_0)  \neq 0} \frac{E(t)}{E(t_0)},
\end{align}
which corresponds to the peak energy over all disturbances and all time.
The flow perturbation $\evec{x}_o^{\text{opt}}$ that results in this MTEG is called the \emph{worst-case disturbance} or the \emph{optimal perturbation}~\cite{ButlerPOF1992}.
% , the maximum peak in the energy $E(t)$ for all time is called the maximum transient energy growth~(MTEG). The MTEG is represented by $\Theta_{max}$ and is mathematically defined as follows:

%\msh{move to intro, or remove (repetitive with statements already made.}
%
%Linear quadratic regulation~(LQR) has been a commonly employed strategy for transition control;
%however, LQR controllers do not explicitly minimize---nor necessarily even reduce---the MTEG.
%
%MTEG-minimizing controllers have been proposed to directly target
%the non-modal mechanisms that can give rise to sub-critical transition~\cite{WhidborneIEEE2007}.
%
%
% The LQR and MTEG-minimizing controller synthesis problems are then described in further detail in Section~\ref{sec:Control_design}.
% 
%\mshnote{updated. reshuffled order to output projection setup -> POD modes for energy objectives -> Bal Trunc for control-oriented design. Removed redundant and extraneous equations.}

%\msh{Need to go through this section to clean up missing references and polish language.  Also check all equations for correctness.}

 %, particularly for LMI controllers.%
	%The LMI framework for TEG reduction is developed to minimize the upper bound on $\Theta_{max}$.
	%As stated previously, it is essential to approximate the physical energy of the system to obtain a feasible LMI controller.

%\subsection{Optimal Controller Synthesis}
\label{sec:Control_design}
Owing to the role of transient energy growth in sub-critical transition,
our aim here will be to reduce the MTEG using full-information
feedback control laws of the form $\evec{u}(t)=-K\evec{x}(t)$,
where $K$ is a gain matrix determined by an appropriate synthesis strategy.
In this study, two controller synthesis approaches will be considered: (i)~linear quadratic regulation~(LQR), and (ii)~MTEG minimization via LMI-based synthesis.
% \subsubsection{Linear Quadratic Regulator}
% \label{LQR}
%
LQR is an optimal control
technique that has been commonly employed in a variety of flow control applications,
including transition control.
LQR controllers are designed to minimize the integrated balance between
perturbation energy and control effort; i.e.,
%The LQR minimizes the following cost function
\begin{align}
    \min_{\mathbf{u}(t)} J= \int_{0}^{\infty}(\mathbf{x}^TQ\mathbf{x}+\mathbf{u}^TR\mathbf{u})dt
    \label{eq:LQR}
\end{align}
subject to the linear dynamic constraint in~\eqref{eq:linsys1} and $R>0$.
The feedback control law that minimizes this cost function is given by $\evec{u}=-K\evec{x}$, where the control gain~$K=R^{-1}B^TP$ and $P=P\trans>0$ is determined from the algebraic Riccati equation,
\begin{align}
    A^TP+ PA - PBR^{-1}B^TP+Q = 0.
\end{align}
LQR is a widely used optimal control strategy owing to properties like guaranteed stability margins and robustness to parameter variations. However, it must be noted that
the LQR formulation does not necessarily guarantee reductions in MTEG,
let alone its minimization.
In principle, a controller that minimizes the balance of integrated energies in~\eqref{eq:LQR}
could still yield large energy peaks.
%
%In order to minimize energy peaks,
%it is essential to use controllers that target minimization of the MTEG.

% \subsubsection{LMI synthesis for MTEG minimization}
% \label{LMI}

A state feedback control gain for minimizing the MTEG in a closed-loop system can be found from the solution of a linear matrix inequality~(LMI), as shown in~\citep{WhidborneIEEE2007}.
The solution approach leverages the relationship between MTEG and a system's condition number.
From this, a control law can be devised to minimize the condition number of the closed-loop system in order to minimize the associated $\Theta_{\text{max}}$.
In what follows, it is assumed that an appropriate transformation has been made so that $Q=I$.
A feedback control law that minimizes the 
upper bound $\Theta_u$ of the MTEG 
can be determined from the solution to the LMI generalized eigenvalue problem~\citep{WhidborneIEEE2007}:
\begin{equation}
    \begin{aligned}
        \text{min}~\gamma & \\
        \text{subject to}\qquad & I\leq~P~\leq~\gamma~I\\ 
        & P=P^T>0\\
        & AP+PA^{T}+BY+Y^{T}B^{T}~<0.
    \end{aligned}
    \label{eq:lmi_control}
  \end{equation}
  This LMI problem can be solved using standard convex optimization methods, such as those available in the cvx software package for Matlab~\citep{cvx}.
Here, $\gamma$ upper bounds $\Theta_u$.
Thus, minimizing $\gamma$ also minimizes $\Theta_{u}$,
which consequently minimizes the upper bound on $\Theta_{\text{max}}$.
The resulting full-state feedback control law is given by $\mathbf{u} = -YP^{-1}\evec{x}$, \ak{where $Y$ and $P$ are determined from~\eqref{eq:lmi_control}}.

Standard solution techniques for the LMI problem in~\eqref{eq:lmi_control}
are presented in~\citep{BoydLMI1994}.
However, available algorithms are computationally demanding,  with memory requirements
scaling as $\mathcal{O}(n^6)$~\citep{MartinelliPoF2011},
making controller synthesis intractable for the high-dimensional systems
of interest in flow control.
%This curse of dimensionality is a prohibitive property of the LMI formulation.
%
% \msh{move to intro? remove?}
% In principle, MTEG-minimizing controllers can be computed by solving a linear matrix inequality~(LMI) problem; however, the requisite LMI-problem can be intractable for the high-dimensional problems encountered in fluid flow control applications, in which the state-dimension $n$ is large: computational demand scales as $\mathcal{O}(n^6)$~\cite{MartinelliPoF2011}.
%
The aim of the present study is to investigate the role of reduced-order models~(ROMs)
for facilitating controller synthesis by making the solution of this LMI problem tractable.
Further, constructing reliable control-oriented ROMs
requires consideration of the specific control objective.
Since the LQR and MTEG-minimizing controllers to be studied here are
based on energy-based control objectives,
it stands that approximating the energy $E(t)$ will be
an important consideration for reduced-order modeling.
Section~\ref{sec:COROM} presents a method for obtaining control-oriented ROMs for LQR and MTEG-minimizing controller synthesis.
%
% In Section~\ref{sec:COROM} that follows, we will present a control-oriented
% model-reduction framework that can be used to overcome this computational hurdle,
% while still yielding feedback controllers that achieve
% the original energy-based control objectives.
%
% In section~\ref{sec:COROM}, we discuss an approach for generating reduced-order models that can make these computations tractable for many applications, while still yielding controllers that perform well on the full-order system.
%It must be noted, for the convenience of notation we will continue to represent the states in transformed co-ordinates (where $Q=I$) without the \emph{hat} symbol in the following sections.

%\subsubsection{Controller lifting}
%\msh{is this section needed?  can one not simply implement the reduced order controller directly, via suitable projection of the FOM onto the ROM basis?}
%
%\msh{would it be better to describe how to project ``down'' in order to apply the control? The ``smaller'' gain matrix is more convenient for ctual controller implementation.  A ``large'' gain matrix from lifting is not desirable for large-scale systems.}
%

\subsection{Control-oriented reduced-order models}
\label{sec:COROM}
	Procedures for generating ROMs often rely upon truncating state variables from the system description,
	with the details of the truncation procedure varying with the given application.
	A ROM intended to facilitate controller synthesis should faithfully capture a system's
	input-output dynamics, motivating the use of techniques such as balanced truncation~\citep{Moore1981}.
	On the other hand, within the context of the energy-based control objectives considered in the present study,
	it will also be important for the ROM to faithfully reproduce the system energy $E(t)$.
	POD modes have been shown to optimally capture the energy of a given signal~\cite{Taira2017}. Here, we will combine these two approaches in order to realize ROMs that can facilitate
	the synthesis of controllers with energy-based control objectives.
        Similar approaches have been described in the
        literature~\citep{ilakPOF2008a,DawsonAnnRev2017},
        but have not been leveraged for LMI-based synthesis of MTEG-minimizing controllers.
        In Appendix~\ref{appendix:modal_truncation}, we demonstrate the need for
        tailored ROMs that favor control design.
        % 
	% In what follows, we will assume---without loss of generality---that an appropriate similarity transformation
        % has been made such that the weighting matrix in the definition of
        % energy is~$Q=\mathbf{I}$.
        % %	
        % Indeed, the state transformation $\hat{\evec{x}} = Q^{1/2}\evec{x}$ can always be performed to satisfy this condition. 	
	
	The basic idea underlying the ROMs proposed here is to first append
        an output equation to~\eqref{eq:linsys1} to keep track of the state
        response in terms of $r<n$ POD modes,
	\begin{align}
	\begin{split}
	\label{OP_SYS}
	\dot{\evec{x}} &= A\evec{x} + B\evec{u}\\
	\evec{z} &= \Phi_r^T \evec{x}.
	\end{split}
	\end{align}
	Here, $\evec{z}\in\Re^r$ is a vector of POD coefficients
        and $\Phi_r\in\Re^{n\times r}$ is a matrix whose columns
        are the $r$ dominant POD modes, defined with respect to
        an appropriate inner-product, such
        that $E(t)\approx\tilde{E}(t)=\evec{z}^T(t)\evec{z}(t)$.
	Thus, rather than requiring full-state information to compute energy,
        only the $r$-dimensional
	output of POD coefficients is needed to approximate the energy response.
	This approach is sometimes described as an output projection onto POD modes,
	since the output could also be viewed as a projection of the
        full-state output onto the dominant POD modes~\citep{Rowley2005}. \ak{It should be noted that output projection in~\eqref{OP_SYS} does not change the state dimension $n$ of the model.}
	%
	% Indeed, it is well-established that the optimal subspace for
        % capturing the energy in $\evec{x}(t)$ is
        % achieved using the orthogonal
	% projection $P_r:=\Phi_r \Phi_r^T$ onto an r-dimensional subspace
        % spanned by the dominant POD modes~\citep{Rowley2005}.
	%
	% Noting that POD modes are orthogonal ($\Phi_r^T\Phi_r=I$),
	% %
	% the system energy can be approximated purely in terms of the POD coefficients,
	% as $E(t)\approx\tilde{E}(t)=\evec{z}^T(t)\evec{z}(t)$.
	%

	%
	% Further, since the energy can now be approximated in terms of the system output,
	% the state dimension can be reduced by alternative methods to obtain a control-oriented
	% model that maintains the input-output dynamics needed to facilitate controller synthesis.
	%
	%

	To find the POD modes, we use the snapshot POD approach~\citep{Taira2017}.
        Through heuristics, we found that POD modes obtained from snapshots of
        the system's impulse response matrix provided an adequate basis for the projection.
        %other methods, e.g., using the standard basis as initial conditions to obtain snapshots for the POD procedure.
        Therefore, to compute the POD modes ($\Phi_r$), we obtain the impulse response $G_i(t)$ from the $i^\text{th}$ input to the full-state output for each of the $p$ inputs.
	In generating this data, we first transform the state $\evec{x}$ into a coordinate system in which $Q=I$.
        % , such that the POD modes will capture the physical energy $E(t) = x^T(t)Q
	%x(t)$ of interest for controller synthesis.
	%
	These impulse response data are then collected and arranged within a single snapshot matrix $H:=\left[\begin{array}{cccc}G_1(t)&G_2(t)&\cdots&G_{p}(t)\end{array}\right]$.  The POD modes are then computed by taking
	the singular value decomposition~(SVD) of $H=U\Sigma V\trans$.
	The dominant POD modes are then determined as the leading $r\le n$ left singular vectors---i.e.,~$\Phi_r=\left[\evec{u}_1,\evec{u}_2,\dots,\evec{u}_r\right]$, where $\evec{u}_i$ are columns $i=1,\dots,r$ of $U$.
	We note that the POD modes could also be computed analytically, though the data-driven method of snapshots is most commonly used in practice~\cite{Taira2017}.
	Thus, important parameters for generating the impulse response data will be the length of the simulation~($t_f$)
	and sampling interval~($\delta_t$).
	We address the specifics of these parameters for ROM construction in Section~\ref{sec:Fresp_perf}. %\msh{update section ref.}
	
	Up to this point, we have shown that the system energy can be approximated
        using just $r$ POD modes, with the associated
        POD coefficients tracked as the system output $\evec{z}$.
	However, in order to make the LMI problem for MTEG-minimizing
        controller synthesis computationally tractable,
	we still need to reduce the state dimension of the system.
	In the study here, we will perform a balanced truncation of~\eqref{OP_SYS} and retain only the dominant $s$ balanced modes.
        Such a truncation will ensure that
	the input-output dynamics are preserved, making
        the resulting $s$-dimensional ROM suitable for controller synthesis~\cite{Antoulas2006}.
	In order to perform a balanced truncation, we first apply a balancing
        transformation $\evec{x}=T\bar{\evec{x}}$.
        In the \emph{balanced coordinates} $\bar{\evec{x}}$, the controllability
        Gramian $\bar{W}_c$ and the observability Gramian $\bar{W}_o$ are equal and diagonal.
        To determine the balanced realization, one must first compute
        the controllability Gramian $W_c$ and observability Gramian $W_o$
        from the Lyapunov equations, $AW_c+W_cA^T-BB^T=0$ and
        $W_oA+A^TW_o+C^TC=0$, respectively.
        Once the Gramians are computed, the balancing transformation
        $\evec{x}=T\bar{\evec{x}}$ can be found in three steps~\citep{Laub1987}:
        (i)~compute the lower triangular Cholesky factorizations of
        $W_o = L_oL_o^T$ and $W_c = L_cL_c^T$;
        (ii)~compute the SVD of the products of Cholesky
        factors $L_o^TL_c = \bar{U}S\bar{V}^T$; 
        and (iii)~form the balancing transformation
        $T = L_r\bar{V}S^{\frac{1}{2}}$ and $T^{-1} = S^{\frac{1}{2}}\bar{U}^TL_o^T$.

         In balanced coordinates, each mode's relative contribution to the
         input-output dynamics of the system is clear.
         %; the controllability and observability
%         Gramians are diagonal and equal to the system's Hankel Singular Values~(HSVs),
 %        denoted by $\sigma_i$~\cite{Antoulas2006}. 
	%
         We have $\bar{W}_c=\bar{W}_o=\mathrm{diag}(\bar{\sigma}_1,\dots,\bar{\sigma}_n)$,
         where $\bar{\sigma}_1\ge\bar{\sigma}_2\ge\dots\ge\bar{\sigma}_n> 0$ are
         the system's Hankel Singular Values~(HSVs).
         Since the HSVs relay information about relative contributions
         to the input-output dynamics,
%	\msh{describe how to get the transformation $T$ and/or $T^{-1}$\dots}
	%
         truncating balanced modes with ``small'' HSVs provides a
         convenient strategy for reducing state-dimension while
         preserving the input-output dynamics~\cite{Antoulas2006}.
	%
	% Further, doing so ensures that arbitrary small perturbations do not lead to drastic changes in controllable
	% and observable sub-spaces.
	%
	% An order~($s$), depending on relative contribution of each HSV can be chosen as the order of truncation.
	%
	Upon performing the balanced truncation, the dominant $s$ balanced modes of the system are retained, and the $n-s$ modes corresponding to the lowest HSVs are truncated.
	The resulting state-space realization is given by,
	\begin{align}
	\dot{\bar{\evec{x}}}_s &= \bar{A}_s\bar{\evec{x}}_s + \bar{B}_s\evec{u}\\
	\bar{\evec{z}}_s & = \bar{C}_s\bar{\evec{x}}_s
	\end{align}
	where $\bar{\evec{x}}_s \in R^{s}$ is the reduced state vector and
        the system matrices are defined as $\bar{A}_s:=T^{-1}_sAT_s$,
        $\bar{B}_s:=T^{-1}_sB$, $\bar{C}_s:=\Phi_r^TT_s$.
        Here, $T_s\in\Re^{n\times s}$ denotes a matrix whose $s$ columns are
        the leading $s$ columns of $T$, and $T^{-1}_s\in\Re^{s\times n}$ denotes
        a matrix whose $s$ rows are the leading $s$ rows of $T^{-1}$.
	%  
	%
	% Further, The ROM procedure also provides access to the most controllable and observable modes.
	% To access the most controllable and observable modes the system is internally balanced .i.e., controllability~($W_c$) and observability~($W_o$) Gramians are diagonal and equal to the Hankel Singular Values~(HSV) which are denoted by $\bar{\sigma}_i$.
	% A coordinate transformation $T$ is used to transform the state $\evec{x}$ into an alternate coordinate system which is \textit{internally balanced} i.e., $W_o=W_c$ such that
	% \begin{align*}
	% W_c= W_o =\begin{bmatrix}
	% \bar{\sigma}_1 & & & &\\
	% & \bar{\sigma}_2 & & &\\
	% & & &\ddots  & &\\
	% & & & & \bar{\sigma}_n
	% \end{bmatrix}
	% \end{align*}
	Finally, the perturbation energy $E(t)$ can now be approximated
        from this control-oriented ROM as,
	\begin{align}
	%\bar{E} & = \bar{\evec{z}}^T\bar{\evec{z}}\\
	\bar{E} &= \bar{\evec{x}}_s^T \bar{Q} \bar{\evec{x}}_s,
	\label{Ebar}
	\end{align}
	where $\bar{Q} = T^{T}_s\Phi_r\Phi_r^T T_s\in R^{s \times s}$.
	%
	% Recall that the POD modes in $\Phi_r$ are defined with respect to
	% the weighting matrix $Q$, which was achieved in the impulse
	% response data generation step by transforming the state into
	% a coordinate system in which $Q=I$.
	%
	In this study, we take $s=r$,
	and so will solely report $r$ as the dimension of the reduced-order model.

        In order to assess feedback control performance
        with respect to the MTEG that can be experienced in closed-loop,
        additional care must be taken when computing the optimal
        disturbances in this study.
        \ak{Since MTEG analysis is to be performed on the full order system, feedback controllers designed using the ROMs described here
        are first ``lifted'' to a control gain with compatible dimensions as
        the full order model.  In this way, the optimal disturbance for the full order closed-loop system can be computed directly.}
        % \ak{In particular, feedback control gains designed on the ROMs of dimension $s$; these controllers have to be `lifted' i.e., multiplied by an appropriate transformation matrix to make the computed gain compatible with the full order model (of dimension $n$) when conducting MTEG analysis.}
        %In particular, feedback controllers designed on the ROMs described here
        %are first ``lifted'' to a control gain with compatible dimensions as
        %the full order model when conducting the MTEG analysis.
        % , in order to the ensure that the MTEG calculated
        % will be the actual MTEG for the system.
        %
        Of course, to actually implement the resulting feedback controllers in practice,
        one would simply use the reduced-order gain matrix to take
        advantage of the reduced computational complexity at run-time.

	The procedure outlined here will enable a means of obtaining
        control-oriented reduced-order models that can be used to
        synthesize feedback controllers, especially those aimed at
        achieving certain energy-based objectives---as with
        MTEG-minimization.
	We emphasize that the control-oriented model reduction
        procedure described here and other similar approaches
        have been explored in a number of
        previous studies~\cite{Moore1981,Rowley2005,Willcox2002,ilakAIAA2008,ilakPOF2008a}. %\msh{Moore,Rowley2005,Willcox}.
	 The contribution of the present study is to investigate the
        applicability of such model reduction techniques
	for the purposes of minimizing the MTEG using the LMI-based synthesis
        procedure proposed in~\cite{WhidborneIEEE2007,MartinelliPoF2011}.

        % and described in Section~\ref{sec:Control_design}.
        % \msh{Whidborne, Martinelli}.
	% The ROMs developed in this section enable us to obtain an approximation of the reduced-order energy as well as an excellent reduced-order model that facilitates controller design.
%	We progress to discuss the channel flow model, controller synthesis, and performance of ROMs and the controllers developed using the ROMs proposed here.

% \msh{find a place for this}
% The ROM method enables us to obtain reduced-order control gains using the LMI and LQR control synthesis. Although these control gain can be directly used on the ROM, our aim is to study how these reduced-order controllers perform on the FOM when subjected to an optimal perturbation.

% To use the reduced-order control gains~($\bar{K}$) obtained from the ROMs on the FOM, we use inverse projections to ``lift'' the control gains to the size of the full-order control gain~($K$).
% %Developing the LMI controller on a well tuned ROM will always ensure a stabilizing controller that minimizes the MTEG when compared to the uncontrolled system. 
% Since the LMI frameworks objective is to minimize the MTEG in the system, our goal is to compare the aspect of TEG reduction with another optimal control method- the LQR.
        
%\ak{This section dedicated to ROM. Controller synthesis moved to after channel flow model}

\section{Results}
\label{Results}
% \msh{section intro/summary}
%\msh{Re-write this intro paragraph}

Next, the control-oriented reduced-order modeling approach introduced in
the previous section is investigated within the context of
controlling TEG in a linearized channel flow.
The channel flow system is presented in Section~\ref{sec:model}.
In Section~\ref{sec:Fresp_perf}, the frequency-responses of various ROMs are
analyzed to assess open-loop modeling performance.
Finally, ROM-based controller performance is investigated in Section~\ref{sec:control_perf}, by comparing the MTEG resulting in the closed-loop system response.
The associated flow responses are also studied and discussed.
%
%In this section,  we use energy minimizing control methods developed on the ROM method proposed in this paper.  We illustrate successful minimization of TEG on the linearized channel flow system. First, we introduce the channel flow model in~\ref{sec:model}. Following the preliminary discussions, in section~\ref{sec:Fresp_perf} we discuss the performance of the ROM in comparison to the FOM. In section~\ref{control_perf} we discuss the results obtained by applying LMI based control methods on the ROMs developed in this paper. Additionally, we investigate the performance of the LMI controllers and contrast them to that of the LQR controllers, both of which are developed on the ROMs proposed here.% the following section details the LMI and LQR controller synthesis methods. The latter sections 
%In this section, we introduce the reader to preliminary details on applying the ROM based controllers on the FOM, nomenclature and different flow conditions the controllers have been developed and implemented on. Following the preliminary discussions, in section~\ref{Fresp_perf} we discuss the performance of the ROM in comparison to the FOM. In section~\ref{control_perf} we discuss the results obtained by applying LMI based control methods on the ROMs developed in this paper. Additionally, we investigate the performance of the LMI controllers and contrast them to that of the LQR controllers, both of which are developed on the ROMs proposed here.

\subsection{Channel Flow System}
\label{sec:model}

\begin{figure}[h!]
\centering
%\subfloat[A schematic of channel flow coordinate system ]{\includegraphics[width=0.45\textwidth]{CF_fig/Channel_Flow_Ver2.png}
%\label{fig:CF:Fig1}}
\includegraphics[width=0.5\textwidth]{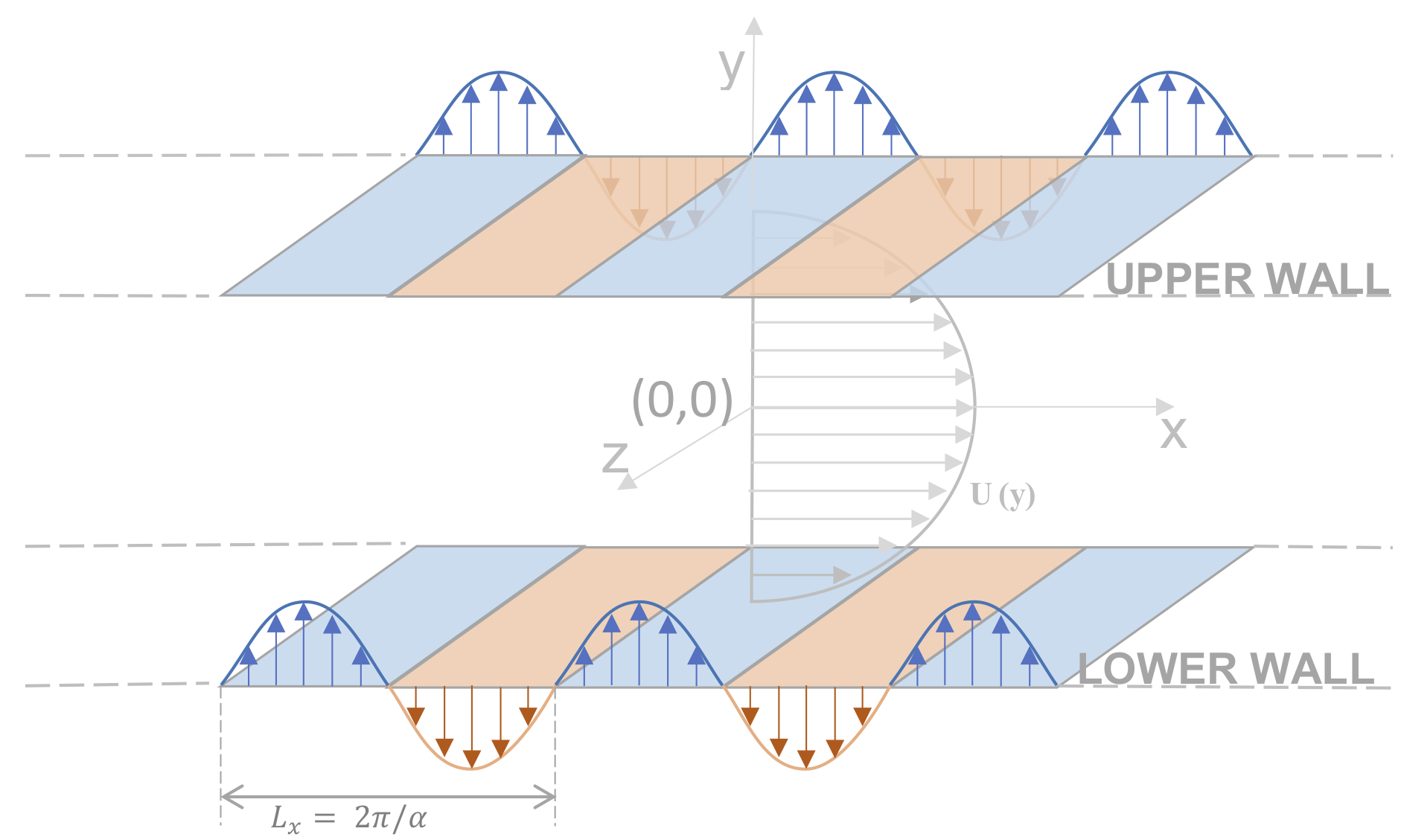}
%\label{fig:CF_Fig2}
\caption{Channel flow setup.}%schematic and visualization for system with streamwise waves ,i.e., $\beta = 0$ }
\label{fig:CF_Fig}
\end{figure}

%\msh{Are all of the details in this section necessary?  Would some be more suitable for the appendix?  It seems we could just report equation \eqref{OSS_Eqn}, and defer everything upstream of it to a citation and maybe a 1--2 sentence summary.}
Consider the flow between two infinitely long parallel plates
separated by a distance of $2h$ (see Figure~\ref{fig:CF_Fig}).
We are interested in the linear evolution of perturbations about a steady laminar
base flow $U(y) = U_{cl} \big(1-\frac{y^2}{h^2}\big)$.
% with a parabolic streamwise profile $U(y) = U_{cl} \big(1-\frac{y^2}{h^2}\big)$ and its evolution is governed by the Navier-Stokes equations.
Thus, we linearize the Navier-Stokes equations about this base flow, and non-dimensionalize
based on the centerline velocity $U_{cl}$ and the channel half-height $h$.
%, pressure by density, and centerline velocity i.e., $\rho U_{cl}^2$. The Reynolds number ($Re$) is ratio of inertia to viscocity($\mu$) and is given as $Re= 
%\frac{\rho U_{cl}h}{\mu}$. The $Re$ used in this study ranges from $1000$ to $10,000$ and will be explicitly stated in the relevant discussions. %The resulting linearized equations describing the evolution of perturbations are given by:
%
%, it is well known that the Navier-Stokes and the continuity equation govern the evolution of the perturbation.
Taking the divergence of the continuity equation and using the $y$-momentum equation to eliminate the pressure term, we obtain an equivalent representation for the evolution equations,
now in terms of the wall-normal velocity~($v$) and the wall-normal vorticity~($\eta$).
A Fourier transform is applied in the streamwise and spanwise directions to obtain 
the well-known Orr-Sommerfeld and Squire equations~\citep{Schmid2001},

\begin{align}
    \begin{split}
        \label{OSS_Eqn}
        \Delta\dot{\tilde{v}} &= \bigg[ i\alpha U \Delta + i \alpha U^{''} +  \frac{\Delta \Delta}{Re} \bigg]\tilde{v},\\
        \dot{\tilde{\eta}} &= \bigg[-i \beta U^{'}\bigg]\tilde{v}+\bigg[-i\alpha U+\frac{\Delta}{Re} \bigg]\tilde{\eta},
    \end{split}
\end{align}
where $\Delta = \frac{\delta}{\delta y^2}-\kappa^2$.
Here, $\kappa^2 = \alpha^2 + \beta^2$, and $(\alpha,\beta)$ denotes
the pair of streamwise and spanwise wavenumbers, respectively.
The no-slip boundary conditions in velocity-vorticity form are $\tilde{v}(y = \pm1) = \frac{\partial \tilde{v}}{\partial y}\mid_{({y=\pm1})} = \tilde{\eta} = 0$.
The wall-normal direction is discretized using Chebyshev
polynomials of the first kind~\citep{JPBoyd2000},
allowing $\tilde{v}$ and $\tilde{\eta}$ to be approximated at $N$ discrete collocation points.
At each point, the approximation uses Chebyshev basis functions $\Gamma_i$
and the respective unknown coefficients $a_{i}$.
The resulting equations of motion can be expressed in
the form $\dot{\evec{x}} = \mathcal{A}\evec{x}$,
where $\evec{x}$ is a vector of Chebyshev coefficients; i.e.,
% ~$\evec{x} = \begin{bmatrix}
% a_{v,0}& \cdots & a_{v,N} & a_{\eta,0} &\cdots & a_{\eta,N}
% \end{bmatrix}^T$.
~$\evec{x} = (a_{v_0}, \dots , a_{v_N} , a_{\eta_0}, \dots, a_{\eta_N})^T$.

%\begin{align}
%    \begin{split}
%        \label{Chebyshev_pol}
%        \tilde{v}(y,t) &= \sum_{n=0}^{N} a_{v,n}(t)\Gamma_n(y)~,\\
%%        \tilde{\eta}(y,t) &= \sum_{n=0}^{N} a_{\eta,n}(t)\Gamma_n(y)~,
%    \end{split}
%\end{align}

%where $a_{n}$ are the unknown coefficients, are the Chebyshev basis functions, and $N$ are the number of points used for discretization. \ak{A new equation is obtained by substituting the Eq.~\eqref{Chebyshev_pol} in Eq.~\eqref{OSS_Eqn}. The newly obtained equation is evaluated at each collocation point while using information of the parabolic profile of base velocity}. Thus resulting in the equation governing the base flow of the form $\dot{\evec{x}} = \mathcal{A}\evec{x}$, here, the system states are the coefficients of  Chebyshev polynomials $\evec{x} = \begin{bmatrix}
%a_{v,0}& \cdots & a_{v,N} & a_{\eta,0} &\cdots & a_{\eta,N}
%\end{bmatrix}^T$ as given in ~\citep{Mckernan2006}.

One of the stages to bypass transition within the channel flow setting is
the transient energy growth~(TEG) of flow perturbations about
the base-flow due to non-modal effects~\citep{Schmid2001}.
%; it is therefore important to define a metric to measure the growth in disturbance.
In accordance with previous studies, we use the kinetic energy density as
a measure of this TEG, % The kinetic energy density $E(t)$ is given by
\begin{align}
E(t) = \frac{1}{V}\int_{V}\rho \frac{{u}^2 + v^2 +w^2}{2} dV~,
\end{align}
where $\rho$ is the fluid density and $V$ is the volume of
a unit streamwise length of channel. 
In relation to earlier discussions, the kinetic energy density can
be re-expressed as $E(t) := \evec{x}^TQ\evec{x}$,  where $Q = Q^T >0$~\citep{McKernan2006,McKernanPhD2006}.
% Also, when acted upon by an optimal perturbation, the maximum peak in the energy $E(t)$ for all time is called the maximum transient energy growth~(MTEG). The MTEG is represented by $\Theta_{max}$ and is mathematically defined as follows:
% \begin{align}
% \Theta_{\text{max}}=\max _{t \geq t_0}\max _{E(t_0)  \neq 0} \frac{E(t)}{E(t_0)}~,
% \end{align}

In order to implement flow control, we introduce wall-normal
blowing and suction actuation at the upper- and lower-walls---\ak{consistent with prior investigations on controlling channel flow~\citep{MartinelliPoF2011,Joshi1999, Bewley2001}}.
% To influence control on the base flow, the system has to be modified correspondingly.
%
Following the modeling procedure in~\cite{McKernan2006},
this can be modeled via the wall-transpiration boundary conditions $\tilde{v}(y = 1) = q_u$, $\tilde{v}(y = -1) = q_l$, $\frac{\partial \tilde{v}}{\partial y}\mid_{y=\pm1} =0$.
Here, $q_u$ and $q_l$ are wall-normal velocities at the upper- and
lower-wall, respectively.
The final system formulation uses $\dot{q}_u$ and $\dot{q}_l$ as
the control inputs, while reclassifying $q_u$ and $q_l$ as system states.
This introduces two integrators---associated with
the controls---within the system model.
Finally, the actuated channel flow system can be expressed in state-space form,
as in~\eqref{eq:linsys1}, with the state vector defined as
% $
% \evec{x}=\begin{bmatrix}
% a_{v,0}& \cdots & a_{v,N} & a_{\eta,0} &\cdots & a_{\eta,N} &\tilde{q}_u & \tilde{q}_l
% \end{bmatrix}^T$
$
\evec{x}=(a_{v_0}, \dots , a_{v_N} , a_{\eta_0} ,\dots , a_{\eta_N}, q_u , q_l)^T$
and the input vector defined as
$\evec{u} = (\dot{q}_u , \dot{q}_l)^T$.
% The resulting state equation for the linearized channel flow system is given by:
% \begin{align}
%     \label{ss_model1}
%     \dot{\evec{x}} = A\evec{x} + B\evec{u}~,
% \end{align}
% where the matrix $A \in \mathbb{C}^{n \times n}$ is the state matrix hosting the dynamics of the system, while the matrix $B \in \mathbb{C}^{n \times p}$ is the input matrix and maps the influence of control inputs on the states. the number of states are represented by $n$ and number of control inputs are given by $p$.
% States of the system are defined as $
% \evec{x}=\begin{bmatrix}
% a_{v,0}& \cdots & a_{v,N} & a_{\eta,0} &\cdots & a_{\eta,N} &\tilde{q}_u & \tilde{q}_l
% \end{bmatrix}^T$, and the control input is given by $\evec{u} = \begin{bmatrix}
% \dot{\tilde{q}}_u & \dot{\tilde{q}}_l
% \end{bmatrix}^T$.
It must be noted that the resulting system here is complex-valued, owing to
the introduction of Fourier transformations in the streamwise and spanwise directions.
As such, a final step is needed to transform the system into an equivalent real-valued state-space realization.
Further modeling details can be found in~\citep{McKernan2006}. %\ak{The primary aim to choose this model and actuation setup relies on the fact that similar control configuration has been used for performing numerous investigations on the Poiseuille flow ~\citep{MartinelliPoF2011,Joshi1999, Bewley2001}}.
%
% for the purpose of implementation these are converted into real matrices using the procedure detailed in ~\citep{McKernanPhD2006}. 
%Further details on modeling can be found in~\citep{Mckernan2006}.

% It must be noted, even though the actuation is introduced using wall-normal blowing and suction, the application of the appropriate rate of transpiration to effectively control the system is accomplished using techniques from modern control theory. The control methods used in this paper are discussed in the section~\ref{sec:Control_design}.

In the remainder, we consider this linearized channel flow model at a sub-critical Reynolds number of $Re=3000$
(unless otherwise stated) for three different wavenumber
pairs $(\alpha,\beta)$: $(1,0)$, $(1,1)$, and $(0,2)$.
In all cases, the number of collocation points $N$ is chosen so that
the resulting state dimension is $n=199$.
The optimal spanwise disturbance to the uncontrolled system 
with $(\alpha,\beta) = (0,2)$
yields the largest MTEG among all other streamwise, spanwise,
and oblique wavenumber configurations.
Hence, this spanwise disturbance is an important case to study
for controller performance evaluation;
The other two wavenumber pairs considered here have also been widely studied in the literature~\ak{\citep{MartinelliPoF2011,Hogberg2003}}.

\subsection{Frequency analysis of control-oriented reduced-order models for the channel flow system}
\label{sec:Fresp_perf}

\begin{figure}[h!]
\centering
\subfloat[$\alpha=1,~\beta=0$]{\includegraphics[width=0.32\textwidth]{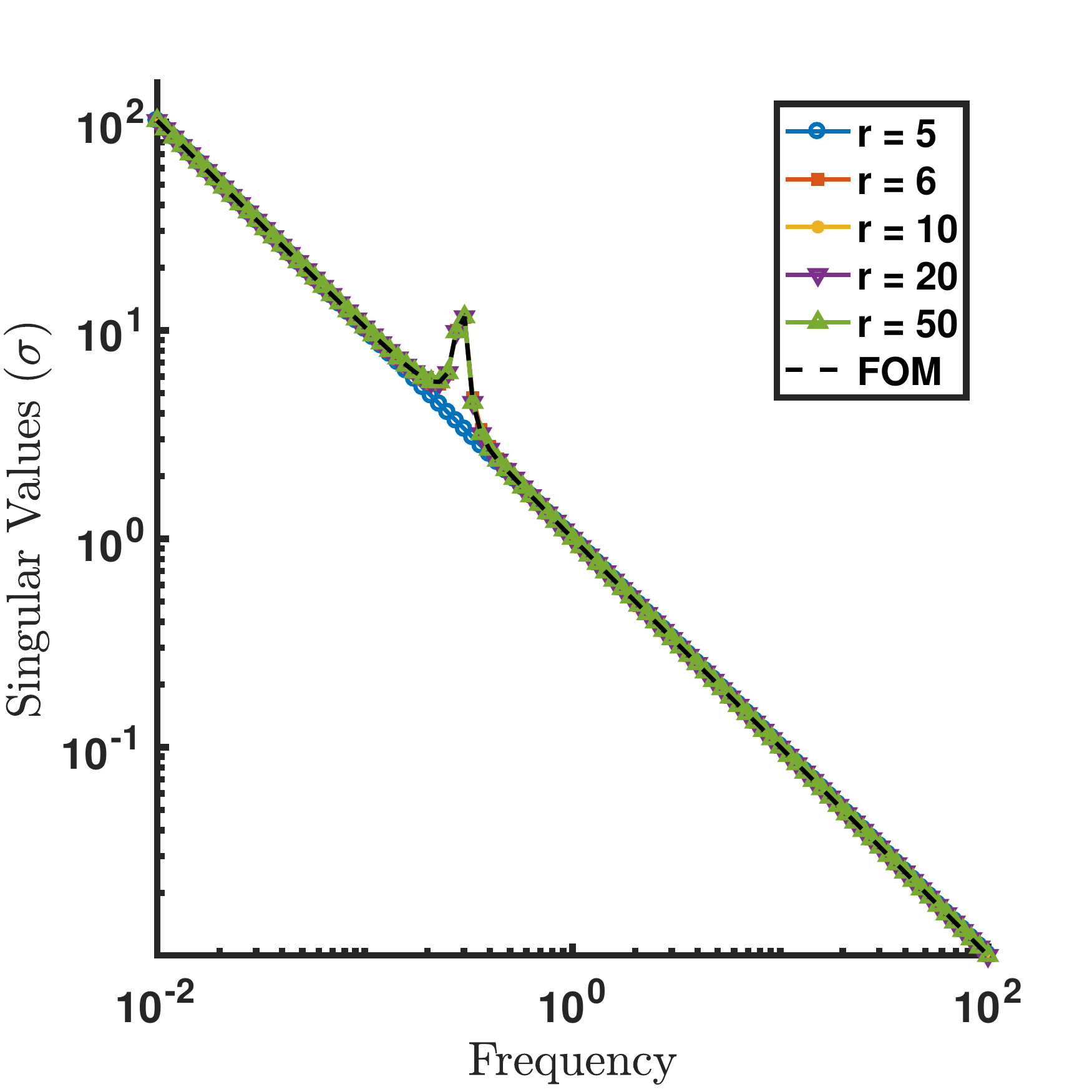}
\label{fig:Fresp_case1}}
\hfill
\subfloat[$\alpha=1,~\beta=1$]{
\includegraphics[width=0.32\textwidth]{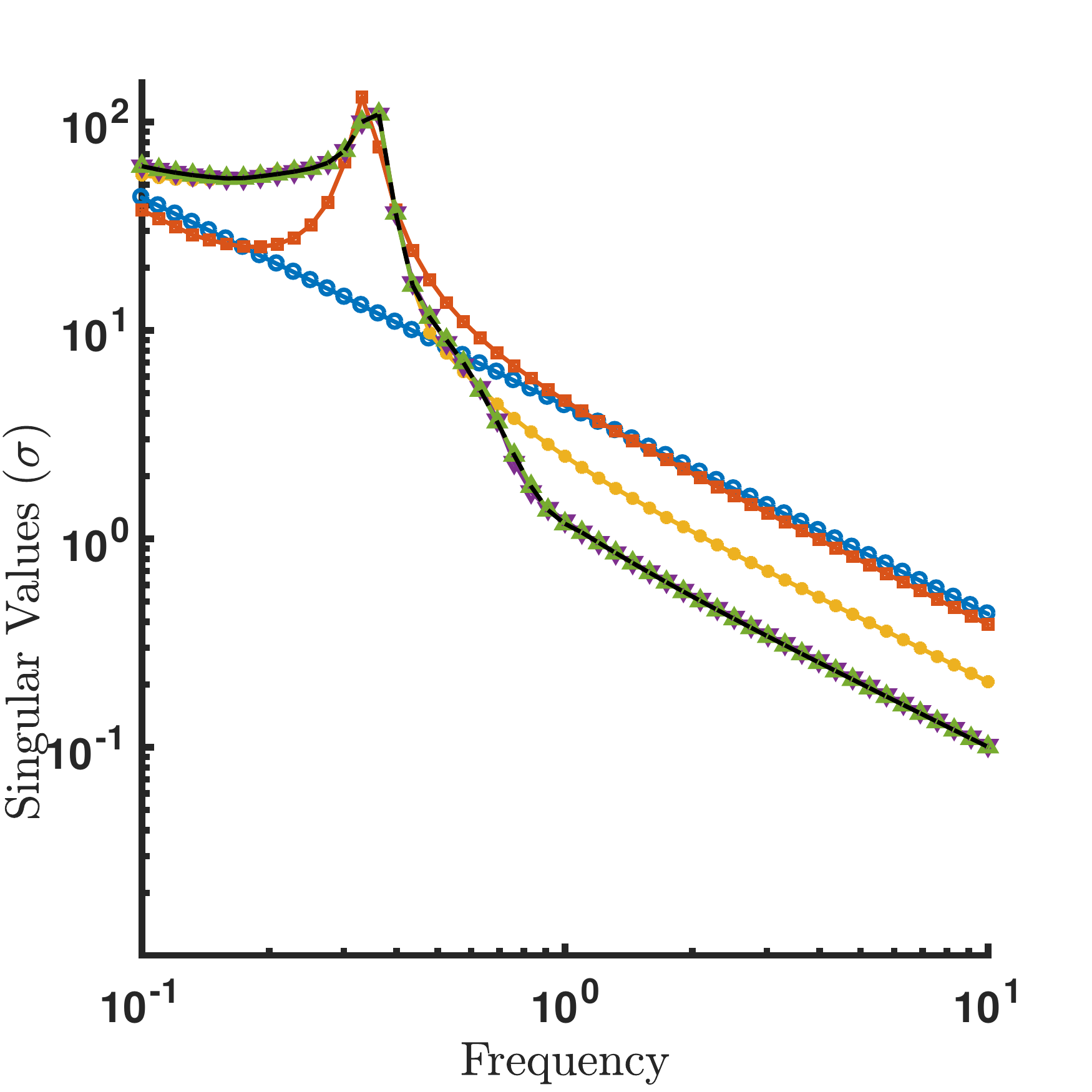}
\label{fig:Fresp_case2}}
\hfill
\subfloat[$\alpha=0,~\beta=2$]{\includegraphics[width=0.32\textwidth]{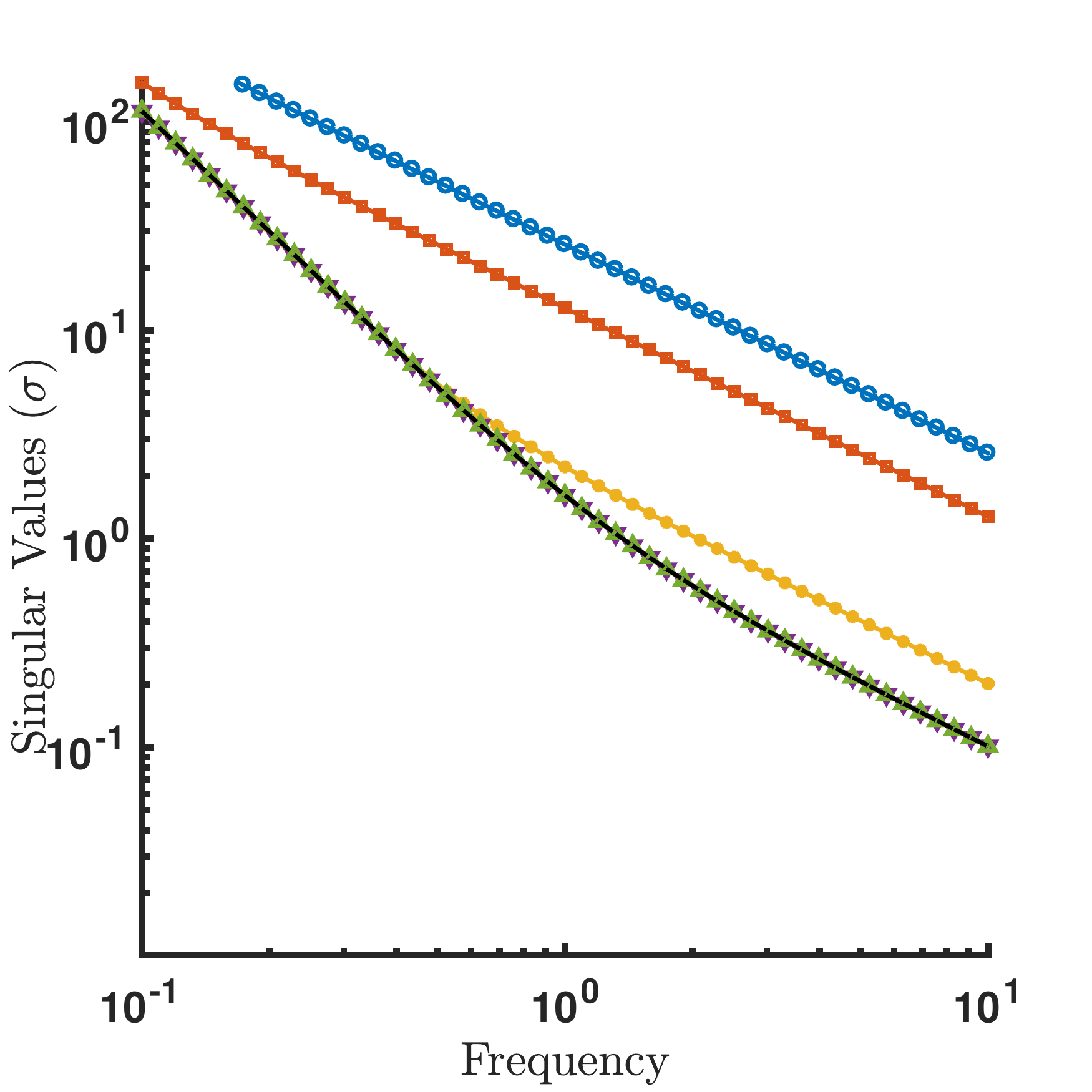}
\label{fig:Fresp_case3}}
\caption{ Frequency response from $\dot{q}_u$ to the flow-state at $Re=3000$.}
%  input channel Of The FOM and ROM are shown for the three wavenumber pairs at $Re=3000$. The figure shows that model of $r \geq 20$ approximate the FOM well in all cases. It can be seen that as $r$ increases the ROM approximate the FOM better.}% \msh{would the data come across more clearly if the $y$-axis was also on a log-scale? Don't forget to to update ``spanwise waves'' etc. to $(\alpha,\beta)=(x,y)$.}}
\label{fig:Fresp_all}
\end{figure}

%\msh{what is the purpose of this section?  we should be sure  to highlight the main takeaways somewhere.}
Here, we examine how the control-oriented ROMs introduced in Section~\ref{sec:COROM}
approximate the dynamics of the full-order model~(FOM) for
the linearized channel flow system, for which $n=199$.
Specifically, we examine whether
the ROMs developed here capture the input-output behavior of the FOM accurately.
%frequency response of the FOM accurately.
%
%We also study the error of the approximation between the ROM and FOM.
% %
% It can be seen that the error in approximation is reduced by increasing the number of modes in the ROM. \msh{point to figure?}
% The ROM framework described in section~\ref{sec:COROM} enables us to capture a low-dimensional representation of the perturbation dynamics. %however,  it is necessary that the ROM used for controller synthesis capture the input-output behavior of the system. Therefore,
%
In multi-input multi-output~(MIMO) systems, 
the singular values $\sigma$ of the system's transfer function over various frequencies provide a means of
characterizing the input-output behavior of the system.
The singular values  provide information on the variation in the system's principal gains
in any of the $p$ input direction~\citep{Skogestad2005}.
% where number of input directions corresponds to
% number of input channels $p$ columns in the input matrix ($B$), these are also called input channels.
%
% It is a standard practice to study the singular values of the system over
% various frequencies to understand the input-output behavior of the system
% in the frequency domain.
Here, we investigate these characteristics as
a function of ROM order $r$ from the input $\dot{q}_u$ to the flow-state (see Figure~\ref{fig:Fresp_all}).
%
%Here, we study the singular values of the system over various frequencies to understand the frequency domain behavior of the ROMs of various order~$r$.
%
\ak{In this work, the scaled frequency is given by
$f^*=ftU_{cl}/h$.} %\msh{if figure axis is updated to $ftU_{cl}/h$, then delete this last sentence.}
%
% This study tells us how the frequency response varies with a change in the order of truncation; 
%This enables us to choose a well approximated ROM.
%
%\msh{Would be nice to be more quantitative here.  Yes, agreement seems ``excellent'' in the ``eye-ball norm'', but that is rather subjective.  What is the actual agreement?  Certainly the ``excellent agreement'' with a small number of modes is not sufficient for controller performance in the next section\dots}
%Here, we study the input-output behavior of the ROMs with different orders~$r$.
%
%
% Figure~\ref{fig:Fresp_all} shows the frequency response of the system from the $\dot{q}_u$ input channel for different model orders $r$.%\msh{which input channel is the first one?  spell it
%out.}
%
The frequency response data reveal that increasing the order $r$ of the ROM
decreases the approximation error with respect to the FOM response. 
Indeed, the trend is more clear from Table~\ref{tab:RSME_Fresp},
which summarizes the root mean square error~(RMSE)
between the ROM and FOM frequency responses.
%
% The RMSE data in Table~\ref{tab:RSME_Fresp} validates the trend
% observed in Fig.~\ref{fig:Fresp_all}.
The acceptable accuracy of a ROM in approximating a FOM
response is mostly application specific.
For the case here, the RMSE metric suggests that ROMs of $r \geq 20$ perform moderately well.
Further, even using the higher-order $r=50$ models---which reflect excellent agreement with the FOM frequency-response---will make LMI-based controller synthesis tractable.
The transfer function from input channel $\dot{q}_l$ to the
flow-state exhibits similar convergence in the frequency-response.
%
% It is also clear from the Fig.~\ref{fig:Fresp_all} and quantified by the results in Table~\ref{tab:RSME_Fresp} that $r = 50$ can be used to obtain an excellent approximation of the FOM.
%
% The results here show that the ROM developed in this paper enables us
% to obtain an accurate low dimensional approximation of the FOM whose order $n=199$.
%
%
%We also examine frequency response for all $Re$
%numbers discussed in section~\ref{sec:control_perf}.\msh{would this fit better in this section?}% 

% We ensure that the ROM framework developed here can be used to obtain well performing low-order approximation for controller synthesis.

%\msh{is this the right analysis to report here? would results of $H_\infty$ norm of error system be more appropriate to report here? again, this comes back to, ``what is the point of this section?''}

%\vspace{0.25 cm}
\begin{table}[h!]
\centering
\begin{tabular}{|cccccc|}
\hline
  &$r=5$ & $r=6$   & $r=10$ & $r=20$  & $r=50$\\
\hline
 $\alpha = 1,\beta = 0$ & 1.079 & 0.049  & 0.0033 & $1.340 \times 10^{-4}$ &  $5.726\times 10^{-6}$\\
\hline
$\alpha = 1,\beta = 1$ &18.765 & 14.135 & 2.990 &0.0891 & $9.603 \times 10^{-5}$ \\
\hline
$\alpha = 0,\beta = 2$ & 119.19 & 61.0560  & 13.332  & 1.705  & 0.0193  \\
\hline
\end{tabular}
\caption{RMSE between the frequency-response of the FOM ($n=199$) and the ROM of order~$r$ at $Re=3000$.}
%The FOM has $n=199$, $Re=3000$ and the input is introduced via first input channel. The RMS Error decreases with increase in $r$.}  %\msh{For all discussions, report $(\alpha,\beta)=(\#,\#)$. Do not refer to these cases as ``streamwise waves'', etc. since that does not provide full information needed to interpret and reproduce the results.}}
\label{tab:RSME_Fresp}
%\end{center}
\end{table}

Finally, it should be noted that, for the ROMs presented here, the data-sampling
parameters $t_f$ and $\delta_t$ in the snapshot POD stage of the
ROM-procedure were tuned to ensure convergence of the ROM response to the FOM response.
\ak{In the $Re=3000$ setting, for streamwise waves and oblique waves $\delta_t = 0.01$ and $t_f = 30 ~\text{and}~ 50$ respectively.} While for spanwise waves we use $\delta_t = 1$ and $t_f = 500$ to obtain adequate models.
All times reported \ak{in} this work are non-dimensionalized, and correspond to non-dimensional
convective time units $\frac{tU_{cl}}{h}$. %\msh{should these be reported in a table?  do we evenm need to report these details here?}
%For each of these cases, we develop ROMs such that an LMI controller synthesis is possible,
%and we further compare TEG suppression of the LMI controller with the LQR controllers.

\subsection{Controller performance}
\label{sec:control_perf}

The synthesis of MTEG-minimizing controllers via the solution of
the LMI-problem in \eqref{eq:lmi_control} is 
enabled by the control-oriented ROMs reported in Section~\ref{sec:Fresp_perf} above.
%
%We design these controllers after various iterations, and careful tuning to ensure the ROMs converged to the FOM.
The resulting controllers will be referred to as \emph{LMI-ROM controllers} here.
%
%Note that the same LMI-based controller synthesis procedure was intractable for the FOM,
%using the same computational resources.
%
For benchmarking, we compare MTEG performance with two sets of
LQR controllers---one designed based on the FOM (``LQR-FOM'')
and one designed based on the same ROM that was used for the LMI-based synthesis (``LQR-ROM'').
Additionally, since the MTEG-minimizing controller in the LMI-ROM was designed without constraining the control input,
we relax the penalty on the input effort within the LQR cost function in~\eqref{eq:LQR} by setting $R=10^{-6} I$, where $I$ denotes the identity matrix.
This is done in an effort 
to make a fair comparison between the LMI-ROM, LQR-ROM, and LQR-FOM controllers.
Note, this differs from the approach taken in~\citep{MartinelliPoF2011},
in which a constrained version of the controller was implemented.
All controllers are applied to the full-order channel flow model described in Section~\ref{sec:model} above, then compared based on the MTEG $\Theta_{\text{max}}$ for each respective closed-loop system.
In general, the optimal perturbation will differ between each closed-loop system,
and will also differ from that of the uncontrolled system. Here we use the optimal perturbation of the respective closed-loop system to perform MTEG analysis.
%
%, i.e., measuring the reduction in MTEG of the corresponding closed-loop system.
%We focus on using the optimal perturbation of the corresponding closed-loop system to evaluate its performance.
In this study, each optimal perturbation is calculated using the algorithm presented in~\citep{Whidborne2011}.

%To study the effectiveness of the LMI controller, we compare the performance of the LMI controller to that of the $\text{LQR}-{\text{FOM}}$ and $\text{LQR}-{\text{ROM}}$ when the system is disturbed by an optimal perturbation.
% To address the transient energy minimization aspect; the controllers are designed and tested on all the three wavenumber pairs listed and for various $Re$~\msh{is this shown here, or later in the section?}.
%

%

% To focus on the aspect of TEG minimization, we study in depth each controllers performance on FOMs of various orders at $Re=3000$. The closed-loop TEG minimization when initialized by an optimal perturbation is shown in Fig.~\ref{fig:Case1_Re3e3}, \ref{fig:Case2_Re3e3} \& \ref{fig:Case3_Re3e3}.  The worst case initial condition also known as the optimal perturbation was used to initialize the closed-loop system for all three cases, i.e., streamwise, oblique and spanwise waves.

\begin{figure}[h!]
\centering
\subfloat[$\alpha=1,\beta=0$ with $r=40$.]{\includegraphics[width=0.3\textwidth]{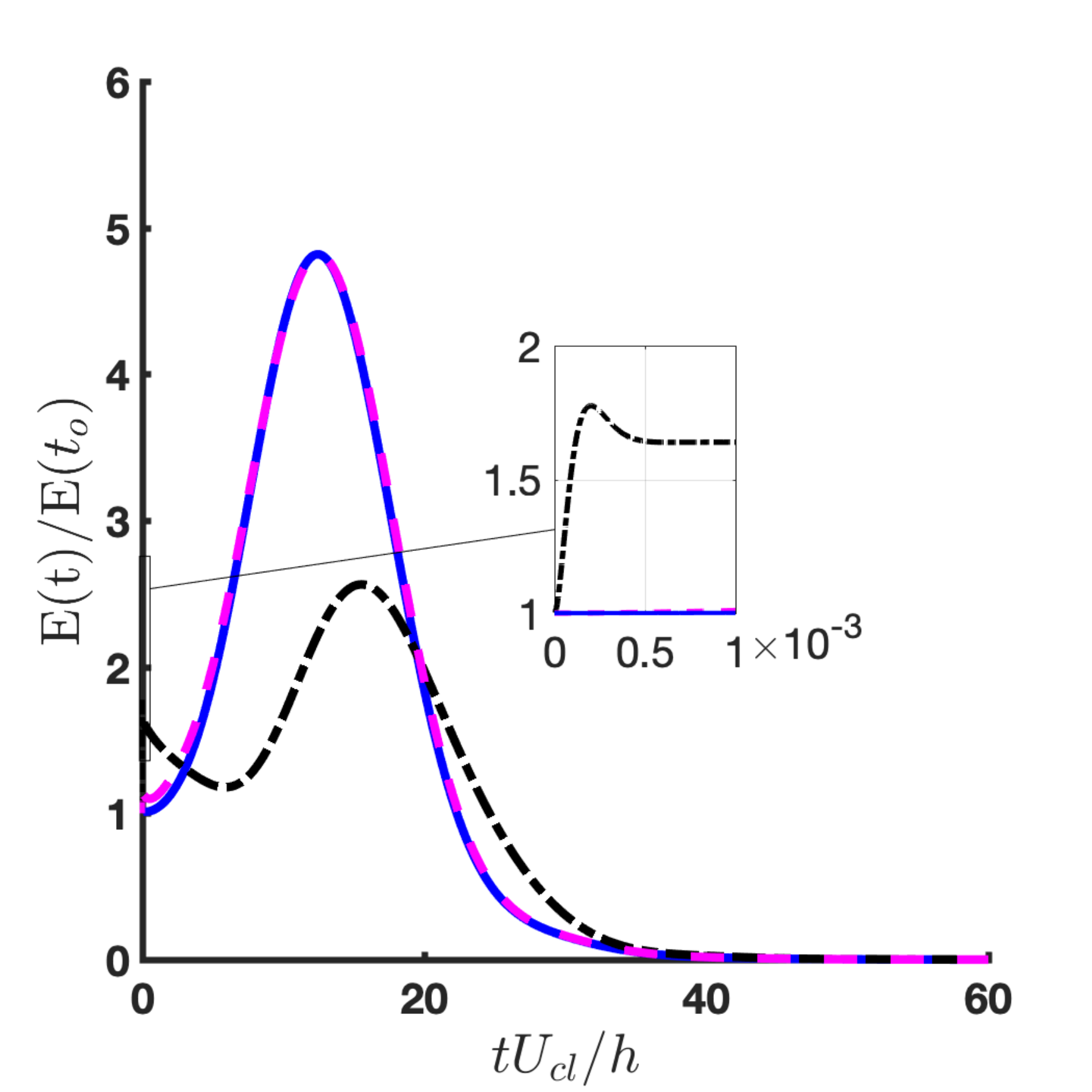}
\label{fig:Case1_Re3e3}}
\hfill
\subfloat[ $\alpha=1,\beta=1$ with $r=58$]{\includegraphics[width=0.3\textwidth]{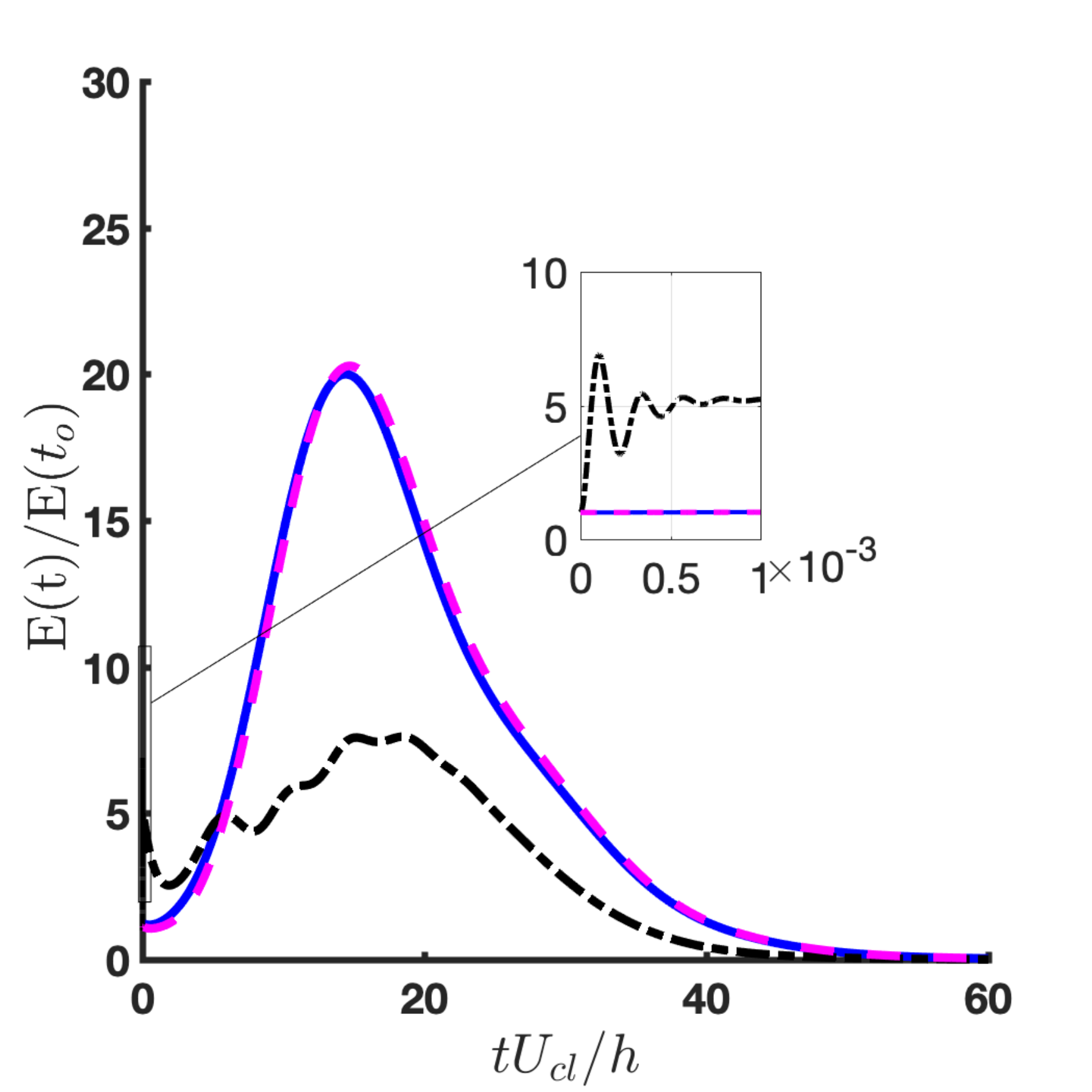}
\label{fig:Case2_Re3e3}}
\hfill
\subfloat[ $\alpha=0,\beta=2$ with $r=40$]{\includegraphics[width=0.3\textwidth]{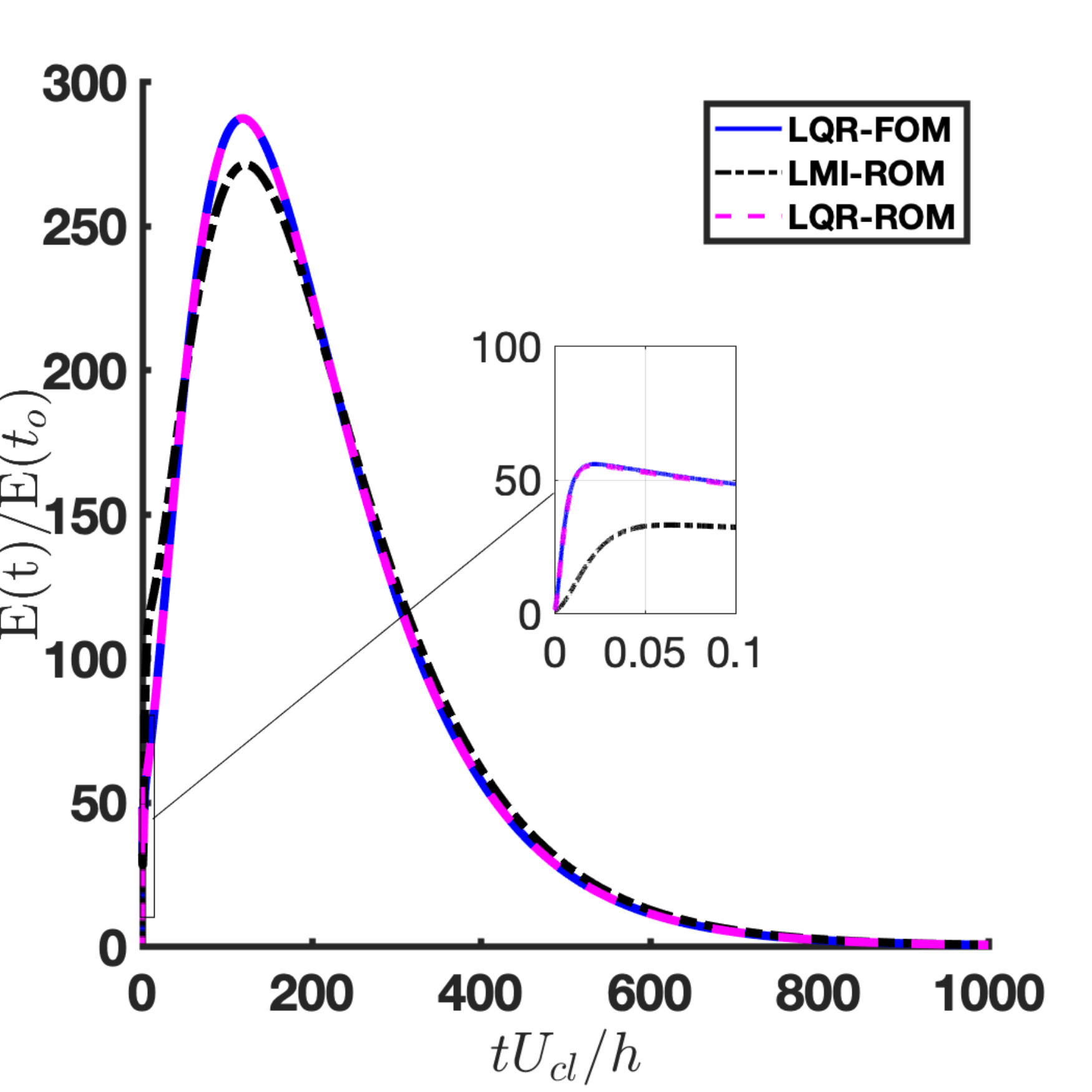}
\label{fig:Case3_Re3e3}}
\caption{Worst-case closed-loop TEG responses at $Re=3000$.}
 % Comparison of closed-loop response for $Re=3000$. The LMI-ROM controller suppresses transient energy growth effectively  while performing better than the $\text{LQR}-{\text{FOM}}$ and $\text{LQR}-{\text{ROM}}$ in \cref{fig:Case1_Re3e3,fig:Case2_Re3e3}, while in fig.~\ref{fig:Case3_Re3e3} the LMI-ROM controller only performs slightly better than the LQR controllers. The inset in (a), (b) \& (c) shows changes observed in transient energy.}
\label{fig:cntrl_all_case}
\end{figure}

%As shown in section.~\ref{sec:Control_design}, the LMI control gain enables the minimization of MTEG in the system.
As shown in Figure~\ref{fig:cntrl_all_case}, the LMI-ROM controller
reduces the MTEG in the system relative to each of the LQR-based controllers
for each of the wavenumber pairs considered.
For $(\alpha,\beta)=(1,0)$, the LMI-ROM controller reduces $\Theta_{\text{max}}$ by a factor of $\approx 1.88$ in comparison with the LQR controllers (see Figure~\ref{fig:Case1_Re3e3}).
Similarly, the LMI-ROM reduces MTEG by a factor of $\approx 2.6$ relative to the LQR controllers for $(\alpha,\beta)=(1,1)$ (see Figure~\ref{fig:Case2_Re3e3}).
%
% in  we see the closed-loop TEG plots for oblique waves setting; the LMI-ROM controller outperforms both the LQR controllers by a factor of $\approx 2.6$ and suppresses the $\Theta_{\text{max}}$ by a factor of $\approx 41$ in comparison with the uncontrolled system.
%
Finally, for $(\alpha,\beta)=(0,2)$, the difference in MTEG reduction between the LMI-ROM controller and the LQR controllers is only marginally greater (See Figure~\ref{fig:Case3_Re3e3}).
This result is consistent with the findings in~\cite{MartinelliPoF2011},
for which wall-normal blowing/suction was found to be less effective for TEG control than spanwise blowing/suction at the walls.
% performs only slightly better
% than either of the LQR controllers, as seen in . 
%
For all controllers and all wavenumber pairs considered here, MTEG was reduced relative to
the uncontrolled flow.
These results are summarized in Table~\ref{tab:TEG_Table}.
%
% Note also that the LMI-ROM controller
% reduces the MTEG by a factor of $\approx 7.9$ when compared to the uncontrolled flow.

% To study the effectiveness of the LMI controller, we compare the performance of the LMI controller to that of the $\text{LQR}-{\text{FOM}}$ and $\text{LQR}-{\text{ROM}}$ when the system is disturbed by an optimal perturbation. The matrix~$R$ in the LQR objective function given by Eq.~\eqref{eq:LQR} penalizes the control input. In the present work, since we use the  LMI controller that does not constrain the control input, we set the $R$ matrix in the LQR controller to be a minimal value (R $\approx 10^{-6} I$) to obtain a fair comparison between the LMI and LQR controllers. To address the transient energy minimization aspect; the controllers are designed and tested on all the three wavenumber pairs listed and for various $Re$. We design these controllers after various iterations, and careful tuning to ensure the ROMs converged to the FOM. To focus on the aspect of TEG minimization, we study in depth each controllers performance on FOMs of various orders at $Re=3000$. The closed-loop TEG minimization when initialized by an optimal perturbation is shown in Fig.~\ref{fig:Case1_Re3e3}, \ref{fig:Case2_Re3e3} \& \ref{fig:Case3_Re3e3}.  The worst case initial condition also known as the optimal perturbation was used to initialize the closed-loop system for all three cases, i.e., streamwise, oblique and spanwise waves.

\begin{table}[h!]
\centering
\begin{tabular}{c|ccccc|}
\cline{2-6}
%\rowcolor{Gray}
\cellcolor{white}& &\multicolumn{4}{|c|}{$\Theta_{\text{max}}$} \\ \cline{3-6}
& \multirow{-2}{*}r  & \multicolumn{1}{|c}{Uncontrolled} & LMI-ROM& LQR-FOM &LQR-ROM\\ \cline{1-6}
\multicolumn{1}{|c|}{$\alpha = 1,\beta = 0$}&40 &20.31 & 2.56 & 4.81 & 4.81\\ \cline{1-6}
\multicolumn{1}{|c|}{$\alpha = 1,\beta = 1$}& 58& 107.00& 7.58&20&20.28\\ \cline{1-6}
\multicolumn{1}{|c|}{$\alpha = 0,\beta = 2$}&40& 1762&271.2&287.1&287.1\\ \cline{1-6}
\end{tabular}
\caption{MTEG $\Theta_{\text{max}}$ for the uncontrolled and controlled channel flow system at $Re=3000$.} %\msh{report $r$ for each row of the table.}}
\label{tab:TEG_Table}
\end{table}

\begin{figure}[!htbp]
\subfloat[$\alpha=1,\beta=0$.]{\includegraphics[scale=0.45]{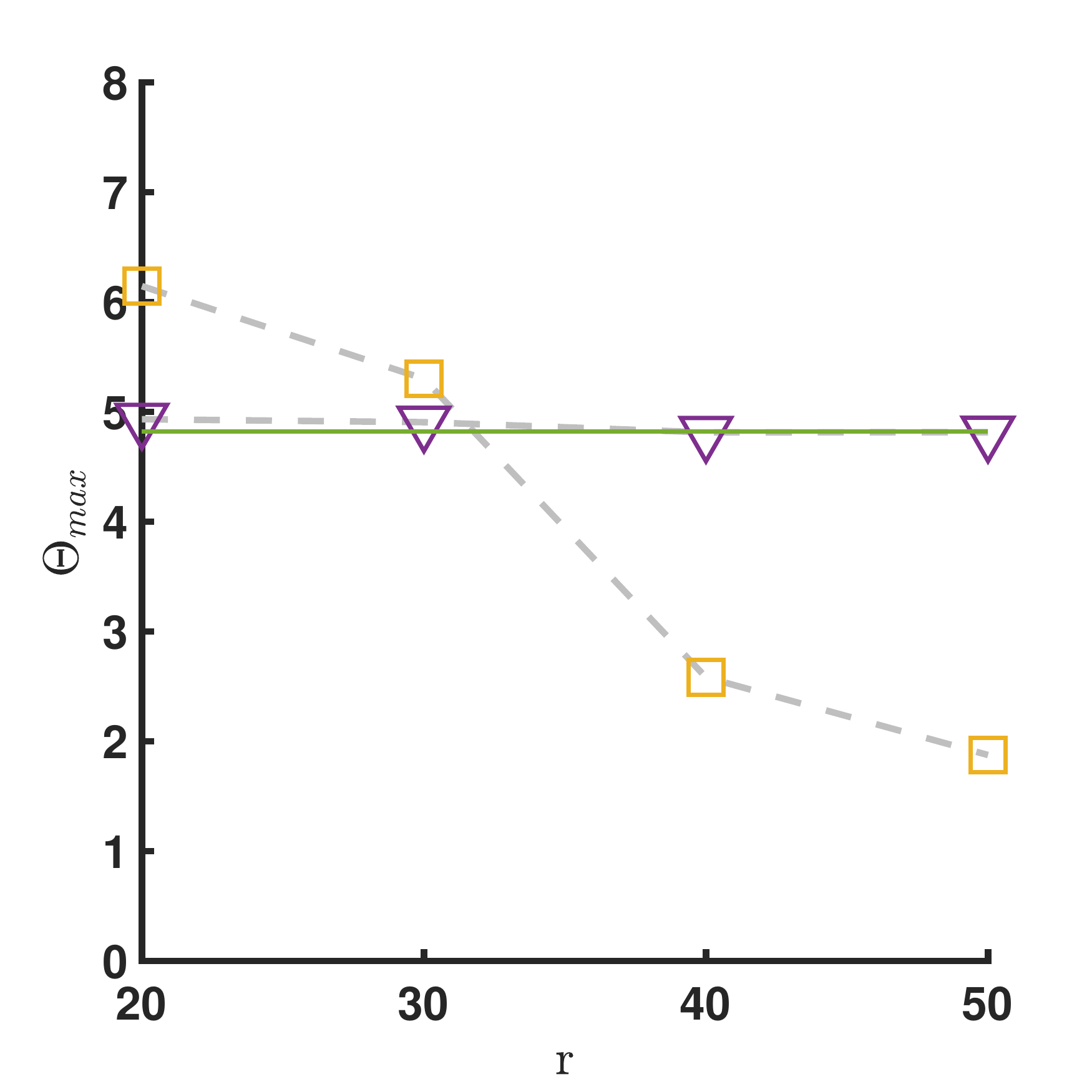}
\label{fig:MTEG_Case1}}
\hfill
\subfloat[ $\alpha=1,\beta=1$]{\includegraphics[scale=0.45]{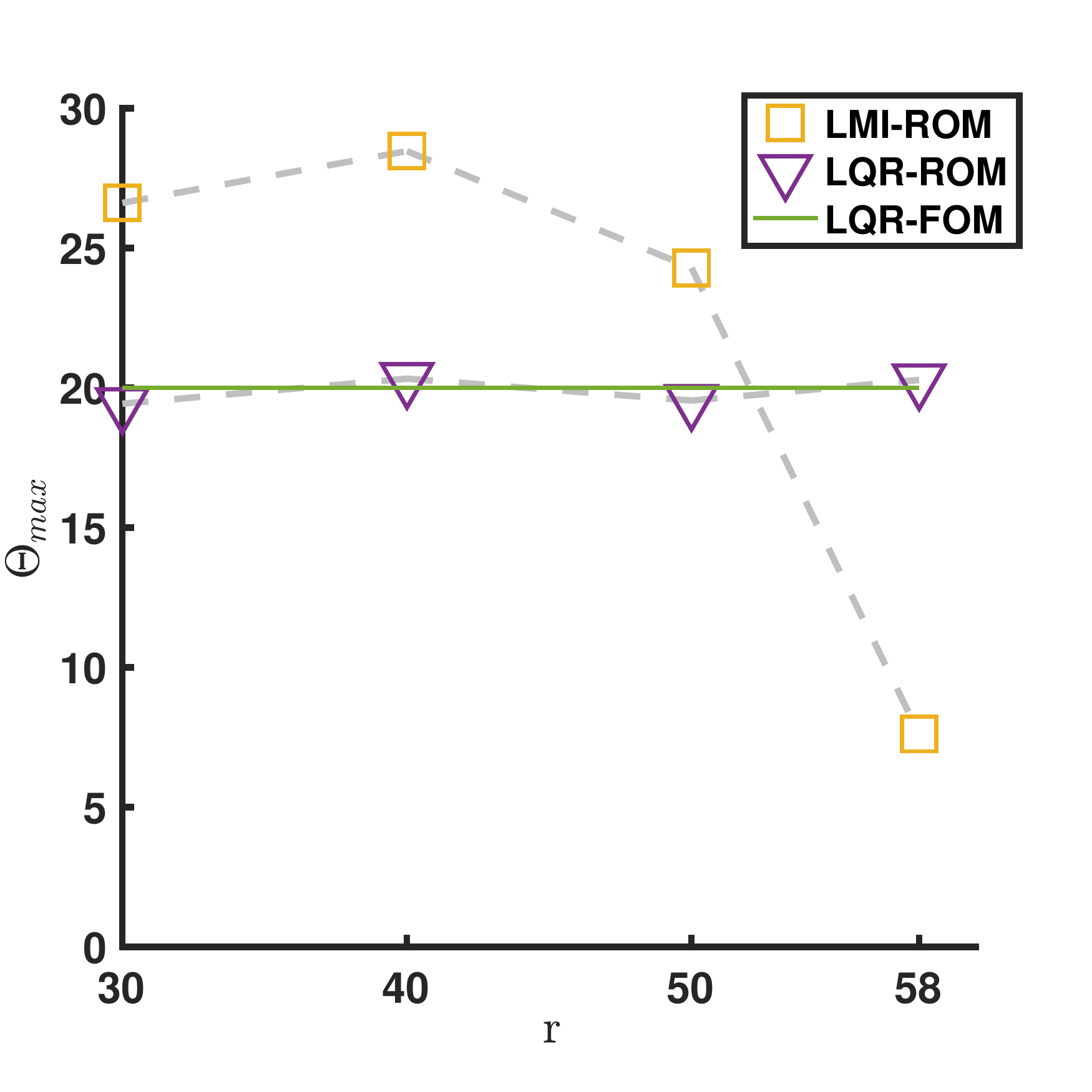}
\label{fig:MTEG_Case2}}
\caption{Closed-loop MTEG $\Theta_{\text{max}}$ as a function of ROM order $r$ at $Re=3000$.}
  %The $\Theta_{\text{max}}$ for LMI-ROM controller is significantly lower than LQR at sufficiently high $r$.} %\msh{why not consider fewer than 20 modes in the plots here?}}
\label{fig:MTEG_chart}
\end{figure}

Although the LMI-ROM controller outperforms
both of the LQR controllers in reducing
the MTEG ($\Theta_{\text{max}}$) in the results above, 
these results depend on the order $r$ of the ROM used
for controller synthesis. %on the FOM.
%
% As such, we next study how the performance of the controllers vary at different values of $r$.
To investigate this influence, we vary $r$ in the ROM design,
then study controller performance on the FOM for streamwise and oblique waves---$(1,0)$ and $(1,1)$, respectively---still with $Re= 3000$.
We only report values of $r$ for which the open-loop ROMs successfully converge to
the FOM dynamics, and so do not consider $r$ smaller than these values here.
%
%$\Theta_{\text{max}}$ for each controller is shown on the y-axis for a corresponding value of $r$ on the x-axis of Fig.~\ref{fig:MTEG_chart}.
%
Figure~\ref{fig:MTEG_chart} shows that $\Theta_{\text{max}}$ decreases
with increasing ROM order $r$ using the LMI-ROM controller.
The same is true for the LQR-ROM controller, but the convergence is more rapid.
MTEG for the LQR-ROM converges to that of the LQR-FOM for $r=20$ for streamwise and spanwise waves, and for $r = 25$ for oblique waves. %\msh{determine this value.  we really shoudl be plotting lower $r$ values in the figures.}
We found that for streamwise and oblique wavenumber pairs considered,
there exists a ROM order $r$ such that the LMI-ROM controller
will outperform both of the LQR controllers for MTEG reduction.
We note that although the LMI-ROM yields a reduction in MTEG relative to both LQR-based controllers in the spanwise wave case, this case was also found
to be relatively insensitive to the ROM order.
Again, this finding is consistent with the results reported in~\cite{MartinelliPoF2011},
which found that spanwise blowing/suction actuation was required to
achieve meaningful reductions in MTEG.
Since the current investigation is focused on control-oriented ROM,
we continue to focus on wall-normal blowing/suction actuation.
Lastly, note that the fact that the LMI-ROM requires a larger $r$
than the LQR-ROM for MTEG performance to converge is to be expected;
TEG is a phenomenon intimately related to
$\dot{E}(t)$, whose approximation requires
higher-precision estimates of $E(t)$.
%
% We also note that in the spanwise waves, we find an LMI-ROM controller that outperforms the LQR controllers as shown in Fig.~\ref{fig:Case3_Re3e3}; however, it was hard to obtain a decreasing trend line for $\Theta_{\text{max}}$ (while increasing $r$). This difficulty can be attributed to the sensitivity of the LMI-ROM controller to the ROM methods and also to the fact that spanwise actuation enables the most reduction in TEG for spanwise waves~\citep{MartinelliPoF2011}; however, we consider only wall-normal actuation in our study.

\begin{figure}[!htbp]
\centering
\subfloat[LMI-ROM control]{\includegraphics[width=0.3\textwidth]{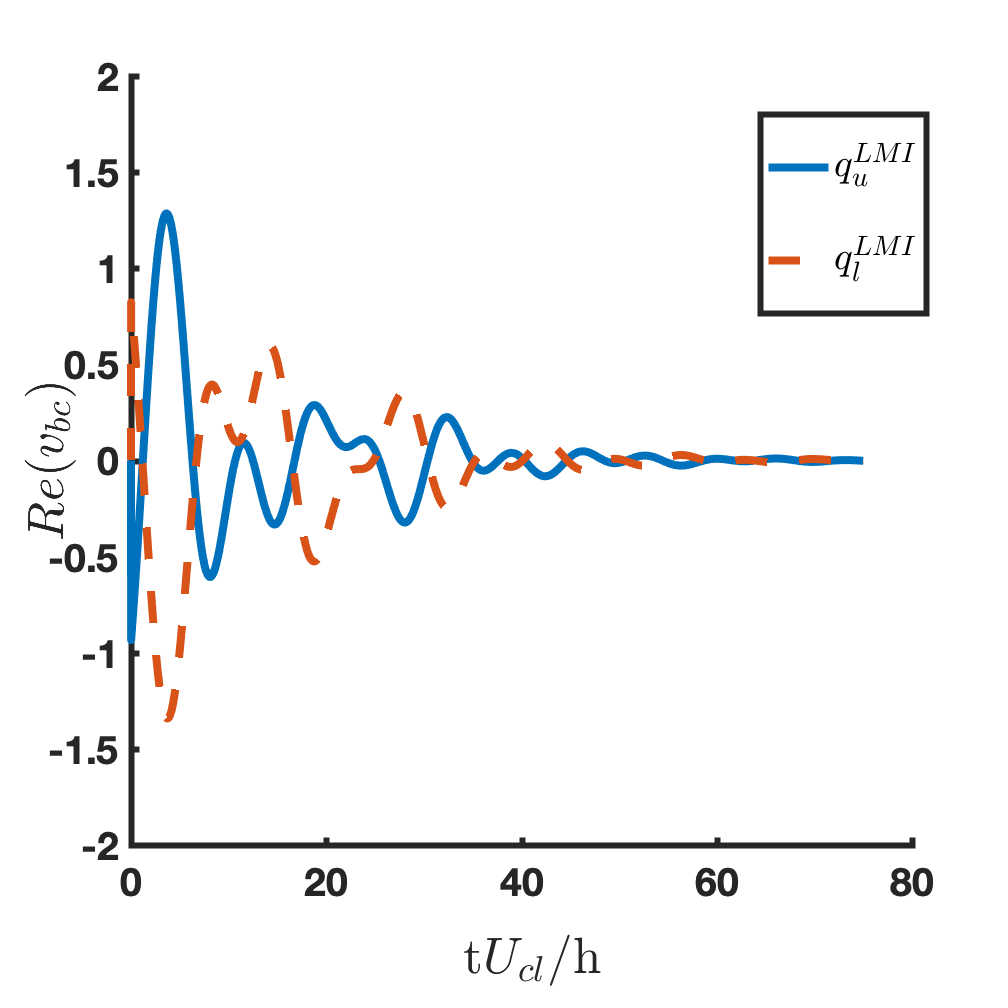}
\label{fig:Case1_vlmi}}
\hfill
\subfloat[LQR-FOM control]{\includegraphics[width=0.3\textwidth]{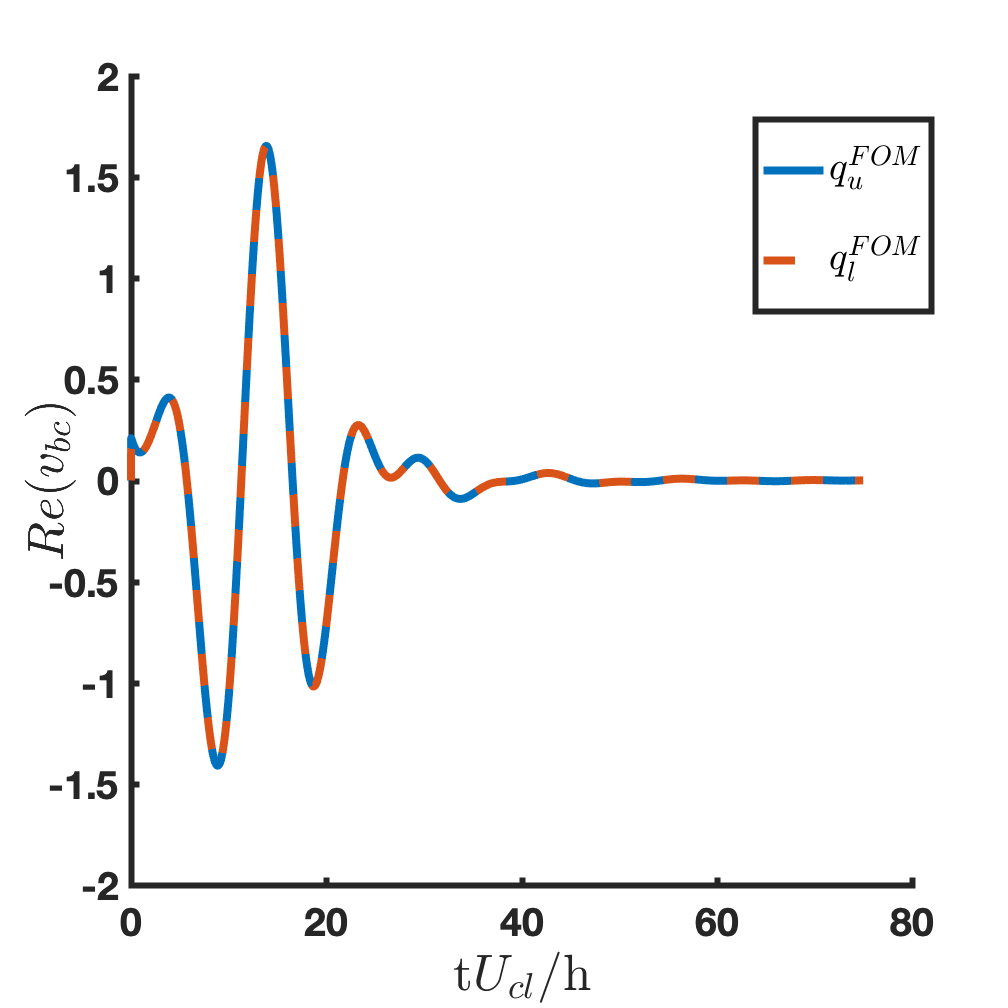}
\label{fig:Case1_vfom}}
\hfill
\subfloat[LQR-ROM control]{\includegraphics[width=0.3\textwidth]{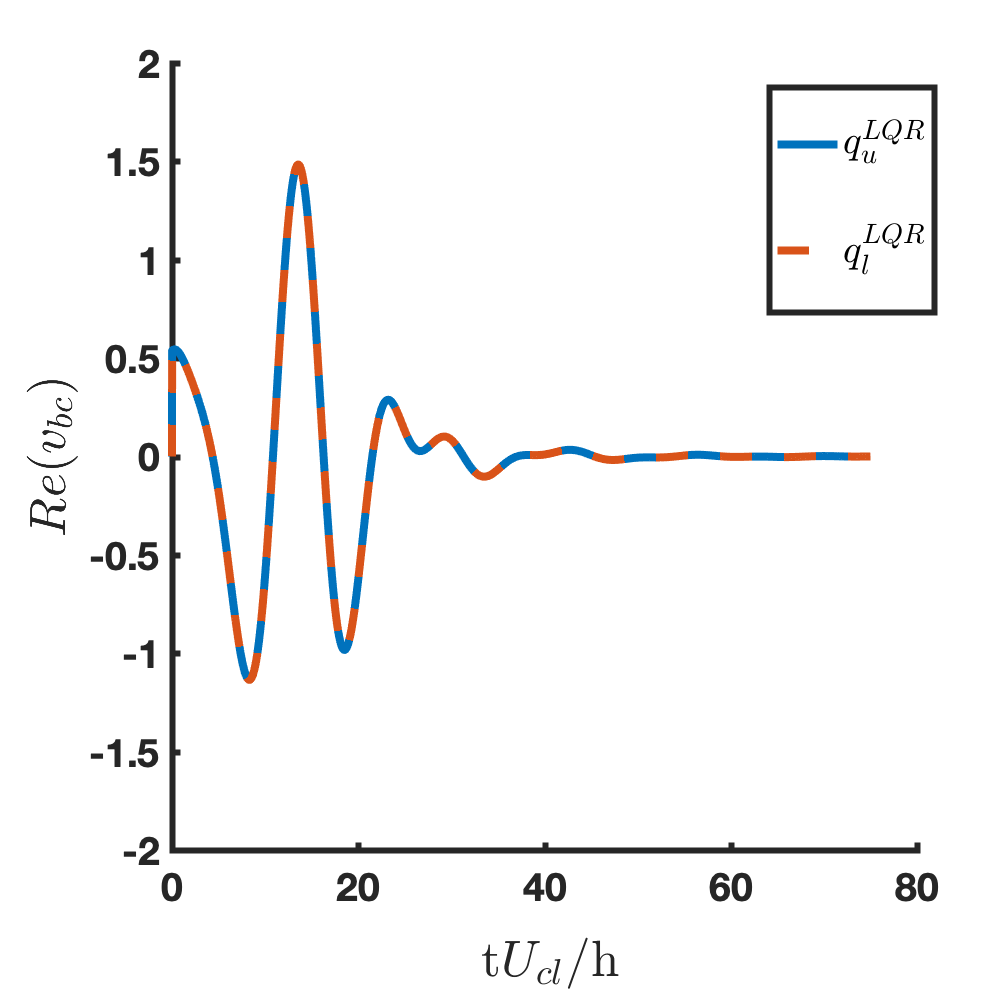}
  \label{fig:Case1_vlqr}}
\caption{Actuated wall-normal velocity for $r=40$, $(\alpha,\beta)=(1,0)$, and $Re=3000$.}
%\caption{Magnitude of the real part of wall-normal velocity ($Re(v)$) for the closed loop system with wavenumber pair $\alpha=1,\beta=0$. The actuation for upper wall and lower wall are given by $q_u$ \& $q_l$ respectively. The actuators in the LMI-ROM controller ($q^{LMI}$) at the upper wall and lower wall provide symmetric actuation commands while the upper wall and lower wall actuators in the LQR controllers~($q^{FOM}$ \& $q^{LQR}$) provide very similar actuation in both the cases. The LMI-ROM controller is active for a longer time in contrast to the LQR controllers}
\label{fig:Case1_vbc_all}
\end{figure}

It is clear that the LMI-ROM controller in the above cases outperforms
both of the LQR controllers.
In order to investigate this further, we proceed to analyze
the details of the wall-normal actuation and its influence on the flow perturbations.
%
% The controllers provide the rate of change in wall-normal velocity as the input, correspondingly, the actuators at the upper and lower walls produces a wall-normal velocity that perturbs the flow. The actuation is given by $q_u$, and $q_l$ where the subscript $u~\text{and}~l$ represent actuators on the upper and lower walls of the channel.
Figure~\ref{fig:Case1_vbc_all} shows the actuated wall-normal velocity at the upper- and lower-walls---$q_u$ and $q_l$, respectively---for each controller.
The case of $(\alpha,\beta)=(1,0)$, shown in Figure~\ref{fig:Case1_vlmi},
reveals that LMI-ROM controller results in blowing and suction at the upper-
and lower-walls that are \ak{equal in magnitude, but opposite in direction}. %exactly out of phase with each other. %\msh{confirm that this is true.}
 In contrast, each of the LQR-ROM and the LQR-FOM controllers produces almost
identical~(\ak{both in magnitude and direction}) blowing and suction at the upper- and lower-walls.
Interestingly, the LQR-ROM controls differ from the LQR-FOM controls for a short period at the beginning of the response, which is an expected artifact of the modal truncation.
The 5\% settling time on the actuation for the LMI-ROM controllers is ~$\approx 43$ convective time units;
the LQR controllers settle in $\approx 34$ convective time units,
indicating a shorter duration of control.

% streamwise waves : controller input for each controller

% oblique waves : controller input for each controller
\begin{figure}[!htbp]
\centering
\subfloat[LMI-ROM control]{\includegraphics[width=0.3\textwidth]{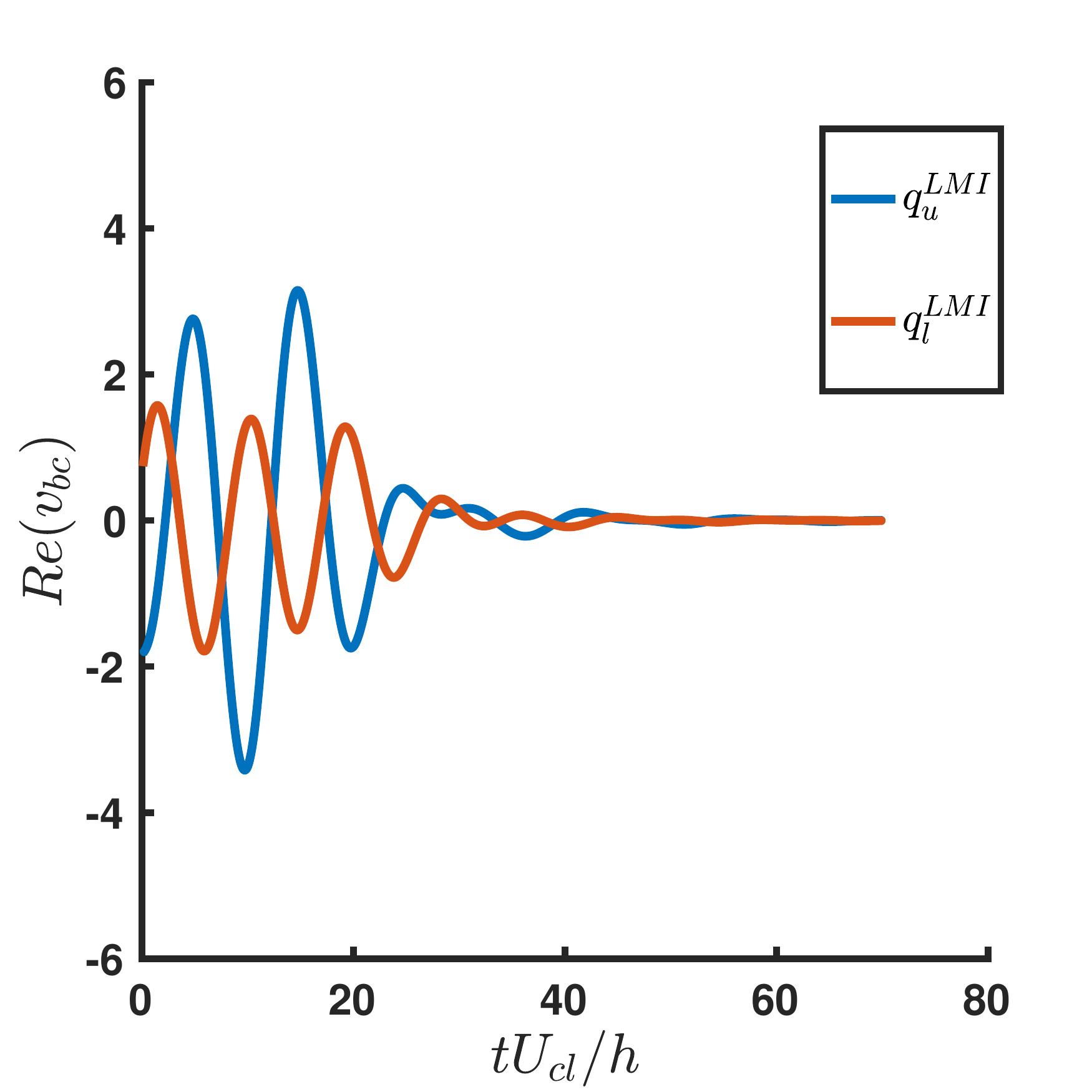}
\label{fig:Case2_vlmi}}
\hfill
\subfloat[LQR-FOM control]{\includegraphics[width=0.3\textwidth]{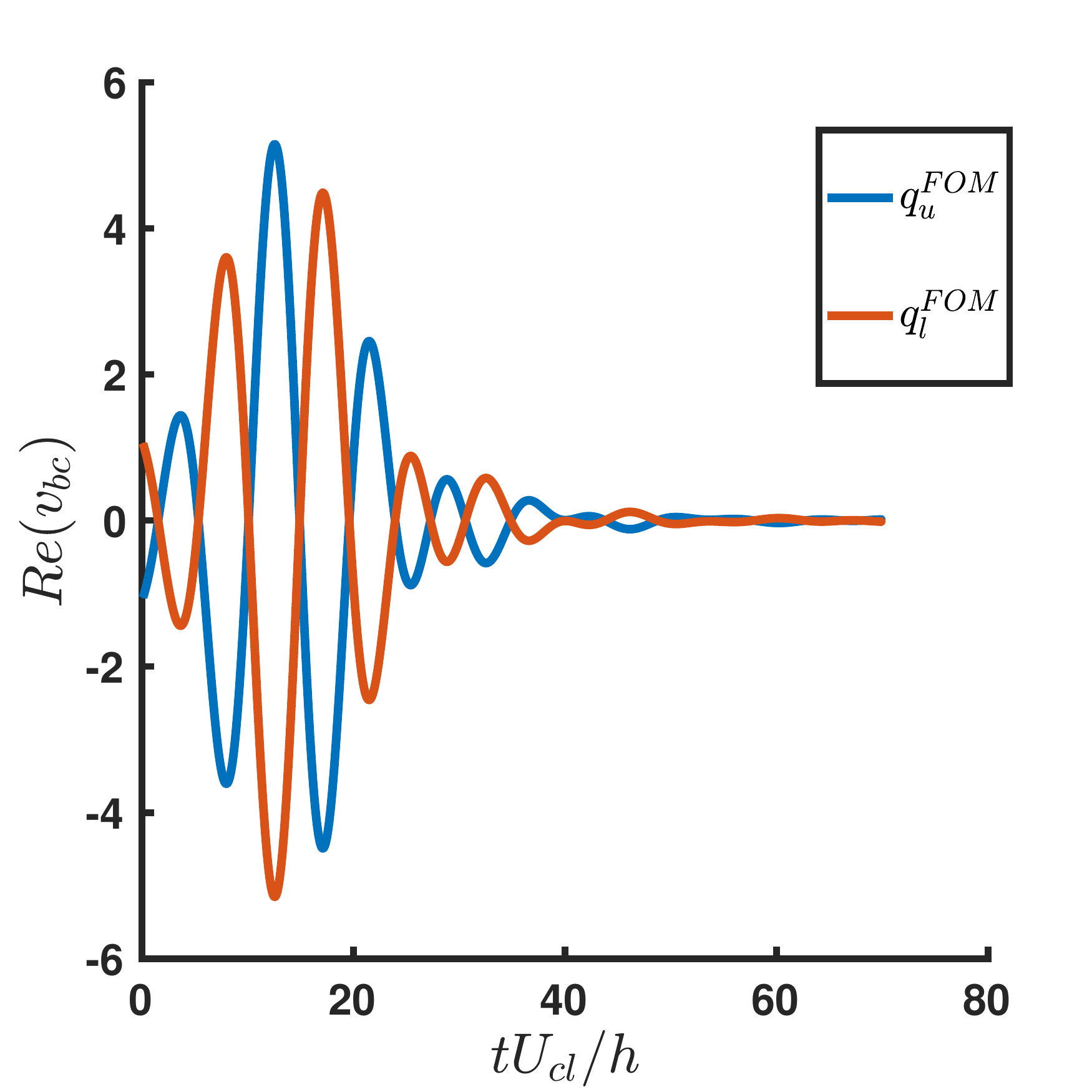}
\label{fig:Case2_vfom}}
\hfill
\subfloat[LQR-ROM control]{\includegraphics[width=0.3\textwidth]{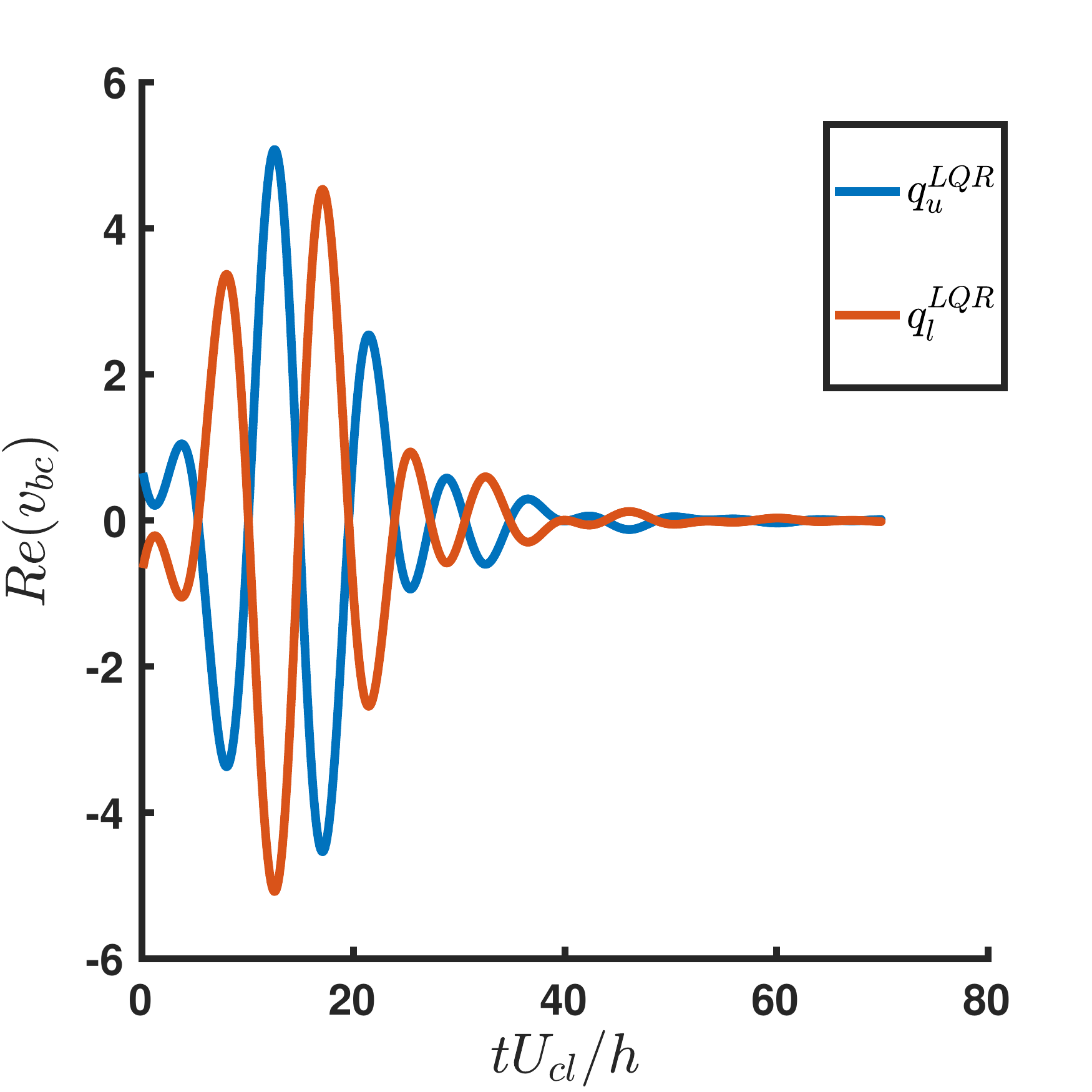}
\label{fig:Case2_vlqr}}
\caption{Actuated wall-normal velocity for $r=58$, $(\alpha,\beta)=(1,1)$, and $Re=3000$.}
  %Magnitude of real part of the wall-normal velocity~($Re(v)$) of the closed loop system for wavenumber pair $\alpha=1,\beta=1$. The actuation for upper wall and lower wall are given by $q_u$ \& $q_l$ respectively. LMI-ROM controller~($q^{LMI}$) in ~\ref{fig:Case2_vlmi} has a lower magnitude of actuation while also having a larger settling time compared to the LQR controllers~($q^{FOM}$ \& $q^{LQR}$) in~\ref{fig:Case2_vfom},~\ref{fig:Case2_vlqr}}.
\label{fig:Case2_vbc_all}
\end{figure}

\begin{figure}[!htbp]
\centering
\subfloat[LMI-ROM control]{\includegraphics[width=0.3\textwidth]{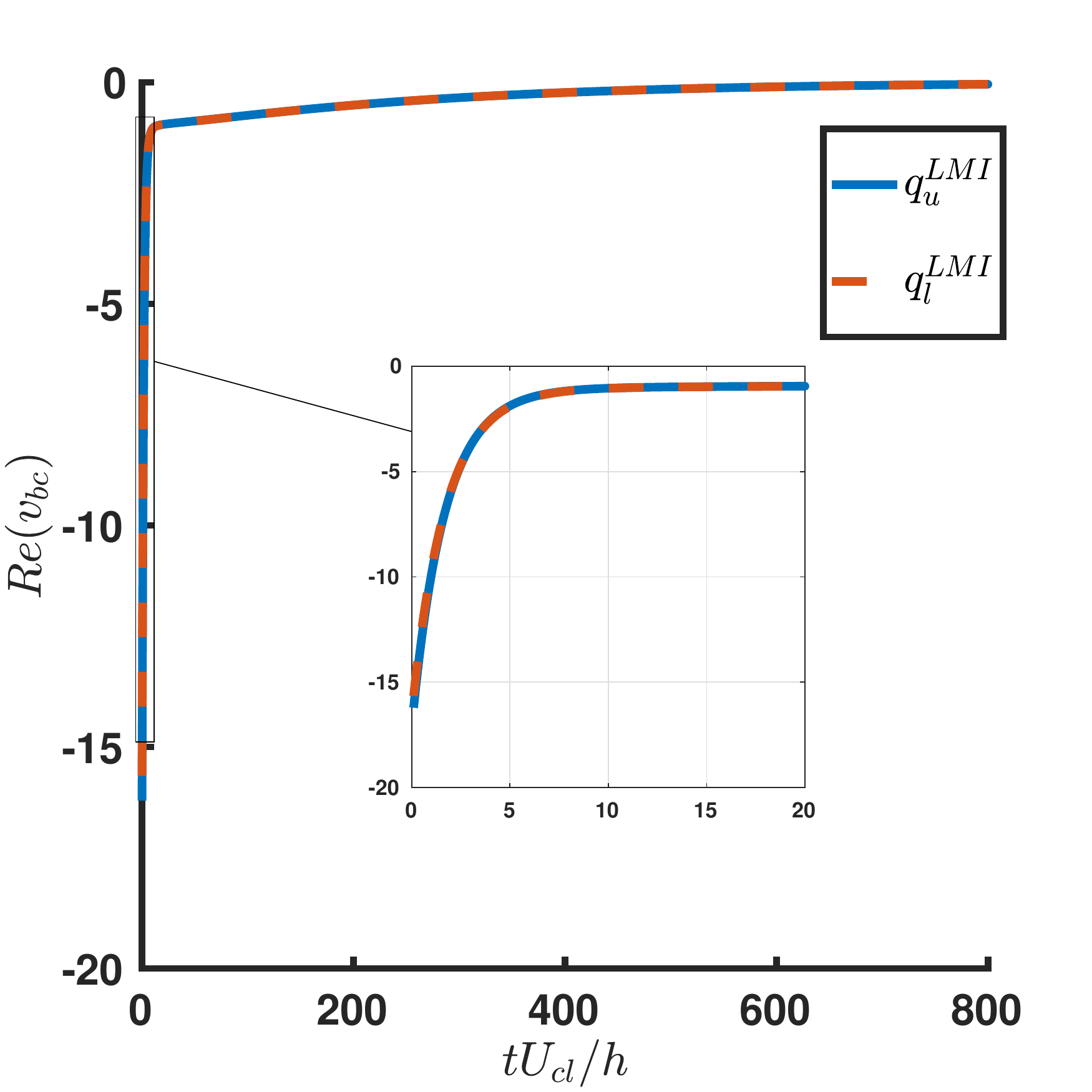}
\label{fig:Case3_vlmi}}
\hfill
\subfloat[LQR-FOM control]{\includegraphics[width=0.3\textwidth]{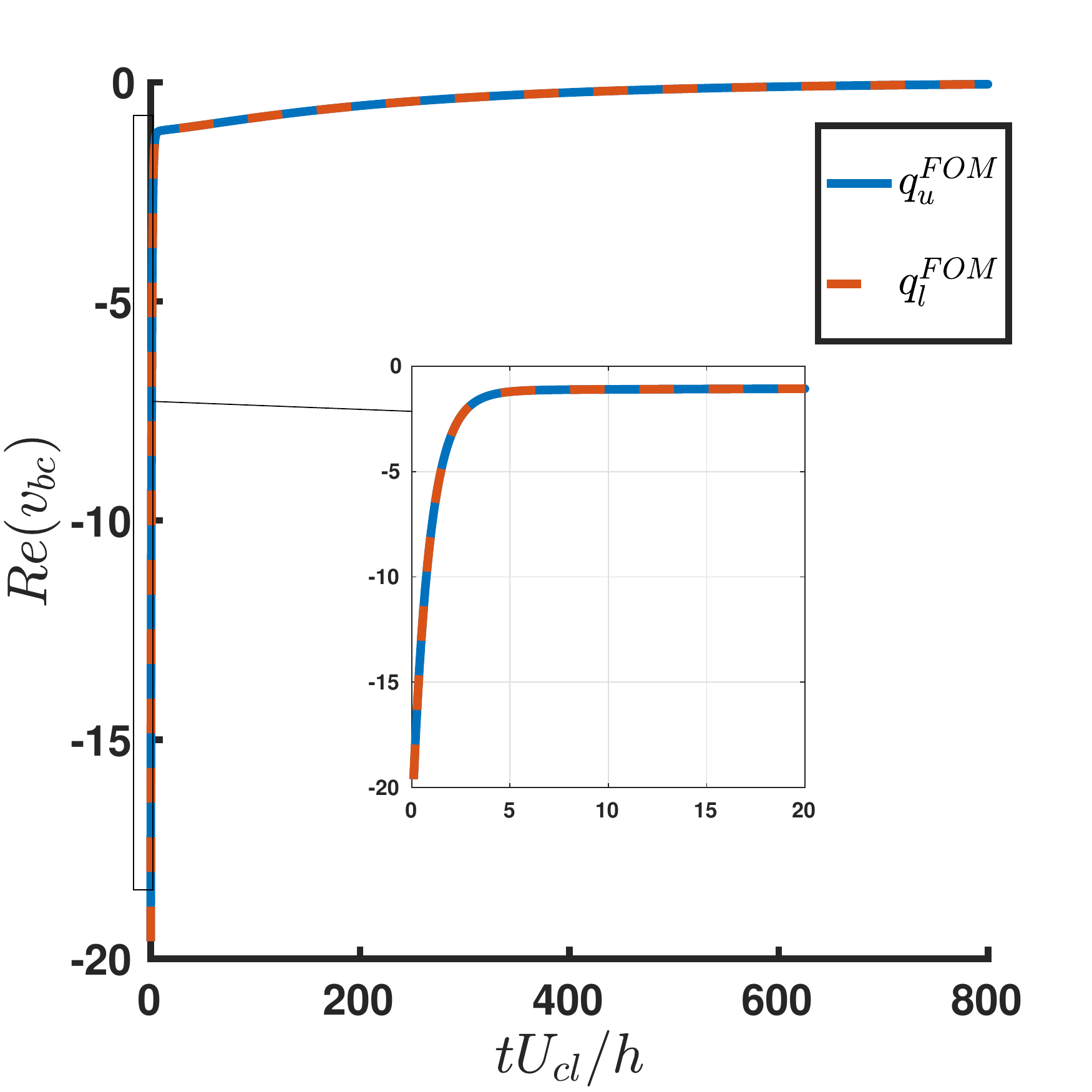}
\label{fig:Case3_vfom}}
\hfill
\subfloat[LQR-ROM control]{\includegraphics[width=0.3\textwidth]{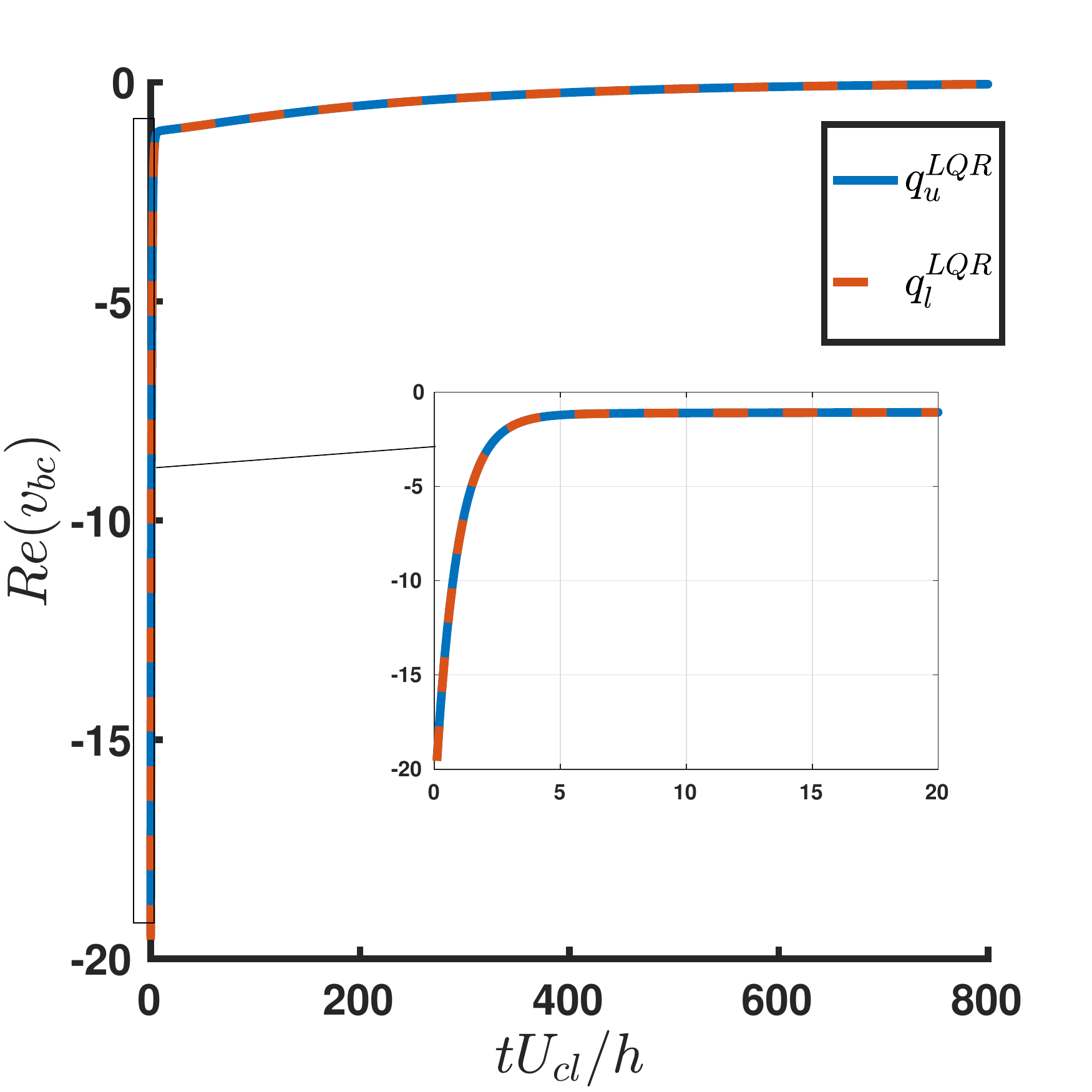}
\label{fig:Case3_vlqr}}
\caption{Actuated wall-normal velocity for $r=40$, $(\alpha,\beta)=(0,2)$, and $Re=3000$.}
  %The actuation for upper wall and lower wall are given by $q_u$ \& $q_l$ respectively. LMI-ROM controller~($q^{LMI}$) in ~\ref{fig:Case3_vlmi} has a lower magnitude of actuation compared to the LQR controllers~($q^{FOM}$ \& $q^{LQR}$) in~\ref{fig:Case3_vfom},~\ref{fig:Case3_vlqr}. All the controllers provide similar wall-normal velocity profile at the walls.}%\msh{comparison of results would be easier if they were all in a single plot.}}
\label{fig:Case3_vbc_all}
\end{figure}

In the oblique wave case of $(\alpha,\beta)=(1,1)$, shown in Figure~\ref{fig:Case2_vlmi},
the LMI-ROM controller results in a maximum wall-normal
blowing/suction velocity that is $\approx 1.6$ times lower than either of the LQR controllers.
From Figure~\ref{fig:Case2_vlmi} we find that the $q_u$ produces a control input of larger magnitude compared to $q_l$, i.e., the actuator on the upper-wall is inducing a larger velocity compared to the actuator on the lower-wall.
In Figures~\ref{fig:Case2_vfom} and \ref{fig:Case2_vlqr} the upper-wall and lower-wall actuators produce a velocity of similar magnitude, but \ak{opposite directions relative to each other}. In the oblique wave case, the $5\%$ settling time of the actuation signal for the LMI-ROM controller is $\approx 41$ convective time units, while the LQR-FOM and LQR-ROM each have a settling time of $\approx 37$ convective time units.
In the spanwise waves setting, shown in Figure~\ref{fig:Case3_vbc_all},
the LMI-ROM controller actuates the system similarly to both of the 
%$v$ profile at the wall is very similar to the actuation of the
LQR controllers, but with a lower initial magnitude.
%, compared to that of the LQR controllers~($\approx -20$).
For all controllers, wall-normal transpiration at the upper-wall is
identical to that at the lower-wall (see Figure~\ref{fig:Case3_vbc_all}).

% spanwise waves : controller input for each controller

\begin{figure}[!htb]
\centering
\subfloat[LMI-ROM control]{\includegraphics[scale=0.3]{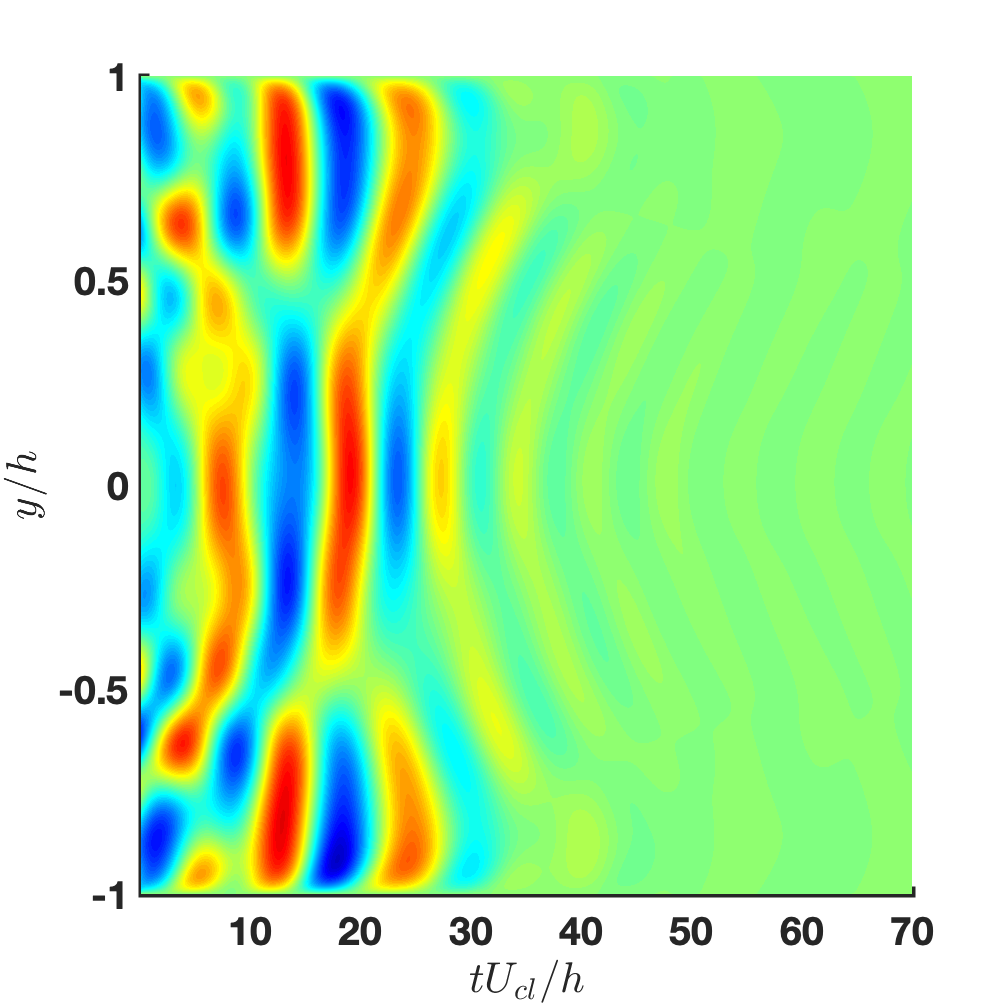}
\label{fig:Case1_ulmi}}
\hfill
\subfloat[LQR-FOM control]{\includegraphics[scale=0.3]{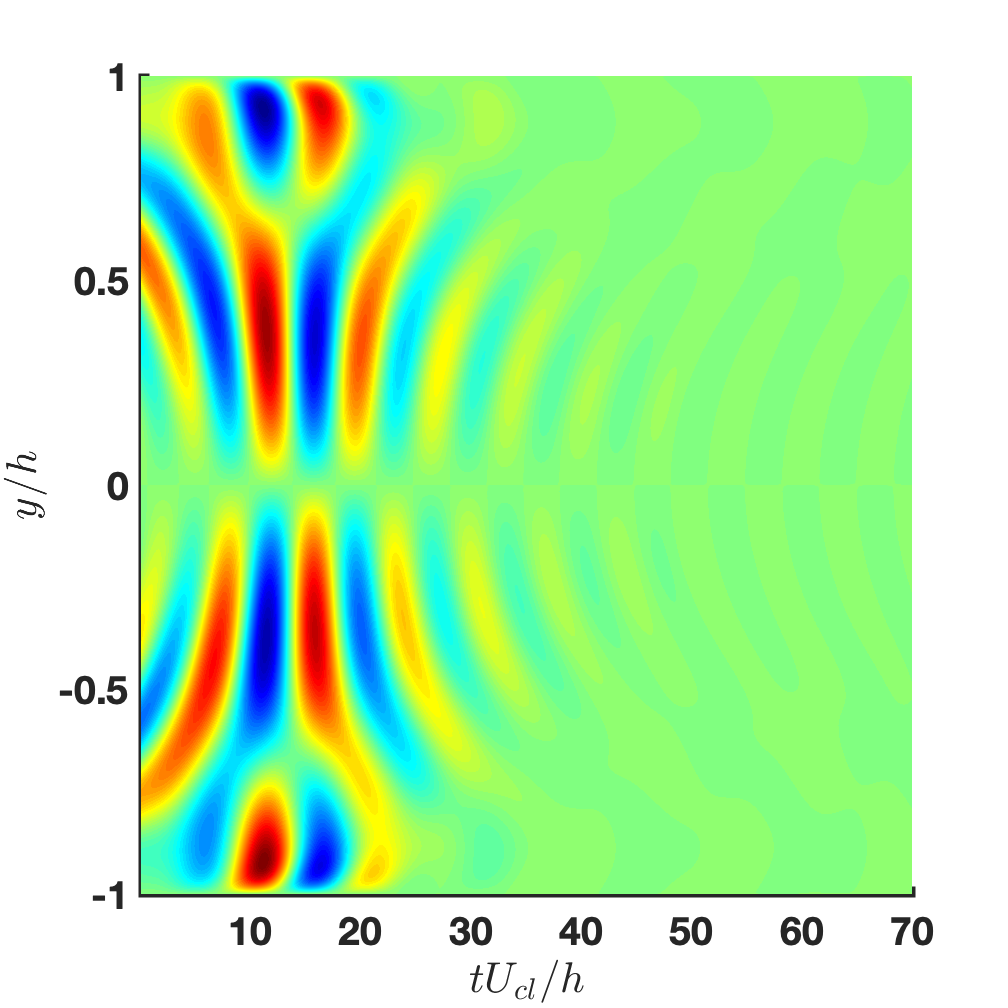}
\label{fig:Case1_ufom}}
\hfill
\subfloat[LQR-ROM control]{\includegraphics[scale=0.3]{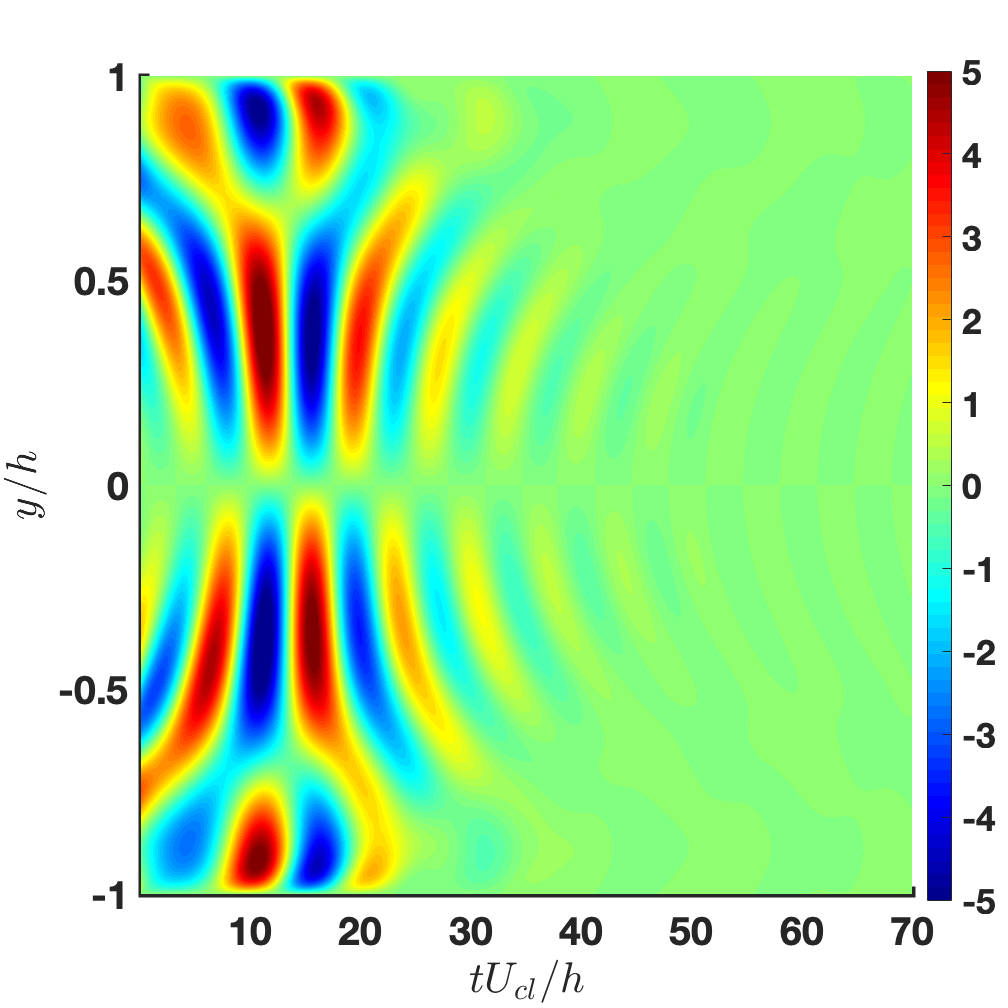}
\label{fig:Case1_ulqr}}\\
\subfloat[Uncontrolled]{\includegraphics[scale=0.3]{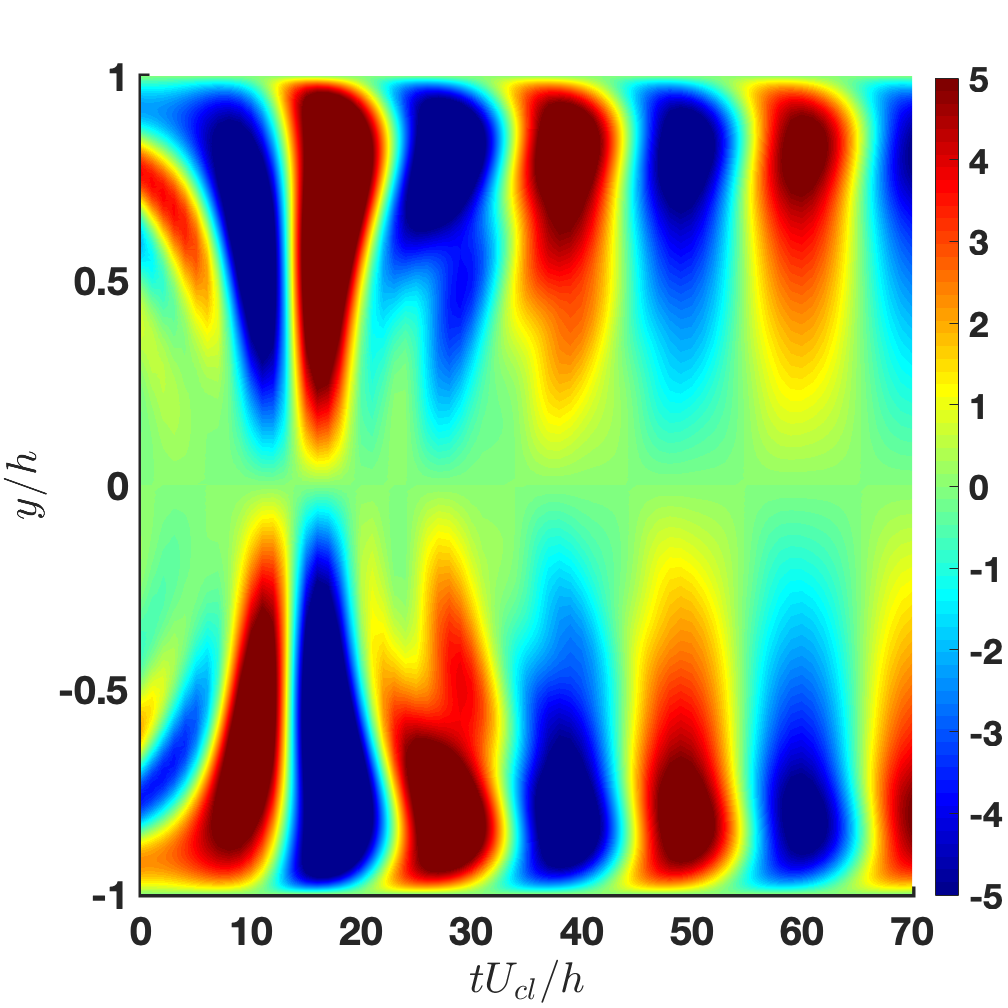}}
\caption{Evolution of streamwise velocity perturbations~($u$) for $r=40$, $(\alpha,\beta)=(1,0)$, and $Re=3000$.}
  %perturbations. The time evolution of the real part of $u$ is shown against the channel height on the y-axis. The initial conditions of the three systems is the respective closed loop optimal perturbation. In Fig.~\ref{fig:Case1_ufom},\ref{fig:Case1_ulqr} controllers perform similar to each other. The $u$ profile of the LMI-ROM controller is qualitatively different from that of the two LQR controllers. Results for $\alpha=1,\beta=0$ at $Re=3000$ and ROM of size $r=40$}
\label{fig:Case1_uall}
\end{figure}

\begin{figure}[!htbp]
\centering
\subfloat[LMI-ROM control]{\includegraphics[scale=0.3]{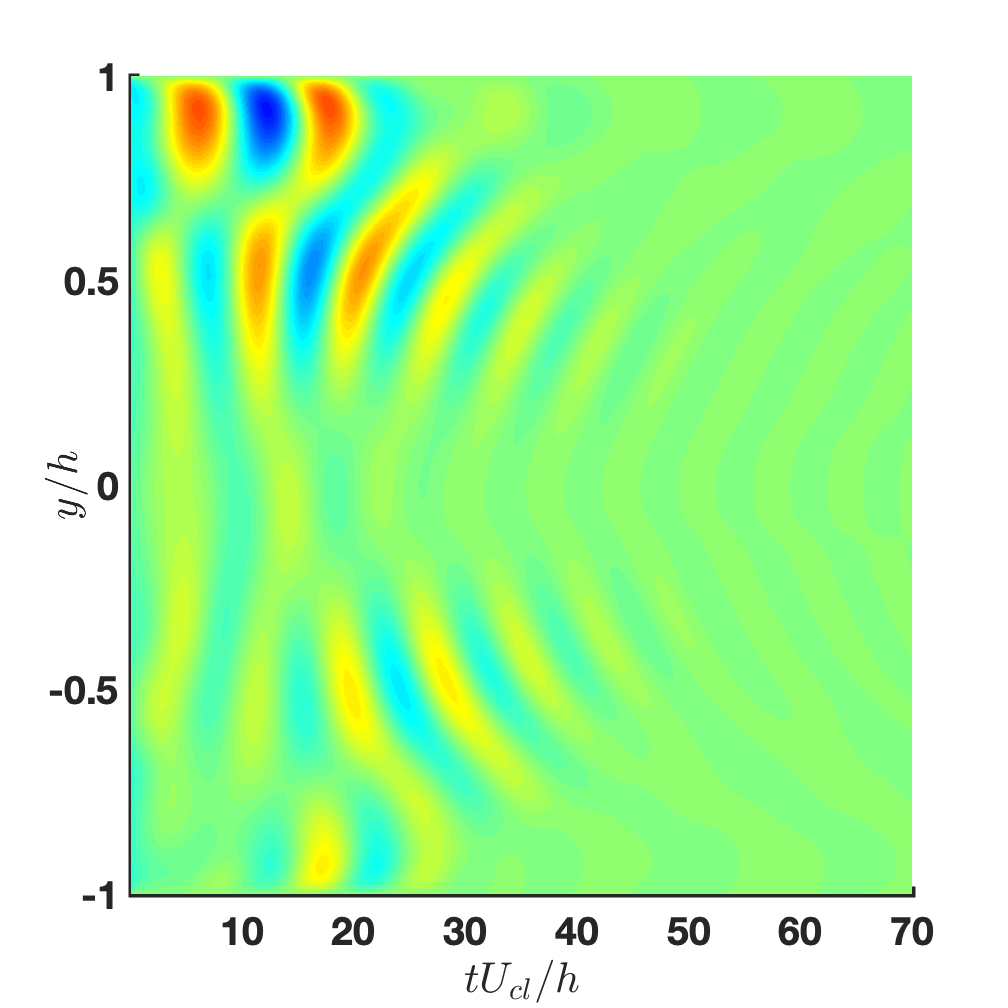}
\label{fig:Case2_ulmi}}
\hfill
\subfloat[LQR-FOM control]{\includegraphics[scale=0.3]{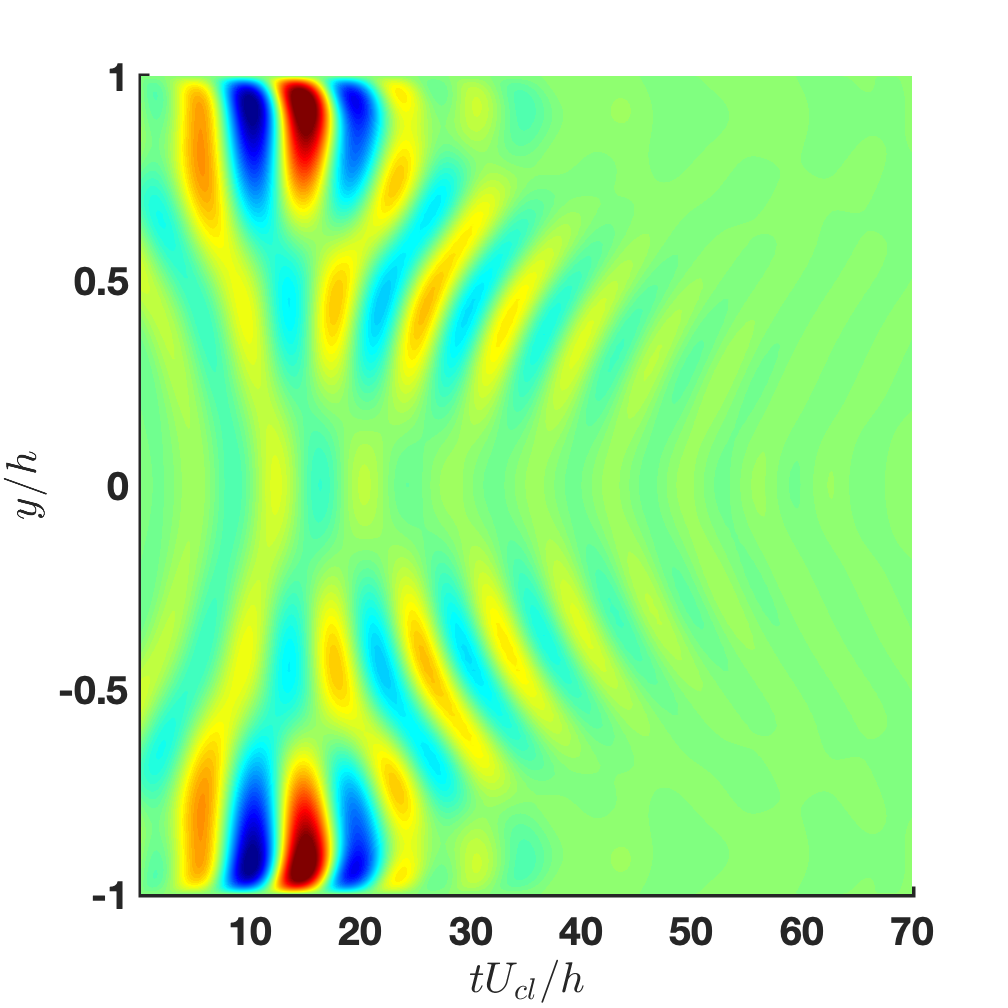}
\label{fig:Case2_ufom}}
\hfill
\subfloat[LQR-ROM control]{\includegraphics[scale=0.3]{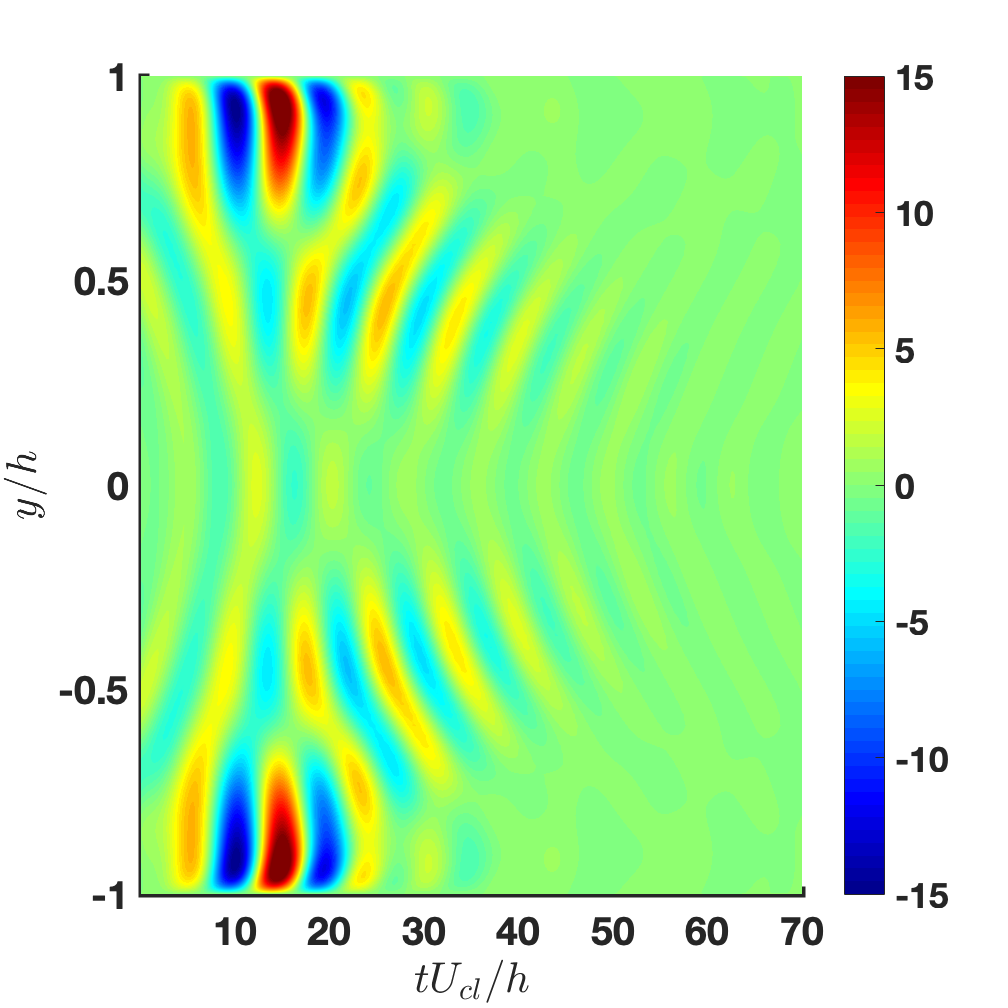}
\label{fig:Case2_ulqr}}\\
\subfloat[Uncontrolled]{\includegraphics[scale=0.3]{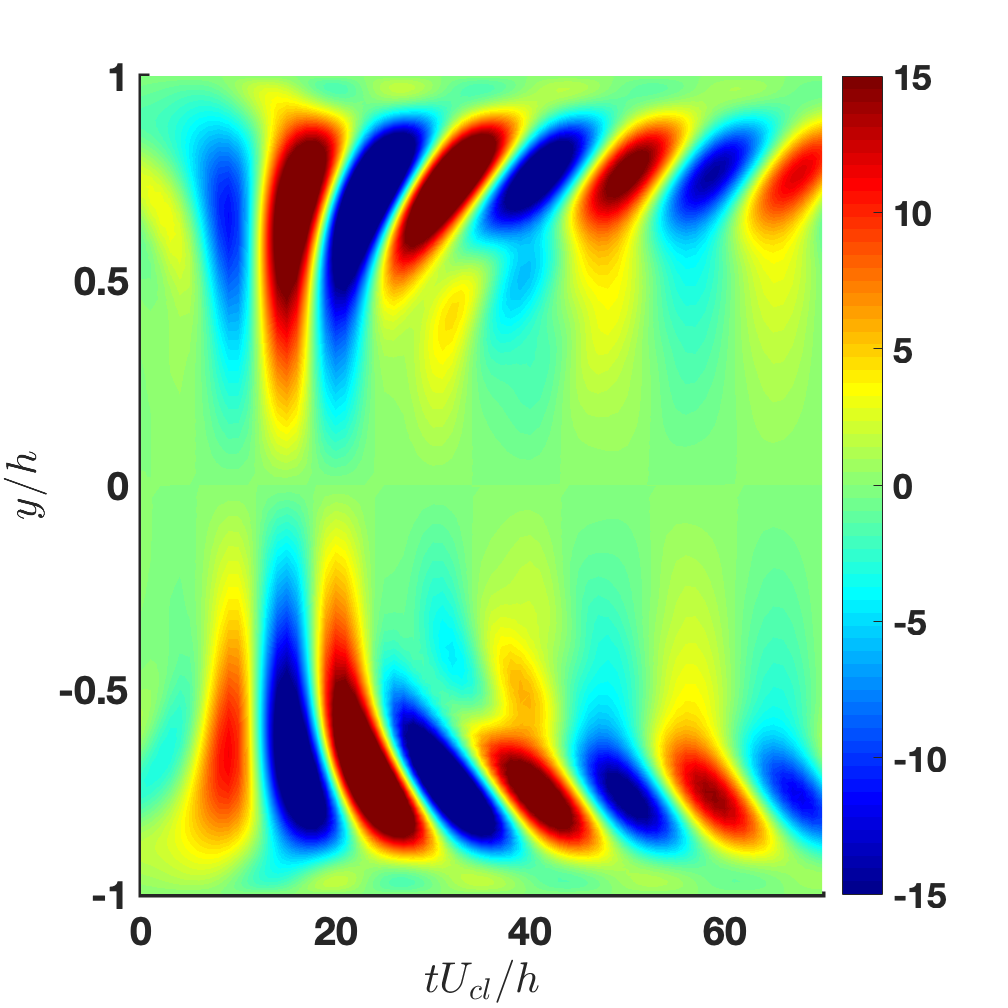}}
\caption{Evolution of streamwise velocity perturbations~($u$) for $r=58$, $(\alpha,\beta)=(1,1)$, and $Re=3000$.}
%  The time evolution of the real part of $u$ is shown against the channel height on the y-axis. The initial conditions of the three systems is the respective closed loop optimal perturbation. In Fig.~\ref{fig:Case2_ufom},\ref{fig:Case2_ulqr} controllers perform similar to each other. The magnitude of $u$ given by the LMI-ROM controller shows a reduction in settling time and a decrease in intensity of $u$ at the lower wall. Results for $\alpha=1,\beta=1$ at $Re=3000$ and ROM of size $r=58$.}
\label{fig:Case2_uall}
\end{figure}

\begin{figure}[!htbp]
\centering
\subfloat[LMI-ROM control]{\includegraphics[scale=0.3]{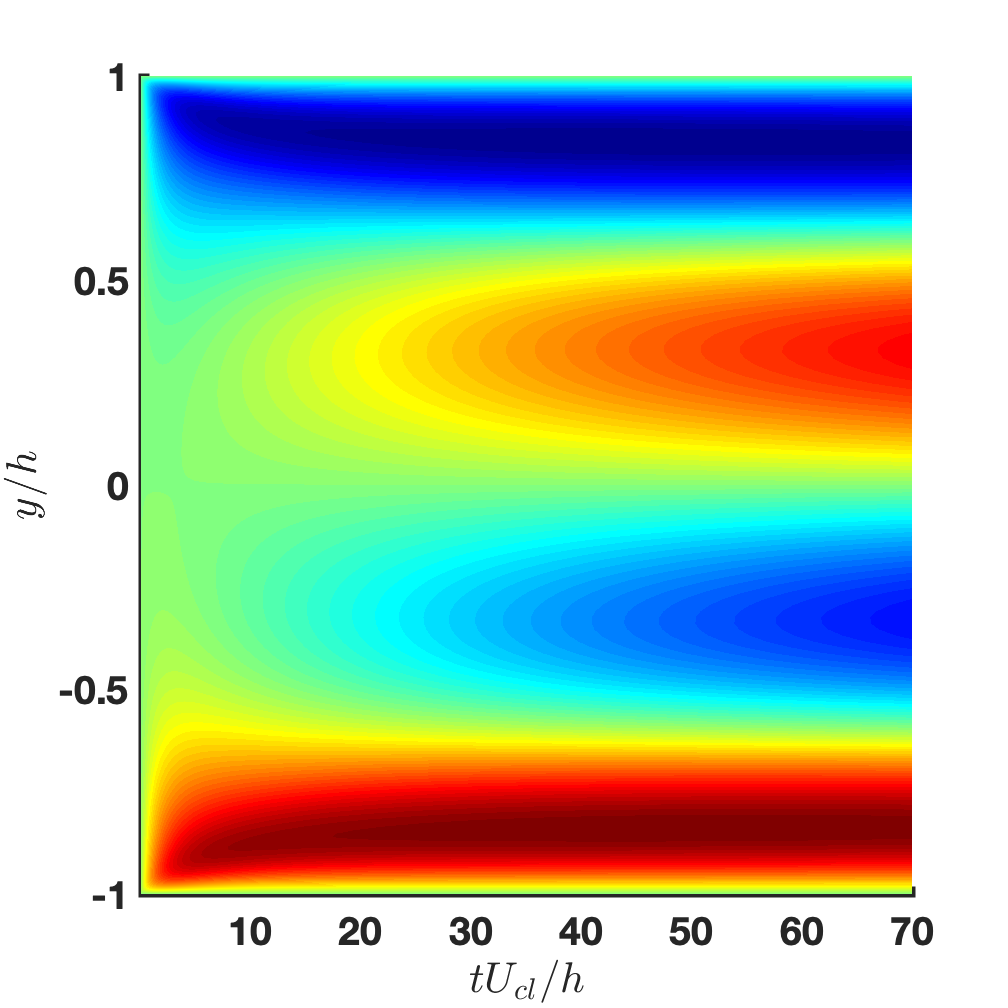}
\label{fig:Case3_ulmi}}
\hfill
\subfloat[LQR-FOM control]{\includegraphics[scale=0.3]{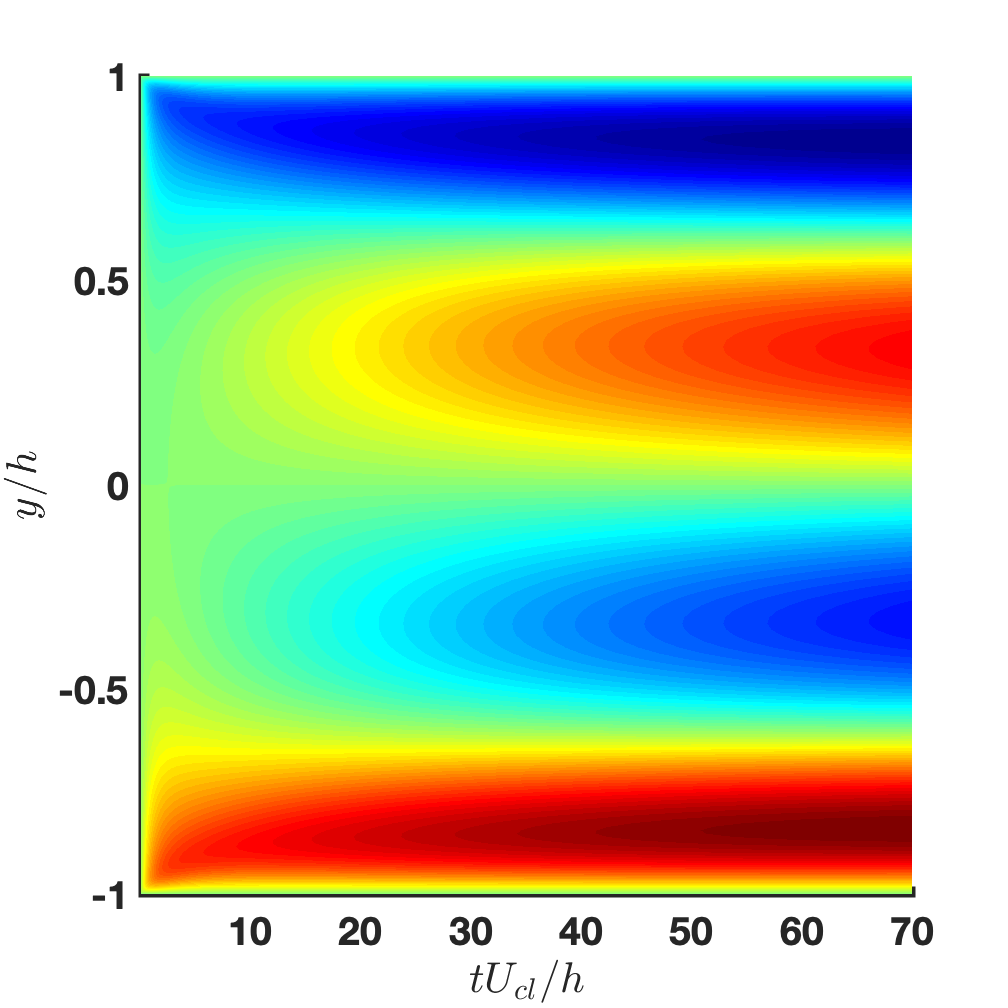}
\label{fig:Case3_ufom}}
\hfill
\subfloat[LQR-ROM control]{\includegraphics[scale=0.3]{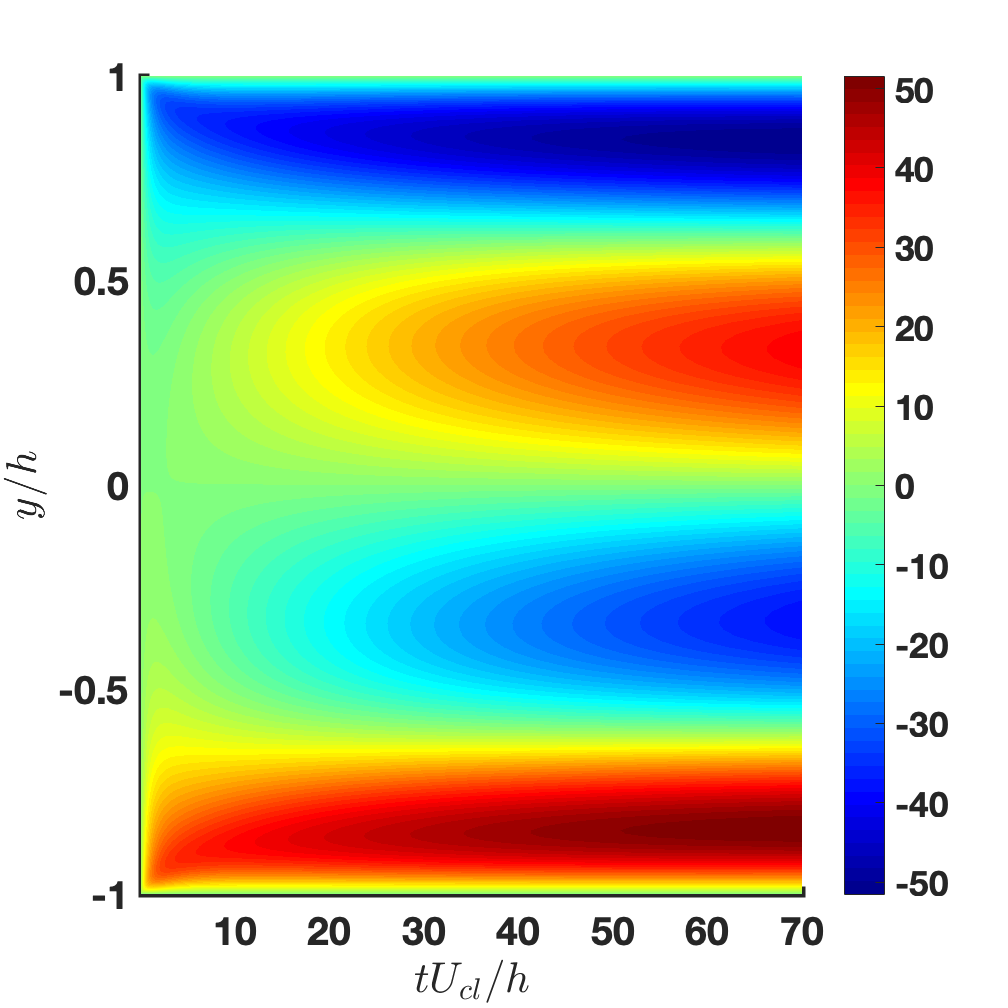}
\label{fig:Case3_ulqr}}\\
\subfloat[Uncontrolled]{\includegraphics[scale=0.3]{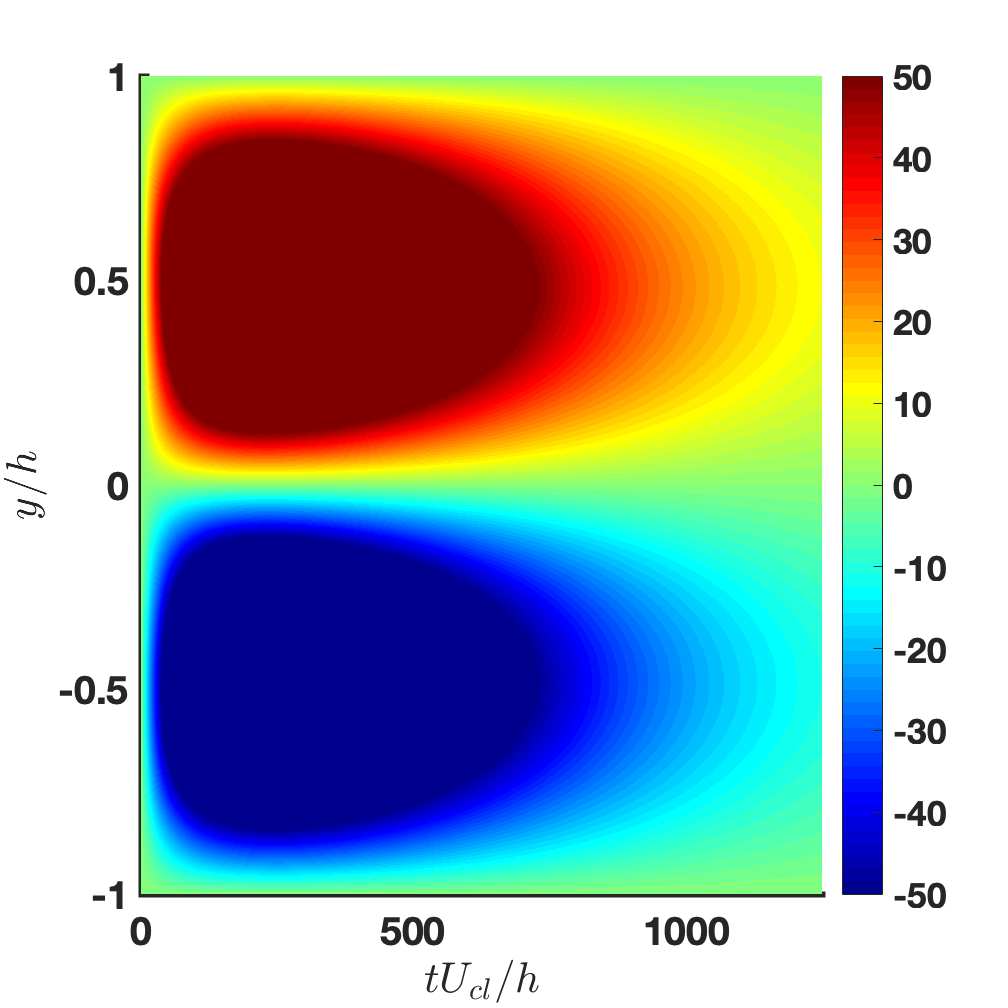}}
\caption{Evolution of streamwise velocity perturbations~($u$) for $r=40$, $(\alpha,\beta)=(0,2)$, and $Re=3000$.}
  %The real part of $u$ is shown against the channel height on the y-axis. In Fig.~\ref{fig:Case3_ufom},\ref{fig:Case3_ulqr} controllers perform similar to each other. The $u$ for the three controller shows a similar profile qualitatively. Results for $\alpha=0,\beta=2$ at $Re=3000$ and ROM of size $r=40$}
\label{fig:Case3_uall}
\end{figure}

%Next, we examine the effect of each control method on the flow. %streamwise velocity~($u$) perturbations. \msh{why $u$?  why not other quantities as well?}\ak{The goal is to show how the perturbations evolve/decay in time and in that case u,v,w would tell the same story. So I choose u, moreover, I do plot v velocity at the boundary~(control input) because it is more intuitive to interpret the results.}
Next, we examine the evolution of perturbations in streamwise velocity~($u$)
to analyze the effect of the different controllers on
the flow response. %(see Figures~\ref{fig:Case1_uall}--\ref{fig:Case3_uall}). 
 \ak{Figures~\ref{fig:Case1_uall} -- \ref{fig:Case3_uall} show the evolution of normalized $u$ perturbations}. In these figures, the initial optimal perturbation has been normalized such that $E(0)=1$.
%\ak{In these figures, the magnitude of $u$ is relative to $u_0$, where $u_0$ is the magnitude of velocity such that TEG at $t = 0$ is normalized to $E_0=1$.}
%
%that the real part of the ~$u$'s time evolution is plotted against the channel height.
Figure~\ref{fig:Case1_uall} shows the response for the case of streamwise waves.
Note that the LMI-ROM controller has a streamwise perturbation profile different
from that of the LQR-ROM and the LQR-FOM.
The $u$ perturbations die out faster with the LMI-ROM controller than with either of the LQR controllers. % in Figures~\ref{fig:Case1_ulqr} and \ref{fig:Case1_ufom}.
The LQR-FOM and the LQR-ROM result in a similar evolution of $u$ perturbation,
as is to be expected from the similarity in TEG profiles we
noted earlier in Figure~\ref{fig:Case1_Re3e3}.
In the case of oblique waves (see Figure~\ref{fig:Case2_uall}),
the LQR-ROM and LQR-FOM again yield a similar response in $u$-perturbations.
In contrast to these responses, the LMI-ROM controller results in a
reduced magnitude of $u$ at the lower walls of the channel.
Again, the LMI-ROM results in a faster decay of $u$ perturbations than the LQR-ROM and LQR-FOM controllers.
Finally, in the case of spanwise waves (see Figure~\ref{fig:Case3_uall}),
all the three controllers yield similar responses in $u$.

%---streamwise waves: u streamwise velocity for closed loop 

%---------------------------------
\begin{comment}
\begin{figure}[!h]
    \centering
    \includegraphics[width=\textwidth]{Case_1/combined_u.png}
    \caption{Time evolution of the closed loop systems streamwise velocity~($u$) perturbations. The time evolution of the real part of $u$ is shown against the channel height on the y-axis. The initial conditions of the three systems is the respective closed loop optimal perturbation. In Fig.~\ref{fig:Case1_ufom},\ref{fig:Case1_ulqr} controllers perform similar to each other. The $u$ profile of the LMI controller is qualitatively different from that of the two LQR controllers. Results for $\alpha=1,\beta=0$ at $Re=3000$ and ROM of size $r=40$}
    \label{fig:my_label}
\end{figure}
\end{comment}
%---------------------------------
%oblique waves:u streamwise velocity for closed loop 

%spanwise waves:u streamwise velocity for closed loop 

\begin{figure}[h!]
\centering
\subfloat[$Re= 1000$ ($r=40$)]{\includegraphics[width=0.3\textwidth]{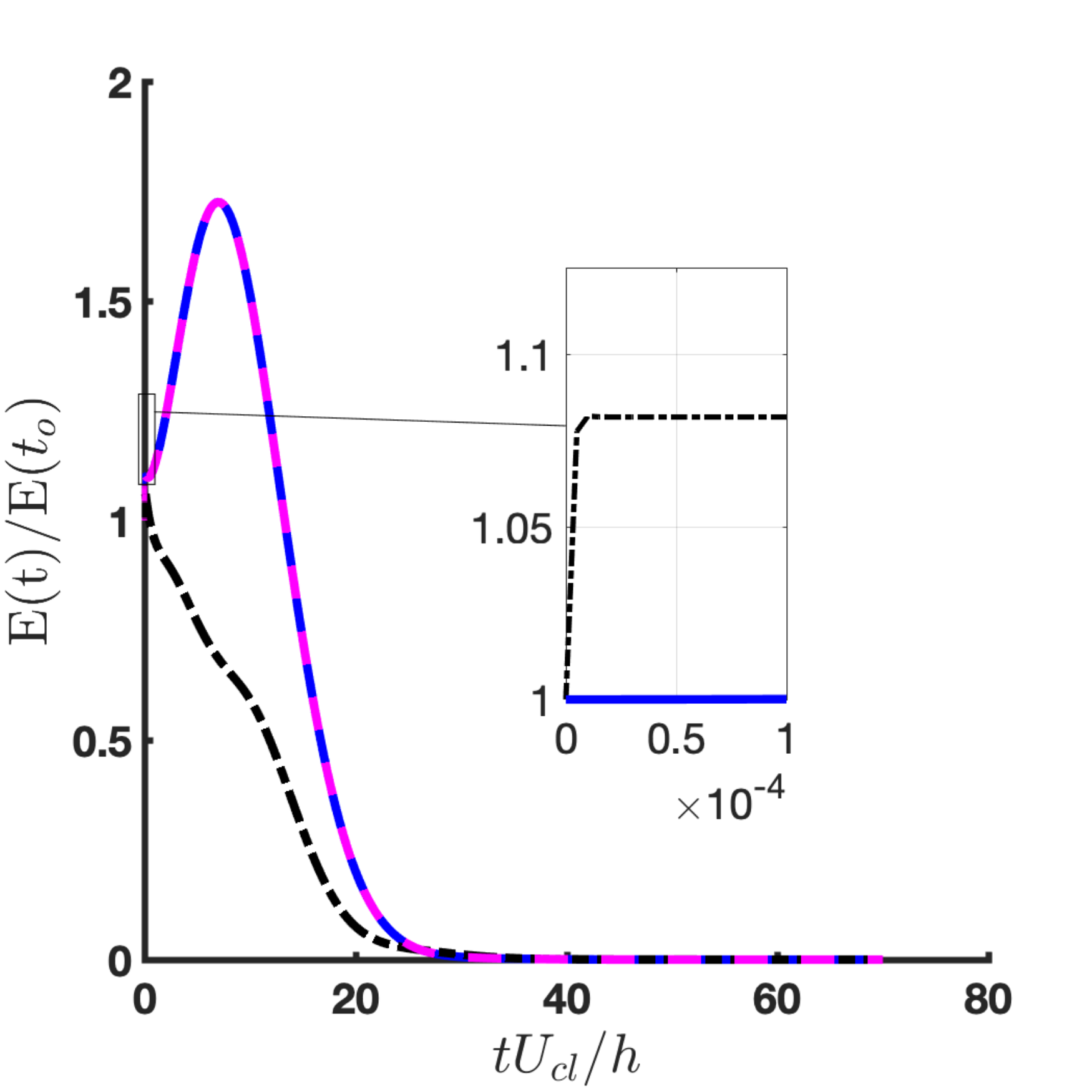}
\label{fig:Case1_Re1e3}}
\hfill
\subfloat[$Re=5000$ ($r=50$)]{\includegraphics[width=0.3\textwidth]{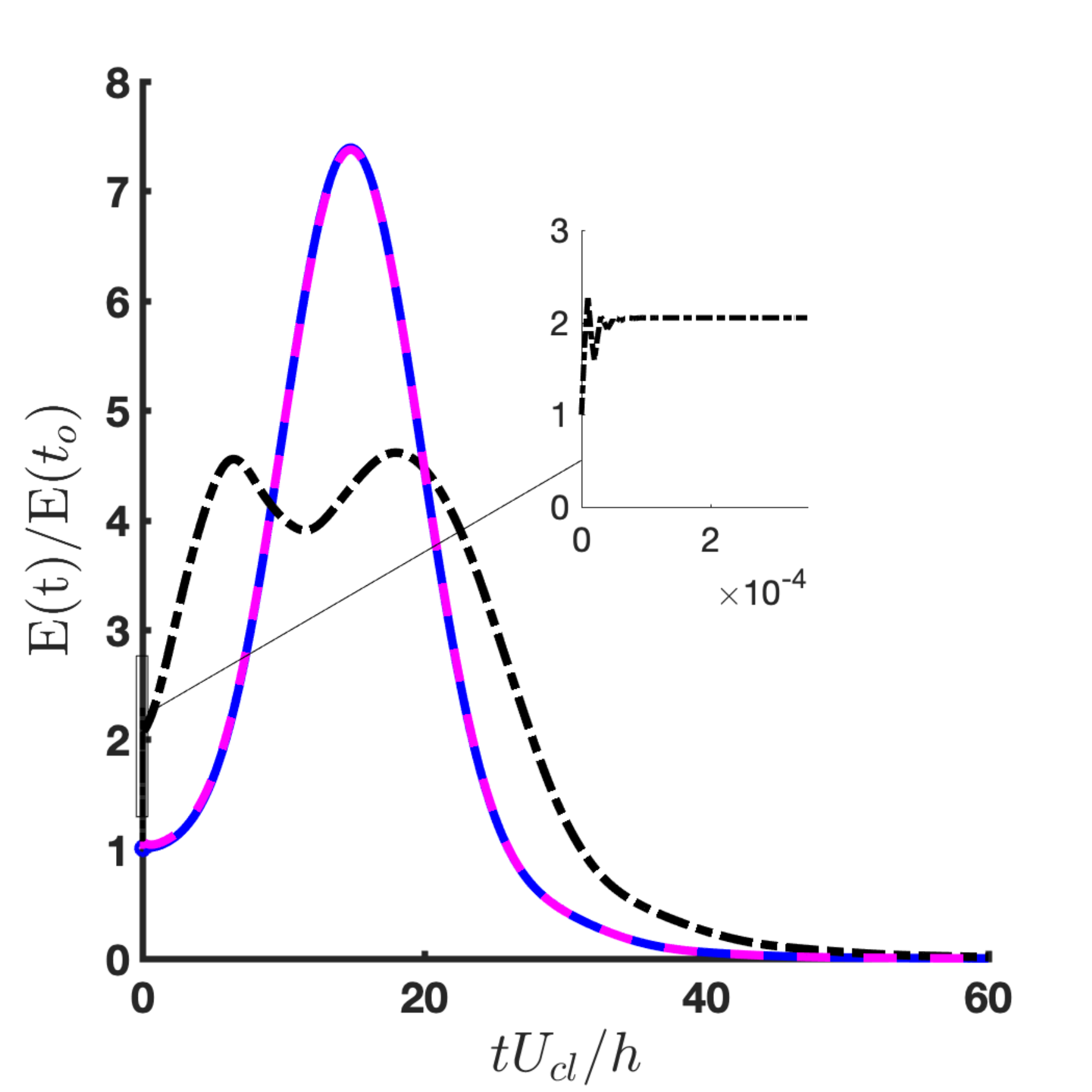}
\label{fig:Case1_Re5e3}}
\hfill
\subfloat[ $Re=10,000$ ($r=58$)]{\includegraphics[width=0.3\textwidth]{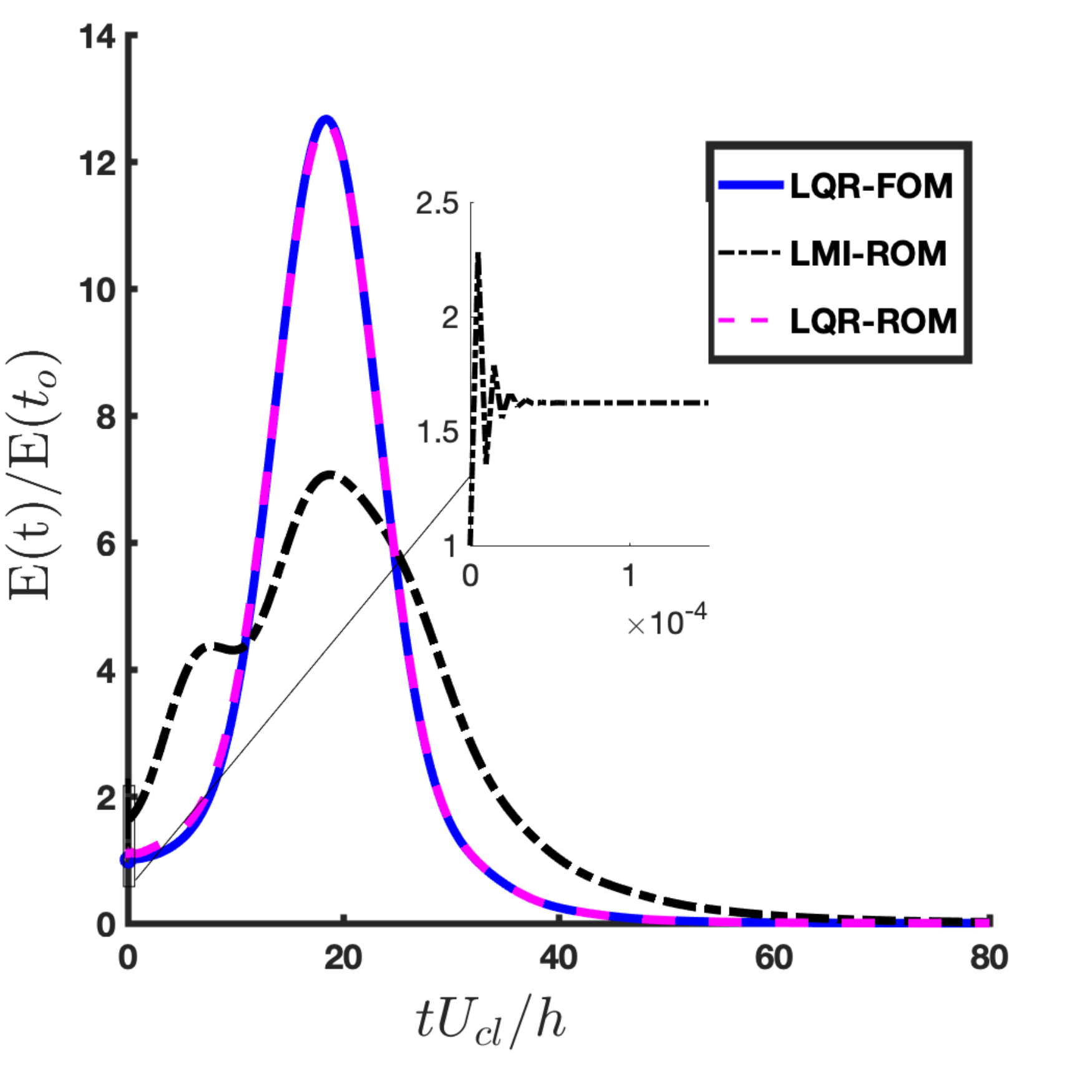}
\label{fig:Case1_Re1e4}}
\caption{Worst-case closed-loop TEG responses for $(\alpha,\beta)=(1,0)$.}
%  The LMI-ROM controller suppress TEG effectively in comparison with the $\text{LQR}-{\text{FOM}}$ and $\text{LQR}-{\text{ROM}}$ controllers. The figure shows the controller performance for wave-number pair ($\alpha=1,\beta=0$) in the sub-critical regimes ($Re= 1000,5000$) and the unstable regime ($Re=10000$). The inset in~\ref{fig:Case1_Re1e3},\ref{fig:Case1_Re5e3} \& \ref{fig:Case1_Re1e4} shows changes observed on a transient time scales}
\label{fig:Case1_Re_Sweep}
\end{figure}

\begin{figure}[!htbp]
\centering
\subfloat[$Re=1000$ ($r=40$)]{\includegraphics[width=0.3\textwidth]{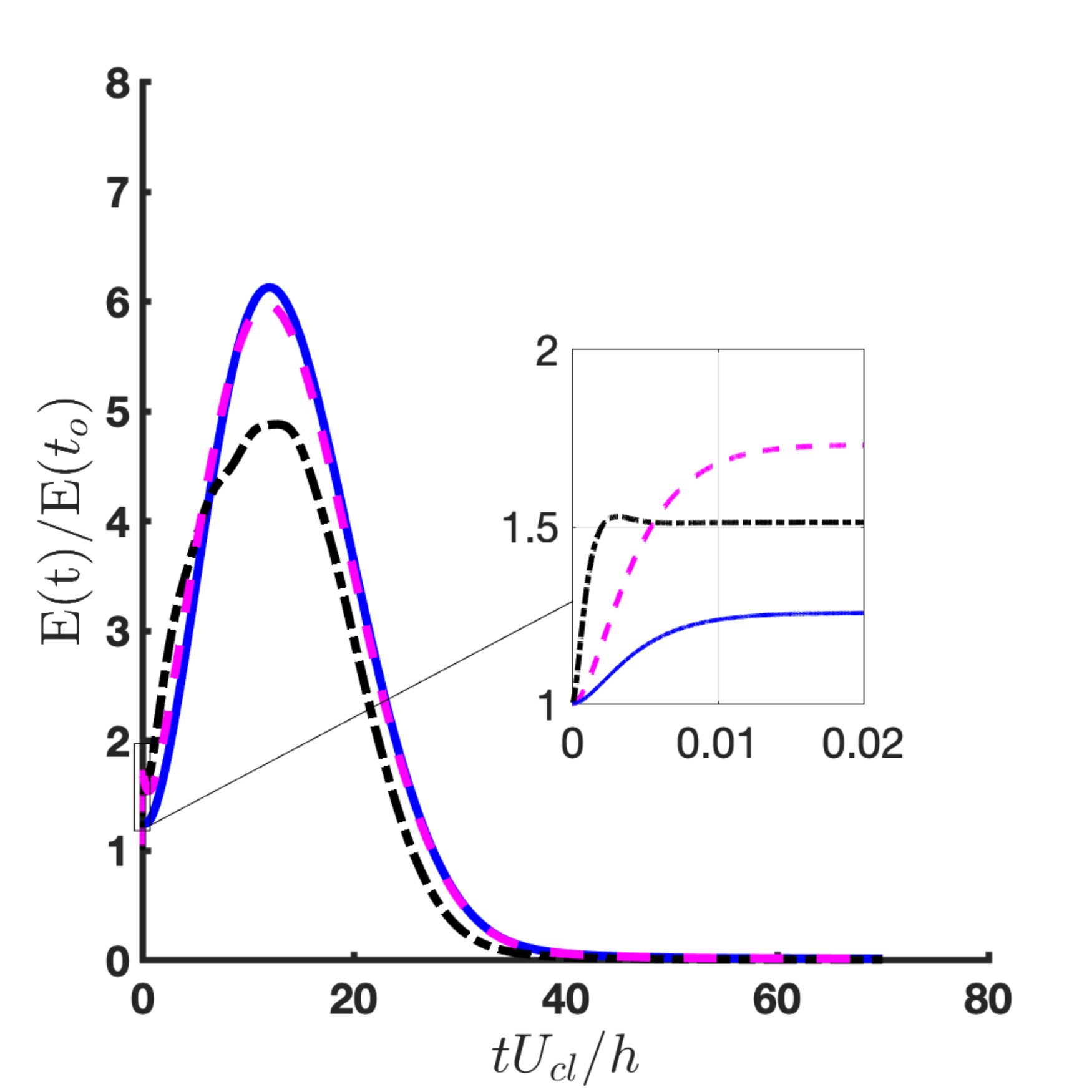}
\label{fig:Case2_Re1e3}}
\hfill
\subfloat[$Re=5000$ ($r=58$)]{\includegraphics[width=0.3\textwidth]{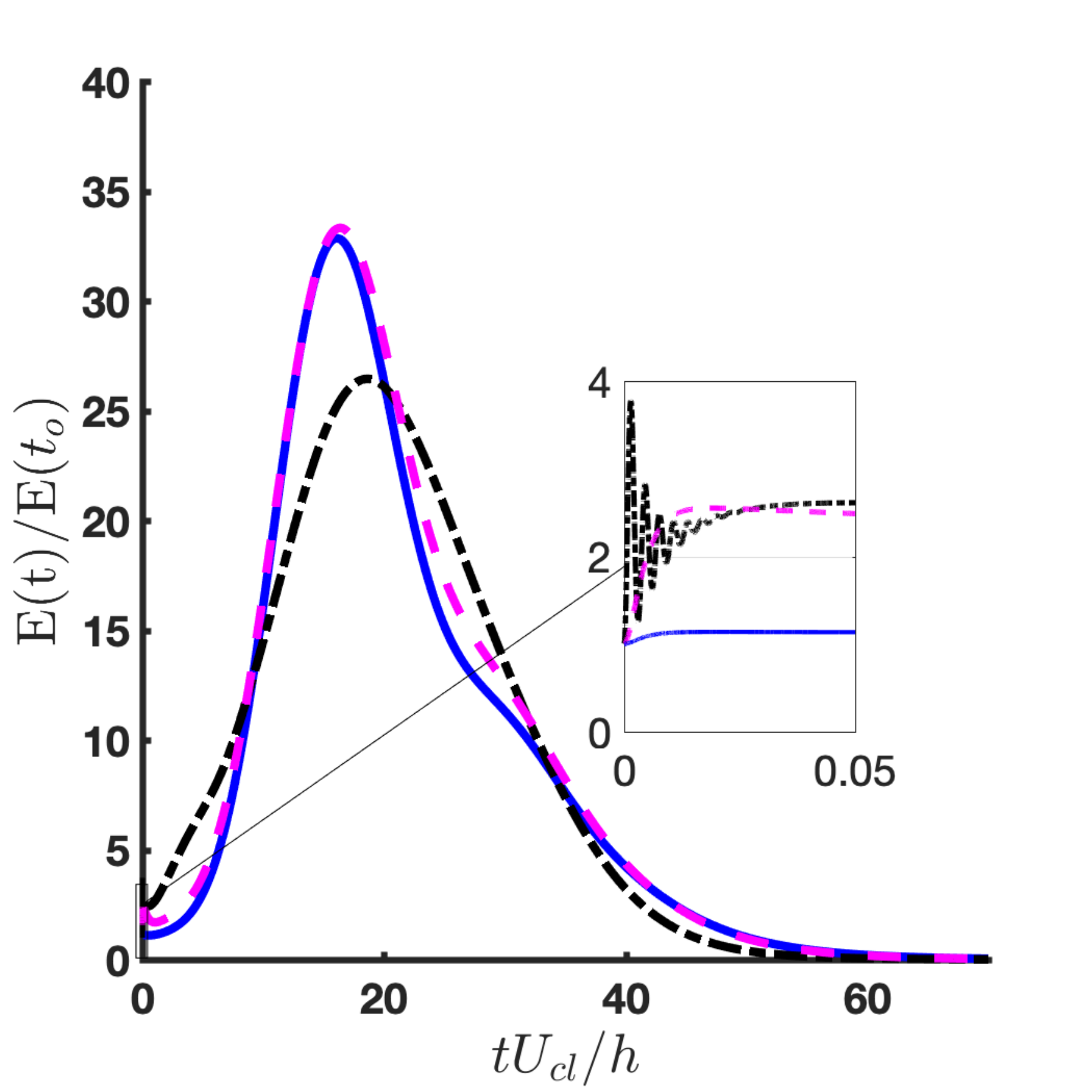}
\label{fig:Case2_Re5e3}}
\hfill
\subfloat [$Re=10,000$ ($r=58$)]{\includegraphics[width=0.3\textwidth]{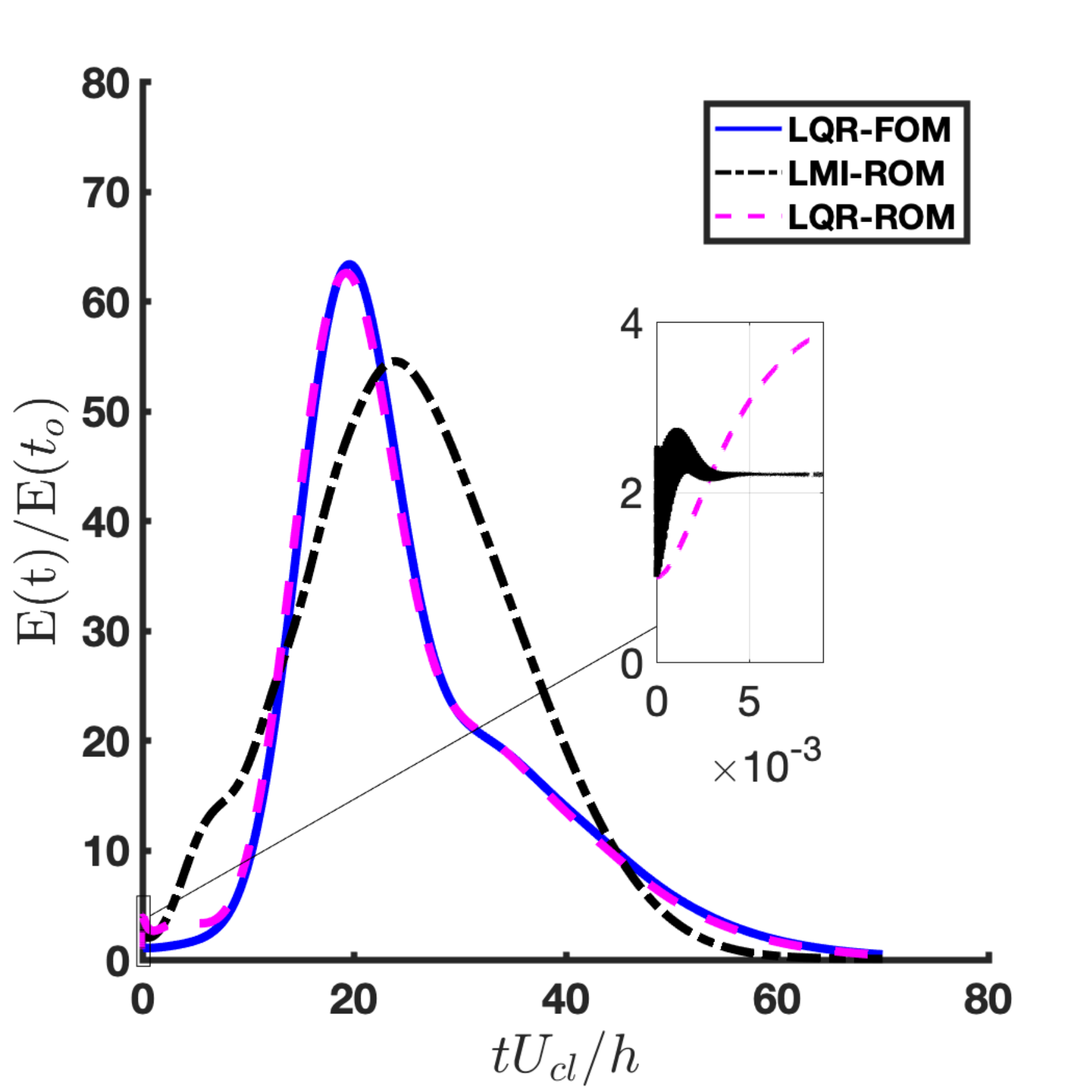}
\label{fig:Case2_Re1e4}}
\caption{Worst-case closed-loop TEG responses for $(\alpha,\beta)=(1,1)$.}
%  The LMI-ROM controller suppress TEG effectively in comparison with the $\text{LQR}-{\text{FOM}}$ and $\text{LQR}-{\text{ROM}}$ controllers. The figure shows the controller performance for wave-number pair ($\alpha=1,\beta=1$) in the sub-critical regimes ($Re= 1000,5000$) and the unstable regime ($Re=10000$). The inset in~\ref{fig:Case2_Re1e3},~\ref{fig:Case2_Re5e3} \& \ref{fig:Case2_Re1e4} shows changes observed on a transient time scales.}
\label{fig:Case2_Re_Sweep}
\end{figure}

% The above results show the performance of the controllers for a single $Re=3000$. However, for completeness of the study,
Finally, we repeat our study for other $Re$ values to ensure that the ROMs and controllers
can be used in other settings as well.
%
%we examine the controllers designed at different $Re$.
ROMs are developed and tuned to converge for $Re=1000$, $5000$, and $10,000$.
Then, we follow the same procedure for control synthesis as in the $Re=3000$
cases described earlier.
Note that for the case of $Re=10,000$, the flow is linearly unstable;
thus, the balanced truncation procedure for the model reduction is only performed on
the stable subspace of the linearized dynamics.
%
% We show that the LMI-ROM controllers successfully minimizes TEG in comparison to the baseline LQR controllers developed on the FOM and ROM respectively.  For $Re= 1000, 5000$ and $10,000$ the controller performance is shown for streamwise waves and oblique waves.
Figures~\ref{fig:Case1_Re_Sweep} and \ref{fig:Case2_Re_Sweep} report performance results for the cases of $(\alpha,\beta)=(1,0)$ and $(1,1)$, respectively.
The results clearly show that the LMI-ROM controllers reduce TEG to a \ak{greater} extent than
the LQR controllers.
However, with the increase in $Re$,  the closed-loop response to the LMI-ROM control
exhibits oscillations on a transient time scale, on the order of of approximately $ 10^{-2}~\text{to}~10^{-4}$ convective time units.
%
% This oscillation is not an effect of control action itself, as the controller does not expend any control effort in the corresponding small timescales.
%
The transient oscillations appear to be a property of the closed-loop system, which has a set of eigenvalues with large imaginary parts---on the order of $10^4$---that lead to
lightly damped oscillations.
Further investigation is necessary to determine the
underlying cause of these transient oscillations and
the observed spectral properties of the closed-loop system.

%We also perform a study on  system Re

%Case 1: streamwise waves
%case 2: Oblique waves

\section{Conclusions}
\label{sec:conclusion}
In this paper, we have investigated the use of control-oriented ROMs for designing
feedback controllers that minimize the MTEG of flow perturbations within
a linearized channel flow.
The ROMs were formed using an output projection onto POD modes, followed by
a balanced truncation procedure to reduce the state dimension.
POD modes were chosen in order to best approximate the perturbation energy, \ak{as needed for representing the objective function for control}.
\ak{Note that this output projection does not alter the state dimension of the system.
As such, a}
balanced truncation was performed to \ak{reduce the state dimension, retaining only state variables that were
essential to preserving the system's input-output dynamics.}  % \ak{Hence, the combined use of balanced truncation and output projection (onto POD modes) is essential to control-oriented reduced-order modeling for transient energy growth reduction.}
ROMs of the linearized channel flow system enabled the synthesis of MTEG-minimizing controllers through the solution of an LMI problem that would otherwise have been computationally prohibitive.
Specifically, the dimension of the full-order model ($n=199$) was reduced to yield ROMs with order $r=40-60$, depending on the specific configuration.
This constitutes a significant reduction in the computational demand for
the subsequent LMI-based controller synthesis,
since the computational requirements for LMI-based controller
synthesis scale as $\mathcal{O}(n^6)$.
Further, the MTEG-minimizing controllers designed using the proposed ROMs were found to
outperform LQR controllers in suppressing TEG, even when these LQR controllers
were designed based on the full-order system model.
%
%reassuring the fact that control-oriented ROMs are beneficial.
Although not explicitly reported here,
the present investigation revealed that ROM-based MTEG-minimizing controllers
can be sensitive to $Re$ variations, leading to linearly unstable closed-loop
systems when applied in ``off-design'' settings.
As such, future investigations will need to focus on addressing these fragilities.
It is expected that both the model reduction and controller synthesis
approaches will need to be re-formulated to robustly minimize MTEG. 
%

%
 % We also investigate the LMI controllers performance in comparison with the LQR controller, and analyze the controllers and the physics of the flow being altered. Apart from developing computationally expensive LMI controllers, the ROM also enables us to successfully leverage the advantage of retaining the most controllable and observable modes, and the physical energy of the system---to facilitate the control design for reduction in transient energy growth.
% In this work,  we show that using the reduced order model formulated here we can develop LMI controllers that outperform the commonly used LQR controllers. 
 %The control-oriented reduced-order models are shown to facilitate the aspect of TEG reduction control synthesis when compared with the global mode truncation method. 
 
 % \ak{Future work?}
 % The present paper establishes the advantages of using control-oriented ROMs and also examines the effectiveness of the energy minimizing LMI controllers. That being said, additional work has to be done in robustifying the LMI framework to parameter uncertainties. As it was noted in this study that the LMI framework is sensitive to changes in the $Re$.The lack of robustness is a disadvantage for flow control applications since exact models are rare.  Another important aspect to investigate, is the process of obtaining POD modes and its sensitivity to the controller synthesis process. Further, performance of these control schemes using non-linear direct numerical methods can be investigated.

\section{Acknowledgements}
This material is based upon work supported by the Air Force Office of Scientific Research under award numbers FA9550-17-1-0252 and FA9550-19-0034, monitored by Drs. Douglas R. Smith and  Gregg Abate.

\bibliographystyle{aiaa}
\bibliography{Journal_AK}

\begin{thebibliography}{38}
\newcommand{\enquote}[1]{``#1''}
\providecommand{\natexlab}[1]{#1}
\providecommand{\url}[1]{\texttt{#1}}
\providecommand{\urlprefix}{URL }
\expandafter\ifx\csname urlstyle\endcsname\relax
  \providecommand{\doi}[1]{doi:\discretionary{}{}{}#1}\else
  \providecommand{\doi}{doi:\discretionary{}{}{}\begingroup
  \urlstyle{rm}\Url}\fi

\bibitem[{Schmid and Henningson(2001)}]{Schmid2001}
Schmid, P.~J., and Henningson, D.~S., \emph{{Stability and transition in shear
  flows}}, Springer, 2001.
\newblock \doi{10.1115/1.1470687}.

\bibitem[{Patel and Head(1969)}]{PatelJFM1969}
Patel, V.~C., and Head, M.~R., \enquote{{Some observations on skin friction and
  velocity profiles in fully developed pipe and channel flows},} \emph{Journal
  of Fluid Mechanics}, Vol.~38, No.~1, 1969, pp. 181--201.
\newblock \doi{10.1017/S0022112069000115}.

\bibitem[{Reddy and Henningson(1993)}]{ReddyJFM1993}
Reddy, S.~C., and Henningson, D.~S., \enquote{{Energy growth in viscous channel
  flows},} \emph{Journal of Fluid Mechanics}, Vol. 252, No.~-1, 1993, p. 209.
\newblock \doi{10.1017/S0022112093003738}.

\bibitem[{Trefethen et~al.(1993)Trefethen, Trefethen, Reddy, and
  Driscoll}]{Trefethen1993}
Trefethen, L.~N., Trefethen, A.~E., Reddy, S.~C., and Driscoll, T.~A.,
  \enquote{{Hydrodynamic stability without eigenvalues.}} \emph{Science (New
  York, N.Y.)}, Vol. 261, No. 5121, 1993, pp. 578--84.
\newblock \doi{10.1126/science.261.5121.578}.

\bibitem[{Henningson and Reddy(1994)}]{HenningsonPOF1994}
Henningson, D.~S., and Reddy, S.~C., \enquote{{On the role of linear mechanisms
  in transition to turbulence},} \emph{Physics of Fluids}, Vol.~6, No.~3, 1994,
  pp. 1396--1398.
\newblock \doi{10.1063/1.868251}.

\bibitem[{Schmid(2007)}]{Schmid2007}
Schmid, P.~J., \enquote{{Nonmodal Stability Theory},} \emph{Annual Review of
  Fluid Mechanics}, Vol.~39, 2007, pp. 129--62.
\newblock \doi{10.1146/annurev.fluid.38.050304.092139}.

\bibitem[{Chapman(2002)}]{Chapman2002}
Chapman, S.~J., \enquote{{Subcritical transition in channel flows},}
  \emph{Journal of Fluid Mechanics}, Vol. 451, 2002, pp. 35--97.
\newblock \doi{10.1017/S0022112001006255}.

\bibitem[{Bewley(2001)}]{Bewley2001}
Bewley, T.~R., \enquote{{Flow control: New challenges for a new Renaissance},}
  \emph{Progress in Aerospace Sciences}, Vol.~37, No.~1, 2001, pp. 21--58.
\newblock \doi{10.1016/S0376-0421(00)00016-6}.

\bibitem[{Bagheri and Henningson(2011)}]{Bagheri2011}
Bagheri, S., and Henningson, D.~S., \enquote{{Transition delay using control
  theory},} \emph{Philosophical Transactions: Mathematical, Physical and
  Engineering Sciences}, Vol. 369, 2011.
\newblock \doi{10.1098/rsta.2010.0358}.

\bibitem[{Kim and Bewley(2007)}]{KimAnnRev2006}
Kim, J., and Bewley, T.~R., \enquote{A Linear Systems Approach to Flow
  Control,} \emph{Annual Review of Fluid Mechanics}, Vol.~39, No.~1, 2007, pp.
  383--417.
\newblock \doi{10.1146/annurev.fluid.39.050905.110153}.

\bibitem[{Bewley and Liu(2018)}]{BewleyJFM1998}
Bewley, T.~R., and Liu, S., \enquote{{Optimal and robust control and estimation
  of linear paths to transition},} \emph{Journal of Fluid Mechanics}, Vol. 365,
  2018, pp. 305--349.
\newblock \doi{10.1017/S0022112098001281}.

\bibitem[{Ilak and Rowley(2008{\natexlab{a}})}]{ilakAIAA2008}
Ilak, M., and Rowley, C.~W., \enquote{Feedback Control of Transitional Channel
  Flow using Balanced Proper Orthogonal Decomposition,} \emph{AIAA Theoretical
  Fluid Mechanics Conference, Seattle, WA, AIAA paper 2008-4230},
  2008{\natexlab{a}}.

\bibitem[{Martinelli et~al.(2011)Martinelli, Quadrio, Mckernan, and
  Whidborne}]{MartinelliPoF2011}
Martinelli, F., Quadrio, M., Mckernan, J., and Whidborne, J.~F.,
  \enquote{{Linear feedback control of transient energy growth and control
  performance limitations in subcritical plane Poiseuille flow},} \emph{Physics
  of Fluids}, 2011.
\newblock \doi{10.1063/1.3540672}.

\bibitem[{Hogberg et~al.(2003)Hogberg, Bewley, and Henningson}]{Hogberg2003}
Hogberg, M., Bewley, T.~R., and Henningson, D.~S., \enquote{{Linear feedback
  control and estimation of transition in plane channel flow},} \emph{Journal
  of Fluid Mechanics}, Vol. 481, 2003, pp. 149--175.
\newblock \doi{10.1017/S0022112003003823}.

\bibitem[{Joshi et~al.(1999)Joshi, Speyer, and Kim}]{Joshi1999}
Joshi, S.~S., Speyer, J.~L., and Kim, J., \enquote{{Finite Dimensional Optimal
  Control of Poiseuille Flow},} \emph{Journal of Guidance, Control, and
  Dynamics}, Vol.~22, No.~2, 1999.
\newblock \doi{10.2514/2.4383}.

\bibitem[{Hemati and Yao(2018)}]{Hemati2018}
Hemati, M.~S., and Yao, H., \enquote{{Performance Limitations of Observer-Based
  Feedback for Transient Energy Growth Suppression},} \emph{AIAA Journal},
  2018.
\newblock \doi{10.2514/1.j056877}.

\bibitem[{Yao and Hemati(2018)}]{YaoAIAA2018}
Yao, H., and Hemati, M.~S., \enquote{Revisiting the separation principle for
  improved transition control,} \emph{AIAA Paper 2018-3693}, 2018.
\newblock \doi{10.2514/6.2018-3693}.

\bibitem[{Yao and Hemati(2019)}]{YaoAIAA2019}
Yao, H., and Hemati, M.~S., \enquote{{Advances in Output Feedback Control of
  Transient Energy Growth in a Linearized Channel Flow},} \emph{AIAA Paper
  2019-0882}, 2019.
\newblock \doi{10.2514/6.2019-0882}.

\bibitem[{Whidborne and McKernan(2007)}]{WhidborneIEEE2007}
Whidborne, J.~F., and McKernan, J., \enquote{{On the Minimization of Maximum
  Transient Energy Growth},} \emph{IEEE Transactions on Automatic Control},
  Vol.~52, No.~9, 2007, pp. 1762--1767.
\newblock \doi{10.1109/TAC.2007.900854}.

\bibitem[{Butler and Farrell(1992)}]{ButlerPOF1992}
Butler, K.~M., and Farrell, B.~F., \enquote{{Three-dimensional optimal
  perturbations in viscous shear flow},} \emph{Physics of Fluids A: Fluid
  Dynamics}, Vol.~4, No.~8, 1992, pp. 1637--1650.
\newblock \doi{10.1063/1.858386}.

\bibitem[{Jovanovic and Bameih(2005)}]{JovanovicJFM2005}
Jovanovic, M.~R., and Bameih, B., \enquote{Componentwise energy amplification
  in channel flows,} \emph{Journal of Fluid Mechanics}, Vol. 534, 2005, p.
  145–183.
\newblock \doi{10.1017/S0022112005004295}.

\bibitem[{Boyd et~al.(1994)Boyd, {El Ghaoui}, Feron, and
  Balakrishnan}]{BoydLMI1994}
Boyd, S., {El Ghaoui}, L., Feron, E., and Balakrishnan, V., \emph{{Linear
  Matrix Inequalities in System and Control Theory}}, Society for Industrial
  and Applied Mathematics, 1994.
\newblock \doi{10.1137/1.9781611970777}.

\bibitem[{Barbagallo et~al.(2012)Barbagallo, Sipp, and Schmid}]{barbagallo2012}
Barbagallo, A., Sipp, D., and Schmid, P.~J., \enquote{Reduced Order Models for
  Closed Loop Control: Comparison between POD, BPOD, and Global Modes,}
  \emph{Progress in Flight Physics}, Vol.~3, 2012, pp. 503--512.
\newblock \doi{10.1051/eucass/201203503}.

\bibitem[{Jones et~al.(2015)Jones, Heins, Kerrigan, Morrison, and
  Sharma}]{Jones2015}
Jones, B.~L., Heins, P.~H., Kerrigan, E.~C., Morrison, J.~F., and Sharma,
  A.~S., \enquote{{Modelling for robust feedback control of fluid flows},}
  \emph{Journal of Fluid Mechanics}, Vol. 769, 2015, pp. 687--722.
\newblock \doi{10.1017/jfm.2015.84}.

\bibitem[{Taira et~al.(2017)Taira, Brunton, Dawson, Rowley, Colonius, Mckeon,
  Schmidt, Gordeyev, Theofilis, and Ukeiley}]{Taira2017}
Taira, K., Brunton, S.~L., Dawson, S. T.~M., Rowley, C.~W., Colonius, T.,
  Mckeon, B.~J., Schmidt, O.~T., Gordeyev, S., Theofilis, V., and Ukeiley,
  L.~S., \enquote{{Modal Analysis of Fluid Flows: An Overview},} \emph{AIAA
  Journal}, 2017.
\newblock \doi{10.2514/1.J056060}.

\bibitem[{Rowley and Dawson(2017)}]{DawsonAnnRev2017}
Rowley, C.~W., and Dawson, S. T.~M., \enquote{{Model Reduction for Flow
  Analysis and Control},} \emph{Annual Review of Fluid Mechanics}, Vol.~49,
  2017, pp. 387--417.
\newblock \doi{10.1146/annurev-fluid-010816-060042}.

\bibitem[{Rowley(2005)}]{Rowley2005}
Rowley, C.~W., \enquote{{Model reduction for fluids using balanced proper
  orthogonal decomposition},} \emph{International Journal on Bifurcation and
  Chaos}, Vol.~15, No.~3, 2005, pp. 997--1013.
\newblock \doi{10.1142/S0218127405012429}.

\bibitem[{Ilak and Rowley(2008{\natexlab{b}})}]{ilakPOF2008a}
Ilak, M., and Rowley, C.~W., \enquote{{Modeling of transitional channel flow
  using balanced proper orthogonal decomposition},} \emph{Physics of Fluids},
  Vol.~20, 2008{\natexlab{b}}, p. 34103.
\newblock \doi{10.1063/1.2840197}.

\bibitem[{Moore(1981)}]{Moore1981}
Moore, B., \enquote{{Principal component analysis in linear systems:
  Controllability, observability, and model reduction},} \emph{IEEE
  Transactions on Automatic Control}, Vol.~26, No.~1, 1981, pp. 17--32.
\newblock \doi{10.1109/TAC.1981.1102568}.

\bibitem[{Antoulas(2006)}]{Antoulas2006}
Antoulas, A.~C., \emph{{Approximation of Large-Scale Dynamical Systems}},
  Society for Industrial and Applied Mathematics, 2006.

\bibitem[{Willcox and Peraire(2002)}]{Willcox2002}
Willcox, K., and Peraire, J., \enquote{{Balanced Model Reduction via the Proper
  Orthogonal Decomposition},} \emph{AIAA Journal}, Vol.~40, No.~11, 2002.
\newblock \doi{10.2514/2.1570}.

\bibitem[{Grant and Boyd(2014)}]{cvx}
Grant, M., and Boyd, S., \enquote{{CVX}: Matlab Software for Disciplined Convex
  Programming, version 2.1,} \url{http://cvxr.com/cvx}, Mar. 2014.

\bibitem[{Laub et~al.(1987)Laub, Heath, Paige, and Ward}]{Laub1987}
Laub, A.~J., Heath, M.~T., Paige, C.~C., and Ward, R.~C., \enquote{{Computation
  of System Balancing Transformations and Other Applications of Simultaneous
  Diagonalization Algorithms},} \emph{IEEE Transactions on Automatic Control},
  Vol. AC-32, No.~2, 1987, pp. 115--122.
\newblock \doi{10.1109/TAC.1987.1104549}.

\bibitem[{Boyd(2000)}]{JPBoyd2000}
Boyd, J.~P., \emph{{Chebyshev and Fourier Spectral Methods}}, Dover
  Publications, 2000.
\newblock \doi{10.1007/978-0-387-77674-3}.

\bibitem[{McKernan et~al.(2006)McKernan, Papadakis, and
  Whidborne}]{McKernan2006}
McKernan, J., Papadakis, G., and Whidborne, J.~F., \enquote{{A linear
  state-space representation of plane Poiseuille flow for control design: a
  tutorial},} \emph{International Journal of Modelling, Identification and
  Control}, Vol.~1, No.~4, 2006, p. 272.
\newblock \doi{10.1504/IJMIC.2006.012615}.

\bibitem[{McKernan(2006)}]{McKernanPhD2006}
McKernan, J., \enquote{Control of Plane Poiseuille Flow: A Theoretical and
  Computational Investigation,} Ph.D. thesis, Cranfield University, 2006.

\bibitem[{Skogestad and Postlethwaite(2005)}]{Skogestad2005}
Skogestad, S., and Postlethwaite, I., \emph{{Multivariable feedback control :
  analysis and design}}, John Wiley, 2005.

\bibitem[{Whidborne and Amar(2011)}]{Whidborne2011}
Whidborne, J.~F., and Amar, N., \enquote{{Computing the maximum transient
  energy growth},} \emph{BIT Numerical Mathematics}, Vol.~51, No.~2, 2011, pp.
  447--457.
\newblock \doi{10.1007/s10543-011-0326-4}.

\end{thebibliography}

\clearpage
\appendix
\appendixpage
%\appendixtocname
%\section{}

% This appendix aims to convey the fact that ROM methods are not meant to be  ``one approach fits all"---we show that a simple modal truncation may not work for every application---it may often fail on the LMI-based controllers unless we retain carefully selected modes. Therefore certain ROMs perform better in specific applications.
%
% The fundamental principle behind model order reduction methods is to find a low dimensional representation of large systems while capturing essential dynamics.
% %
% To ensure we obtain a beneficial ROM we have to satisfy a few properties; namely, the approximation error is small,  preserving system properties like stability, and the procedure is computationally stable and efficient~\citep{Antoulas2006}.
% %
% In this section, we show that modal truncation methods based on speed and relative contribution in terms of control and observability are not ideal for energy based control and for capturing energy in the system.

\section{Limitations of Modal truncation for MTEG-minimizing control}
\label{appendix:modal_truncation}

Modal truncation based on the eigendecomposition of the matrix $A=V\Lambda V^{-1}$ in \eqref{eq:linsys1}  is a common approach for reduced-order modeling,
primarily owing to its simplicity.
Since the eigenvalues $\lambda_i$ in the diagonal matrix $\Lambda$ provide information
about the relative time-scales of the modal response,
a simple strategy for modal truncation is to omit modes using time-scale arguments.
For example, when long term behavior is of interest, then modes with fast decay rates
and/or high oscillation frequencies can be omitted.
ROMs constructed time-scale based modal truncation will not necessarily
capture the input-output dynamics needed for controller synthesis.
Thus, another common approach considers a measure of
modal controllability/observability,
given by the ratio~\cite{Antoulas2006}
\begin{equation}
  \zeta_i = \frac{\|C_i\|\|B_i\|}{|\mathrm{Re}\{\lambda_i\}|},
  \end{equation}
  where $B_i$ and $C_i$ correspond to the $i^{\text{th}}$ row and column of the modal
  representations of the matrices $B$ and $C$ from \eqref{eq:linsys1}, respectively.
Modes with larger values of $\zeta_i$ contribute more to the input-output dynamics and
should be be retained, whereas those with lower values of $\zeta_i$ can be truncated.

\begin{figure}[!htbp]
\centering
\subfloat[$|\mathrm{Re}\{\lambda_i\}|$ modal truncation ($r=50$)]{\includegraphics[width=0.45\textwidth]{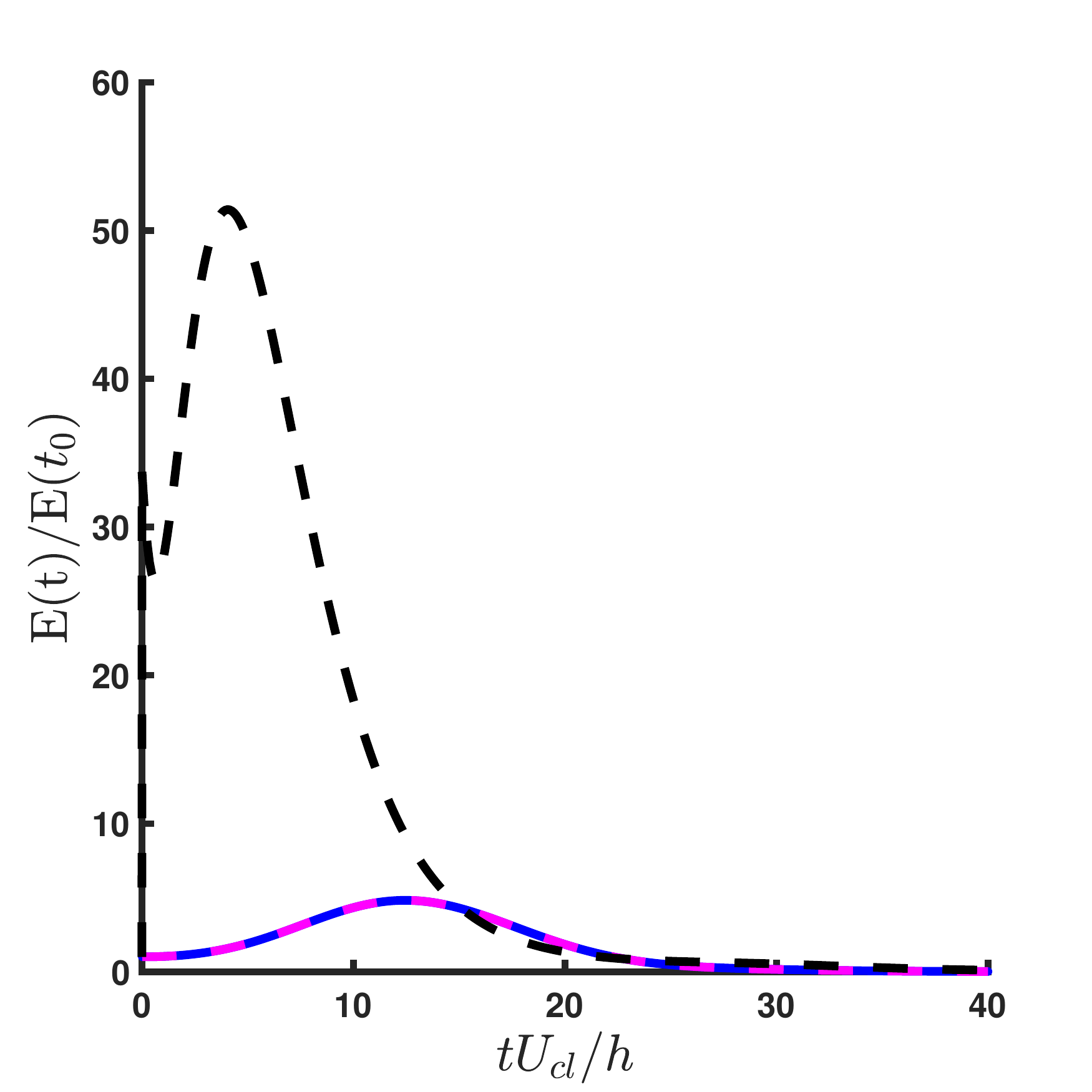}}
\subfloat[$\zeta_i$ modal truncation  ($r=50$)]{\includegraphics[width=0.45\textwidth]{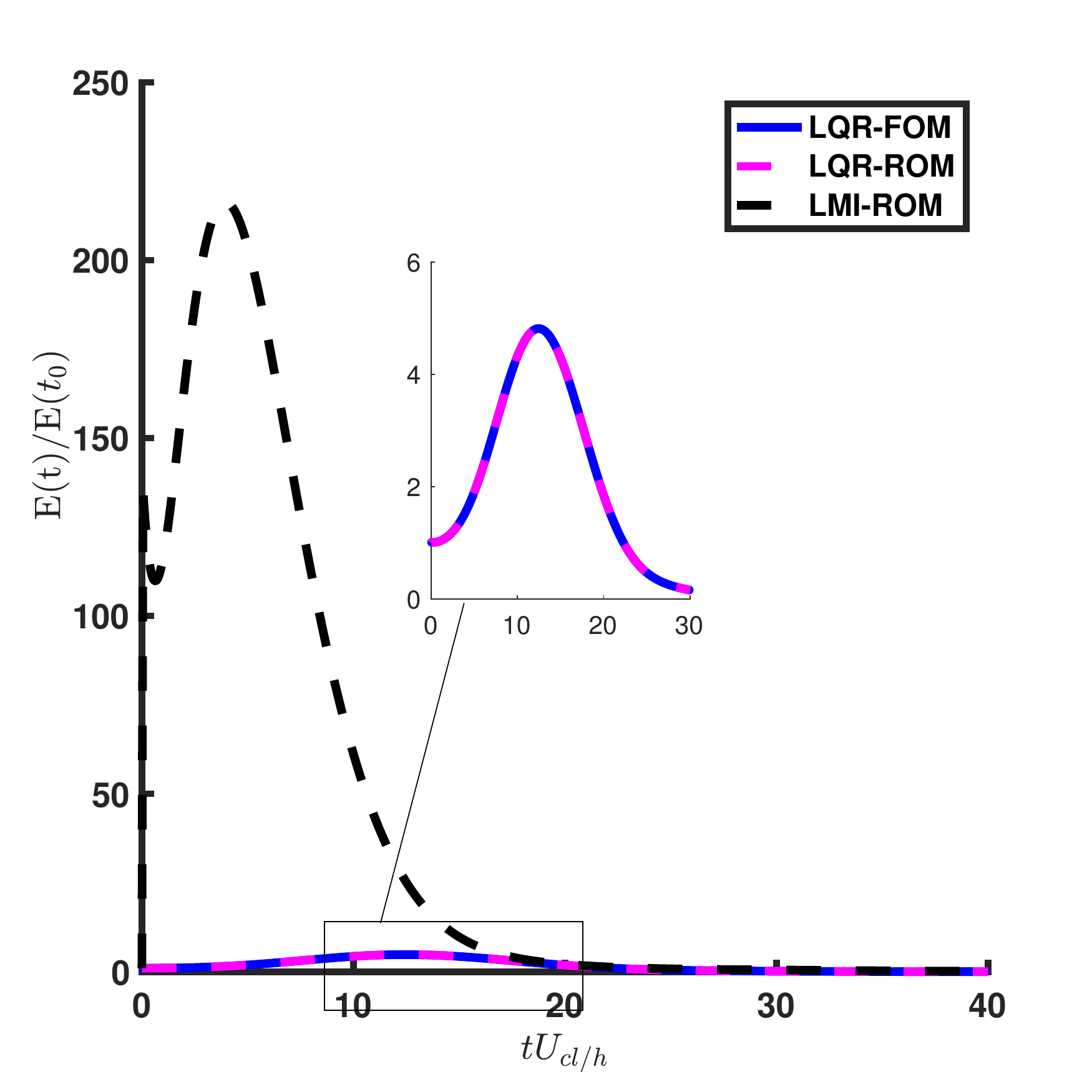}}
\caption{Worst-case closed-loop TEG response for $(\alpha,\beta)=(1,0)$ and $Re=3000$.}
\label{fig:GlobalModes12}
\end{figure}

\begin{figure}[!htbp]
\centering
\subfloat[$|\mathrm{Re}\{\lambda_i\}|$ modal truncation ]{\includegraphics[width=0.45\textwidth]{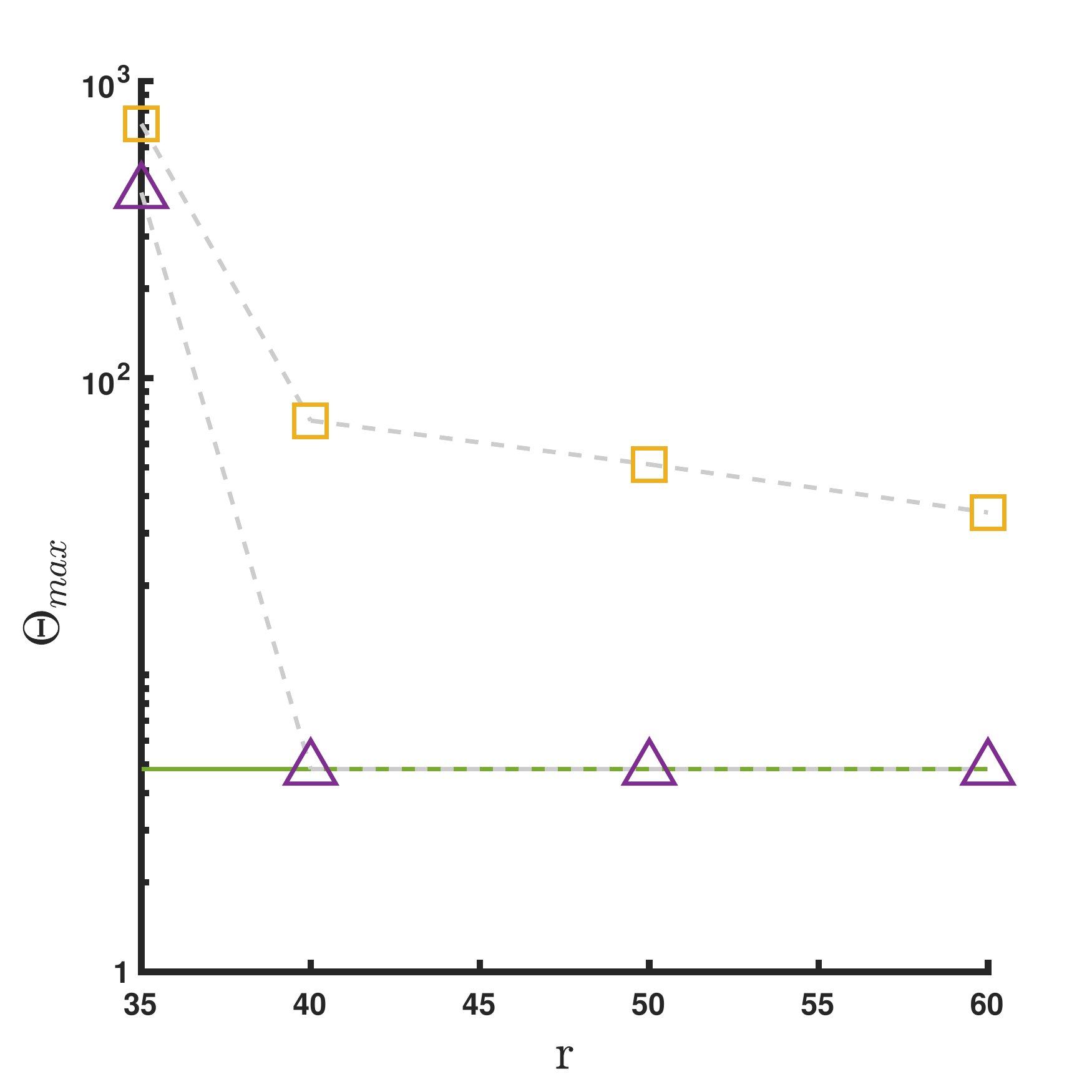}}
\subfloat[$\zeta_i$ modal truncation]{\includegraphics[width=0.45\textwidth]{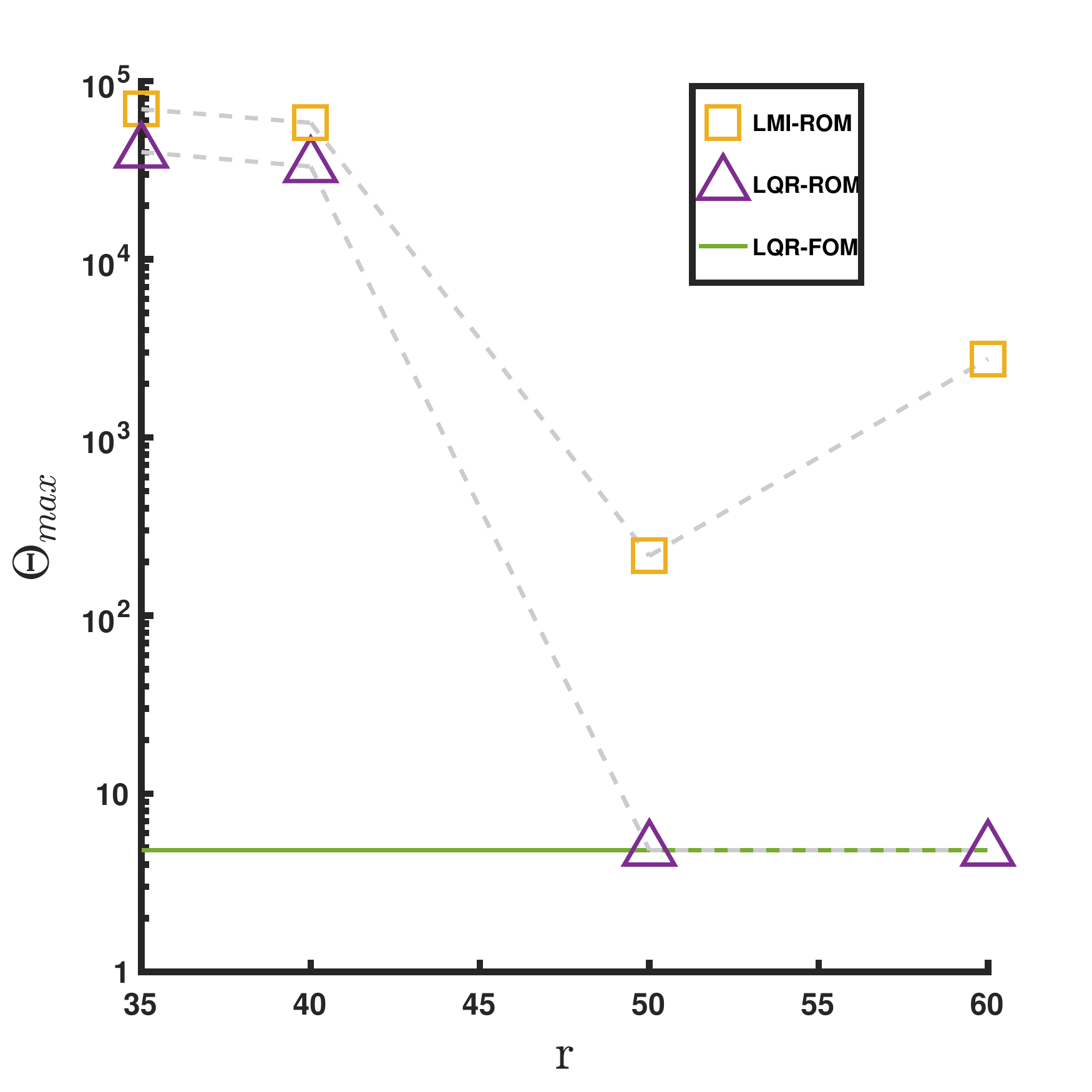}}
  \caption{Closed-loop MTEG $\Theta_{\text{max}}$ as a function of ROM order $r$ for $(\alpha,\beta)=(1,0)$ at $Re=3000$.}
\label{fig:GlobalModes_rstudy}
\end{figure}

Here, we apply both of these modal truncation approaches to arrive at ROMs with
order $r<n$ of the linearized channel flow system with $(\alpha,\beta)=(1,0)$ and $Re=3000$.
For truncation modal truncation by time-scale arguments,
we truncate the ``fastest'' modes (i.e.,~stable modes
with relatively large $|\mathrm{Re}\{\lambda_i\}|$).
For truncation by modal controllability/observability,
we truncate the least controllable/observable modes
(i.e.,~stable modes with relatively small $\zeta_i$).
Each of these ROMs is then used to design the LMI-ROM and LQR-ROM controllers.
The resulting performance of each of these controllers is compared with the LQR-FOM controller
in Figure~\ref{fig:GlobalModes12}.
The model orders here were chosen based on convergence of the ROM open-loop frequency-response relative to the FOM.
We note that for the results reported in the Appendix here, the FOM dimension is set to $n=99$;
without doing so, ROMs based on modal truncation converge with $r>80$, which makes the LMI-based
synthesis procedure intractable with the available computational resources.
Figure~\ref{fig:GlobalModes_rstudy} reports the resulting MTEG $\Theta_{\text{max}}$
as a function of $r$ for ROMs constructed by each of the modal truncation approaches.
Interestingly, the truncation of ``fast modes'' yields improved MTEG performance with the LMI-ROM than
truncation based on $\zeta_i$.
These results indicate that modal truncation is a poor candidate for designing controllers to reduce TEG.
Although in some cases these models can result in effective control laws,
the model orders tend to be higher than the tailored approaches considered in this study.
The short-comings of modal truncation---and other ROM methods---for TEG reduction further motivates the need for
control-oriented model reduction, like the approach presented in Section~\ref{sec:COROM}.

\end{document}